\documentclass{aa} 
\input psfig.tex
\usepackage{graphicx}
\usepackage{natbib}

\usepackage{array}
\usepackage{graphics}
\usepackage{latexsym}
\usepackage{amssymb}
\usepackage{amsmath}
\usepackage{fancyhdr}
\usepackage{morefloats}

\bibpunct{(}{)}{;}{a}{}{,}
\include{hyphe}
\newcommand{\el}[3]{$\mathrm{#1}_{#2}^{#3}$}
\begin{document}
\title{ROBO: a model and a code \\ for studying the interstellar medium}

\author{T. Grassi$^{1}$, P. Krstic$^{2}$, E. Merlin$^{1}$,
U. Buonomo$^{1}$, L. Piovan$^{1}$, C. Chiosi$^{1}$}

\institute{$^1$ Department of Astronomy, Padova University,
  Vicolo dell'Osservatorio 3, I-35122, Padova, Italy\\$^2$ Physics Division, Oak Ridge National Laboratory,
  Oak Ridge, TN 37831-6372, USA\\
  \email{{tommaso.grassi\char64unipd.it  }}}
\date{Received: November 2009; Revised: April 2011; Accepted: May 2011}

\titlerunning{ROBO}
\authorrunning{T. Grassi et al.}
\abstract{We present \textsc{ROBO}, a model and its companion code
for the study of the interstellar medium (ISM). The aim  is to
provide an accurate description of the physical evolution of the ISM
and to set the ground for an ancillary tool  to be inserted in
NBody-Tree-SPH (NB-TSPH) simulations of large-scale structures in the
cosmological context or of the formation and evolution of individual
galaxies. The ISM model  consists of gas and dust. The gas chemical
composition is regulated by a network of reactions that includes a
large number of species (hydrogen and deuterium-based molecules,
helium, and metals). New reaction rates for the charge transfer  in
$\mathrm H^+$ and $\mathrm H_2$ collisions are presented. The dust
contains the standard mixture of carbonaceous grains (graphite
grains and PAHs) and silicates. In our model dust are formed and destroyed by several processes. 
The model accurately treats the cooling process, based on
several physical mechanisms, and cooling functions recently reported
in the literature. The  model is applied to a wide range of the
input parameters and the results for important quantities describing
the physical state of the gas and dust  are presented. The results
are organized in a database suited to the artificial neural networks
(ANNs). Once trained, the ANNs yield  the same results obtained by
ROBO with great accuracy. We plan to develop ANNs suitably tailored
for applications to NB-TSPH simulations of cosmological structures
and/or galaxies.
   }
\maketitle

\section{Introduction}\label{Introduction}
Modeling the gas chemistry is an important step towards
correctly describing the growth of cosmological structures,
the formation and evolution of galaxies, and star formation in general.
For instance, the molecular hydrogen is one of the most
efficient coolants, and its abundance eventually determines the total
amount of stars in the Universe. Structure growth and galaxy
formation and evolution are customarily investigated by means of
large numerical simulations  in which a wide set of
chemical reactions taking place in the ISM should be considered to
get and follow the key molecules (elemental species in general)
eventually governing the efficiency of the star formation and gas
cooling. However, we must face the growing standard
complexity of a typical NB-TSPH model that includes particles of
dark matter, particles of baryonic matter (this in the form of
stars and gas, in turn divided into several thermal and chemical
phases, such as (i) cold, warm and hot, (ii) atomic and molecular, (iii) neutral
and ionized), sources of energy heating and cooling, energy
feedback, and easily many other physical processes. For this reason,
too, a detailed chemical description of the ISM would drastically
reduce the computational performances of any numerical algorithm
(code) that one may adopt to this purpose. This requires a strategy for
optimizing the chemical accuracy of the ISM model and the
computational speed.

In this paper we present a new  model of the ISM  and the associated
code we have developed to  explore the ISM properties over wide
ranges of the physical parameters and, at the same time, to cope with
the above difficulties. The model and companion code are named
\textsc{ROBO}\footnote{The name means ``thing" in some northern Italian
dialects.}.

The model deals with an ideal ISM element of unit volume, containing
gas and dust in arbitrary initial proportions, whose initial
physical conditions are specified by a set of parameters, which is
allowed to evolve for a given time interval. The history leading the
element to that particular initial physical state   is not of
interest here. The ISM element is mechanically isolated from the
host environment; i.e. it does not expand or contract under the
action of large-scale forces. It can, however, be interested by the
passage of shock waves caused by physical phenomena taking place
elsewhere (e.g. supernova explosions). Furthermore, it  neither
acquires nor loses material, so the conservation of total mass
applies, even if its chemical composition can change with time. It
is immersed in a bath of UV radiation generated either by nearby or
internal stellar sources and in a field of cosmic ray radiation. It
can generate its own radiation field by internal processes and so it
has its own temperature, density, and pressure, each related by
an Equation of State (EoS).  If observed from outside, it would
radiate with a certain spectral energy distribution. For the aims of
this study, we do not need to know the whole spectral energy
distribution of the radiation field pervading the element, but only
its UV component. Given these hypotheses and the initial
conditions, the ISM element evolves toward another physical state
under the action of the internal network of chemical reactions
changing the relative abundances of the elemental species and
molecules, the internal  heating and cooling processes, the UV
radiation field, the field of cosmic rays and the passage of shock
waves. In view of the future applications of this model in dynamical
simulations of galaxies, the integration time interval is chosen in
such a way that (i) it is long enough to secure that the
secular evolution of the gas properties is achieved, and (ii) it is
short enough to secure that the physical properties of the ISM at
each instant are nearly independent of the external variations in
the large-scale properties of the system hosting the ISM element. In
other words, \textsc{ROBO} associates a final
state (another point in the same space) to any initial state (a point
in the multidimensional space of the physical parameters) 
through a path. The model is like an operator determining 
the vector field of the local transformations  of the ISM from an
initial state to a final one in the space of the physical parameters. 
This is the greatest merit of this
approach, which secures the wide applicability of the model.

The new ISM model stands on recipes of the internal physical
processes falling  in between those developed by
\citet{GloverJappsen2007} and \citet{Smith2008}. The first one
follows the thermal and chemical evolution of the low-metallicity
gas in large numerical simulations. The chemical network includes a
detailed treatment of H and He but neglects the molecules formed
with heavy elements as the CO molecule. Dust is
included to compute its contribution to the formation of molecular
hydrogen, but its evolution is not calculated. The chemical code
is an on-the-fly routine, running as part of a wider code following
cosmological simulations of structure growth.

The model proposed by \citet{Smith2008} uses the non-equilibrium
treatment  for hydrogen-like species and the standard equilibrium
approximation for all the remaining chemical species. It does not
take any type of dust into account. To calculate the cooling rates,
\citet{Smith2008} use CLOUDY \citep{Ferland98} and get the cooling
rates  as a function of temperature, density, and metallicity. By
doing this, it is possible to include a large chemical network and a
wide set of coolants, but the price to pay is that  several
oversimplifications of the problem are mandatory, e.g. the
assumption of ionization equilibrium.  A similar approach has been
proposed by \citet{Hocuk2010} using the \citet{Meijerink2005} PDR
model instead of  CLOUDY.

Our ISM model and associated code \textsc{ROBO} can not only
describe the  gas evolution in great detail but also includes
large chemical networks and the presence of various types of dust 
which follow the chemistry and the complex interplay between
grain destruction and formation. Many gas and dust components  are
taken into account, among which we recall the molecular hydrogen and
the metal coolants \citep{SantoroShull06,Maio07} or $\mathrm{HD}$
\citep{McGreerBryan08}. To track the evolution  of these components,
the ISM model and \textsc{ROBO} take various physical
processes into account  that may affect the behavior of the whole system. For
instance, dust  is very efficient in forming both $\mathrm H_2$ and
$\mathrm{HD}$ \citep{Cazaux08}. Including dust formation and
destruction  is a formidable task. Among other processes, dust can
be destroyed by  shocks that deserve special care to be properly
modeled. Here we have considered two different approaches. The
first one makes use of a mean shock speed that is assumed to be the
same for all the gas particles, neglecting the effects of the
environment. In reality this is too crude a description. The second
approach starts from the notions that the shocks develop when the
motion of the gas particles becomes turbulent and that the
distribution of turbulent velocities obeys the one predicted by the
Kolmogorov law.  This is significantly better than the previous case,
so it is the approach we prefer.

Finally, we would like to include the gas model (and results of
\textsc{ROBO}) in numerical simulations of galaxies in a simple
way. We suppose  that a numerical code  calculates the
formation and evolution of a model galaxy from an initial stage at
high redshift using the standard  NB-TSPH technique.  At each
time step, it requires an update of the chemical status of the  gas
particles. This is done for every gas particle of the simulation
(typically from $10^4$ up to $10^7$ particles, depending on the
simulation under consideration). We have two methods to our
disposal. The first one is a real-time chemical updater. This
approach must use simple physics and a powerful computer in order to
save computing time. The second one is to use model grids,
calculated in advance for a wide range of the input parameters, in
such a way as to cover the plausible space of the initial conditions.
Since increasing the parameters also increases the space dimensions
of the grids, thus making data handling cumbersome,
we make use of the ANNs technique to get rid of this difficulty.
Once the ANNs are instructed to reproduce the \textsc(ROBO) results
as a function of the parameters, they should replace \textsc{ROBO}
in the NB-TSPH simulator of galaxies. In our case the NB-TSPH model
is \textsc{EvoL} by \citet{MerlinChiosi06, MerlinChiosi07} and
\citet{Merlin10}.  The use of ANNs can greatly improve upon this
point of difficulty. Here we briefly touch upon this problem and
leave the detailed discussion of it to a forthcoming paper
\citep{Grassi11}.

The plan of the paper is as follows. In Section 2 we give a detailed
description of the physics behind the ISM model and \textsc{ROBO}.
This section is divided in three parts: the chemistry of the gas
phase, the presence of dust grains of different types (including their
formation by accretion and destruction), and finally the heating and
cooling processes. In Section 3 we describe the general
characteristics of the code. Section \ref{Results} is dedicated to
describing the results of the calculations run to validate
\textsc{ROBO}. Section \ref{includingROBO} describes how to include
the results of \textsc{ROBO} in the NB-TSPH simulations. Finally,
some concluding remarks are presented in Section \ref{Conclusions}.

\begin{table}
\begin{center}
\caption{Correspondence between elemental species or free particles
and the indices of the differential equations governing the reaction
network. \vspace{1mm} }
\begin{tabular}{|r  l l l  l l l  l| }
\hline
 Ele.s & \el{H}{}{} & \el{H}{}{+}  & \el{H}{}{-} &\el{H}{2}{} & \el{H}{2}{+} &             & \\
Ind.s  & 1          & 2            & 3           & 4          & 5            &             & \\
\hline
 Ele.s & \el{D}{}{} & \el{D}{}{+}  & \el{D}{}{-} &\el{D}{2}{} & \el{HD}{}{} &\el{HD}{}{+} & \\
Ind.s  & 6          & 7            & 8           & 9          & 10           & 11          & \\
\hline
 Ele.s & \el{He}{}{}& \el{He}{}{+} &\el{He}{}{++}&            &              &             & \\
Ind.s  & 12         & 13           & 14          &            &              &             & \\
\hline
 Ele.s & \el{C}{}{} & \el{C}{}{+}  & \el{CH}{}{} &\el{CH}{2}{}& \el{CH}{2}{+}&\el{CH}{3}{+}& \el{CO}{}{} \\
Ind.s  & 15         & 16           & 17          & 18         & 19           & 20          & 21         \\
\hline
 Ele.s & \el{O}{}{} & \el{O}{}{+}  & \el{}{}{}   &\el{}{}{}   & \el{}{}{}    &\el{}{}{}    & \el{}{}{}   \\
Ind.s  & 22         & 23           &             &            &              &             &            \\
\hline
 Ele.s & \el{Si}{}{} & \el{Si}{}{+}& \el{Fe}{}{} &\el{Fe}{}{+}& \el{e}{}{-}   &\el{}{}{ }  & \el{}{}{}   \\
Ind.s  & 24         & 25           & 26          & 27         & 28           &           &            \\

\hline
\end{tabular}
\end{center}
\label{indices}
\end{table}

\section{Physical model of the ISM}\label{Physicalmodel}
In this section we describe the physics of the ISM model  that is
used by \textsc{ROBO}. The problem and associated code are  divided
into three parts, mirrored by the  structure of this section: gas
chemistry, dust, and cooling. All these aspects  are mutually
coupled and overlapped.

\begin{table*}
\begin{center}
\caption{The reaction rates among  hydrogen, deuterium and helium.
References: (a) \citet{GloverSavin09}; (b) \citet{GloverAbel08}.
More details within the text.}
\label{network1} \vspace{1mm}
\begin{tabular*}{165mm}{l l l l}
\hline
    &   Reaction   &Reaction Rate  &  Ref \\
 & &($\mathrm{cm^{3}/s}$)&\\
\hline
 1 & $\mathrm{H_2} + \mathrm{D} \to \mathrm{H} + \mathrm{HD}
$ &
 $\mathrm{if}(T\le2\times10^3\mathrm{K})\ k_1=\mathrm{dex}\left[-56.4737 +5.88886\log(T)\right.$&a\\
 &&$+7.196292\log(T)^2+2.25069\log(T)^3$ &\\
 &&$\left.-2.16903\log(T)^4 +0.317887\log(T)^5\right]$&\\
 &&$\mathrm{if}(T>2\times10^3\ \mathrm{K})\ k_1=3.17\times10^{-10}\exp(-5207/T)$&\\
2 & $\mathrm{HD} + \mathrm{H} \to \mathrm{H_2} + \mathrm{D} $ &
 $\mathrm{if}(T\le200\ \mathrm{K})\ k_2=5.25\times10^{-11}\exp(-4430/T)$ &a\\
 &&$\mathrm{if}(T>200\mathrm{K})\ k_2=5.25\times10^{-11}\exp\left[(-4430/T)+(173900/T^5)\right]$&\\
3 & $\mathrm{D^+} + \mathrm{H_2} \to \mathrm{H^+} + \mathrm{HD} $ &
 $k_3=10^{-9}\left[0.417+0.846\log(T)-0.137\log(T)^2\right]$ &a\\
4 & $\mathrm{H^+} + \mathrm{HD} \to \mathrm{D^+} + \mathrm{H_2} $ &
 $k_4=1.1\times10^{-9}\exp(-488/T)$ &a\\
5 & $\mathrm{H^+} + \mathrm{D} \to \mathrm{D^+} + \mathrm{H} $ &
 $\mathrm{if}(T\le2\times10^{5}\ \mathrm{K})\ k_5=2\times10^{-10} T^{0.402}\exp(-37.1/T)
    -3.31\times10^{-17}T^{1.48}$ &a\\
 &&$\mathrm{if}(T>2\times10^5\mathrm{K})\ k_5=3.44\times10^{-10}T^{0.35}$&\\
6 & $\mathrm{H} + \mathrm{D^+} \to \mathrm{D} + \mathrm{H^+} $ &
 $k_6=2.06\times10^{-10}T^{0.396}\exp(-33/T) +2.03\times10^{-9}T^{-.332}$&a\\
7 & $\mathrm{HD} + \mathrm{D^+} \to \mathrm{D_2} + \mathrm{H^+} $ & $k_7=10^{-9}$ &a\\
8 & $\mathrm{H^+} + \mathrm{D_2} \to \mathrm{D^+} + \mathrm{HD} $ &
 $k_8=2.1\times10^{-9}\exp(-491/T)$&a\\
9 & $\mathrm{H_2} + \mathrm{H^+} \to \mathrm{H_2^+} + \mathrm{H} $ &
$k_9= 10^ A$  with $ A= \sum_{i=0}^7 a_i log(T)^i$, with $a_i$ as in Table \ref{fitkrstic}   &\\
10& $\mathrm{H} + \mathrm{H^+} \to \mathrm{H_2+} + \mathrm{\gamma} $
&
 $k_{10}=\mathrm{dex}\left[-19.38 -1.523\log(T) +1.118\log(T)^2 -0.1269\log(T)^3\right]$ &a\\
11& $\mathrm{H} + \mathrm{H_2^+} \to \mathrm{H_2} + \mathrm{H^+} $ & $k_{11}=6.4\times10^{-10}$&a\\
12& $\mathrm{H} + \mathrm{HD^+} \to \mathrm{H_2} + \mathrm{D^+} $ & $k_{12}=10^{-9}$ &a\\
13& $\mathrm{H^+} + \mathrm{e^-} \to \mathrm{H} + \mathrm{\gamma} $ &
 $k_{13}=2.753\times10^{-14}(315614/T)^{1.5}[1+(1115188/T)^{0.407}]^{-2.242}$&a\\
14& $\mathrm{D^+} + \mathrm{e^-} \to \mathrm{D} + \mathrm{\gamma} $ & $k_{14}=k_{13}$ &a\\
15& $\mathrm{H_2^+} + \mathrm{e^-} \to \mathrm{2H}  $ &$\mathrm{if}(T\le617\ \mathrm K)\ k_{15}=10^{-8}$&a\\
 &&$\mathrm{if}(T>617\ \mathrm K)\ k_{15}=1.32\times10^{-6}T^{-0.76}$&\\
16& $\mathrm{HD^+} + \mathrm{e^-} \to \mathrm{H} + \mathrm{D} $ &$k_{16}=7.2\times10^{-8} T^{-0.5}$ &a\\
17& $\mathrm{H} + \mathrm{e^-} \to \mathrm{H^-} + \mathrm{\gamma} $ &
 $k_{17}=\mathrm{dex}(-17.845 +0.762\log(T) +0.1523\log(T)^2 -0.03274\log(T)^3)$&a\\
18& $\mathrm{D} + \mathrm{e^-} \to \mathrm{D^-} + \mathrm{\gamma} $ & $k_{18}=k_{17}$ &a\\
19& $\mathrm{H} + \mathrm{e^-} \to \mathrm{H^+} + \mathrm{2e^-} $ &
 $k_{19}=\exp\left[-32.71396786 +13.5365560\ln(T_e) -5.73932875\ln(T_e)^2\right.$ &a\\
 &&$+1.56315498\ln(T_e)^3-0.287705600\ln(T_e)^4+3.48255977\times10^{-2}\ln(T_e)^5$&\\
 &&$-2.63197617\times10^{-3}\ln(T_e)^6+1.11954395\times10^{-4}\ln(T_e)^7$&\\
 &&$\left.-2.03914985\times10^{-6}\ln(T_e)^8\right]$&\\
20& $\mathrm{D} + \mathrm{e^-} \to \mathrm{D^+} + \mathrm{2e^-} $ & $k_{20}=k_{19}$ &a\\
21& $\mathrm{H^-} + \mathrm{H} \to \mathrm{H_2} + \mathrm{e^-} $ & $k_{21}=10^{-9}\xi$, $\xi$=[0.65, 5.0], &b\\
22& $\mathrm{D^-} + \mathrm{H} \to \mathrm{HD} + \mathrm{e^-} $ & $k_{22}= 10^{-9}\xi/2$, $\xi$=[0.65, 5.0]&b\\
23& $\mathrm{H^-} + \mathrm{D} \to \mathrm{HD} + \mathrm{e^-} $ & $k_{23}=k_{22}$&b\\
24& $\mathrm{D^-} + \mathrm{D} \to \mathrm{D_2} + \mathrm{e^-} $ & $k_{24}=k_{21}$&b\\
25& $\mathrm{H^+} + \mathrm{D^-} \to \mathrm{HD^+} + \mathrm{e^-} $ &
 $k_{25}=1.1\times10^{-9} (T/300)^{-0.41}$&a\\
26& $\mathrm{D^+} + \mathrm{H^-} \to \mathrm{HD^+} + \mathrm{e^-} $ & $k_{26}=k_{25}$ &a\\
27& $\mathrm{H^-} + \mathrm{e^-} \to \mathrm{H} + \mathrm{2e^-} $ &
 $k_{27}=\exp\left[-18.01849334 +2.36085220\ln(T_e) -0.282744300\ln(T_e)^2\right.$ &b\\
 &&$+1.62331664\times10^{-2}\ln(T_e)^3-3.36501203\times10^{-2}\ln(T_e)^4$&\\
 &&$+1.17832978\times10^{-2}\ln(T_e)^5-1.65619470\times10^{-3}\ln(T_e)^6$&\\
 &&$\left.+1.06827520\times10^{-4}\ln(T_e)^7 -2.63128581\times10^{-6}\ln(T_e)^8\right]$&\\
\hline
\end{tabular*}
\end{center}
\end{table*}

\begin{table*}
\begin{center}
\caption{Continuation of Table \ref{network1}.
References: (a) \citet{GloverSavin09}; (b) \citet{GloverAbel08}; (c)
\citet{Abel97}.} \label{network2} \vspace{1mm}
\begin{tabular*}{165mm}{l l l l}
\hline
 &Reaction&Reaction Rate &Ref \\
 & & (cm$^3$/s)& \\
\hline
28& $\mathrm{H^-} + \mathrm{H} \to \mathrm{2H} + \mathrm{e^-} $ &
 $\mathrm{if}(T_e\le0.1\ \mathrm{eV})\ k_{28}=2.5634\times10^{-9}T_e^{1.78186}$ &b\\
 &&$\mathrm{if}(T_e>0.1\ \mathrm{eV})\ k_{28}= \exp\left[-20.372609 +1.13944933\ln(T_e)\right.$ &\\
 &&$-0.14210135\ln(T_e)^2+8.4644554\times10^{-3}\ln(T_e)^3$&\\
 &&$-1.4327641\times10^{-3}\ln(T_e)^4+2.0122503\times10^{-4}\ln(T_e)^5$&\\
 &&$+8.6639632\times10^{-5}\ln(T_e)^6-2.5850097\times10^{-5}\ln(T_e)^7$&\\
 &&$\left.+2.4555012\times10^{-6}\ln(T_e)^8 -8.0683825\times10^{-9}\ln(T_e)^9\right]$ &\\
29& $\mathrm{H^-} + \mathrm{H^+} \to \mathrm{2H} $ &
 $\mathrm{if}(T\le10^4\ \mathrm{K})\ k_{29}=2.4\times10^{-6}T^{-0.5}(1+2\times10^{-4}T)$& b\\
30& $\mathrm{H_2^+} + \mathrm{H^-} \to \mathrm{H_2} + \mathrm{H} $ & $k_{30}=5\times10^{-7}\sqrt{100/T}$&b\\

31& $\mathrm{H^-} + \mathrm{H^+} \to \mathrm{H_2^+} + \mathrm{e^-}$ &
 $\mathrm{if}(T_e\ge1.719\ \mathrm{eV})\ k_{31}=8.4258\times10^{-10}T_e^{-1.4}\exp(-1.301/T_e)$ &b\\
 &&$\mathrm{if}(T_e<1.719\ \mathrm{eV})\ k_{31}=2.291\times10^{-10}T_e^{-0.4}$&\\
32& $\mathrm{H_2} + \mathrm{e^-} \to \mathrm{2H} + \mathrm{e^-} $ &
 $k_{32}=5.6\times10^{-11}\sqrt{T}\exp(-102124/T)$ &b\\
33& $\mathrm{H_2} + \mathrm{H} \to \mathrm{3H}$ &
 $k_{33}=1.067\times10^{-10}T_e^{2.012}\exp\left[-(4.463/T_e)(1+0.2472\,T_e)^{3.512}\right]$ &b\\
\hline

34& $\mathrm{He} + \mathrm{e^-} \to \mathrm{He^+} + \mathrm{2e^-} $
&
 $k_{34}=\exp\left[-44.09864886 +23.91596563\ln(T_e) -10.7532302\ln(T_e)^2\right. $ &b\\
    &&$+3.05803875\ln(T_e)^3-0.56851189\ln(T_e)^4 +6.79539123\times10^{-2}\ln(T_e)^5$&\\
    &&$-5.00905610\times10^{-3}\ln(T_e)^6+2.06723616\times10^{-4}\ln(T_e)^7 $&\\
    &&$\left.-3.64916141\times10^{-6}\ln(T_e)^8\right]$&\\
35& $\mathrm{He^+} + \mathrm{e^-} \to \mathrm{He} + \gamma $ & $k_{35} = k_{35}^d + k_{35}^e$ &c\\
  &                     &$k_{35}^\mathrm{d}=1.544\times10^{-9}T_e^{-1.5}\exp(-48.596/T_e)
                                \times \left[0.3+\exp(-8.1/T_e)\right]$ & \\
  &                     &$
  k_{35}^\mathrm{e}=3.925\times10^{-13}T_e^{-0.6353}$\\
36& $\mathrm{He^+}+\mathrm{e^-} \to \mathrm{He^{++}}+\mathrm{2e^-} $ &
 $k_{36}=\exp\left[-68.71040990 +43.93347633\ln(T_e) -18.4806699\ln(T_e)^2\right.$ &b\\
    &&$+4.70162649\ln(T_e)^3-0.76924663\ln(T_e)^4 +8.113042\times10^{-2}\ln(T_e)^5$&\\
    &&$-5.32402063\times10^{-3}\ln(T_e)^6+1.97570531\times10^{-4}\ln(T_e)^7 $&\\
    &&$\left.-3.16558106\times10^{-6}\ln(T_e)^8\right]$&\\
37& $\mathrm{He^{++}}+\mathrm{e^-} \to \mathrm{He^+}+\gamma $ &
 $k_{37}=3.36\times10^{-10}T^{-0.5}(T/1000)^{-0.2}\left[1+(T/10^6)^{0.7}\right]^{-1}$&b\\
\hline
\end{tabular*}
\end{center}
\end{table*}

\begin{table*}
\begin{center}
\caption{Reaction rates for the processes where the metals are
involved and where cosmic rays (CR) interact with the ISM. 
References: (c)
\citet{VernerFerland96}; (d) \citet{Voronov97}; (e)
\citet{GloverJappsen2007}; (f) \citet{Zhao2004}; (g)
\citet{Walmsley04}. More details within the text.} \label{network3} \vspace{1mm}
\begin{tabular*}{148mm}{l l l l}
\hline
 &Reaction&Reaction Rate  &Ref \\
 & &(cm$^3$/s) &\\
\hline
38& $\mathrm{C^+}+\mathrm{e^-} \to \mathrm{C}+\gamma $ &
 $k_{38}=\Phi_\mathrm{rec}(T,6.85\times10^{-8},0,11.3,0.193,0.25)$&c\\
39& $\mathrm{C}+\mathrm{e^-} \to \mathrm{C^+}+\mathrm{2e^-} $ &
 $k_{39}=\Phi_\mathrm{col}(T_e,6.556\times10^{-10},65.23,2.446\times10^7,0.7567)$ &d\\
40& $\mathrm{Si^+}+\mathrm{e^-} \to \mathrm{Si}+\gamma $ &
 $k_{40}=\Phi_\mathrm{rec}(T,3.59\times10^{-8},0,13.6,0.073,0.34)$ &c\\
41& $\mathrm{Si}+\mathrm{e^-} \to \mathrm{Si^+}+\mathrm{2e^-} $ &
 $k_{41}=\Phi_\mathrm{col}(T_e,8.616\times10^{-10},119.1,4.352\times10^7,0.7563)$ &d\\
42& $\mathrm{O^+}+\mathrm{e^-} \to \mathrm{O}+\gamma $ &
 $k_{42}=\Phi_\mathrm{rec}(T,1.88\times10^{-7},1,8.2,0.376,0.25)$ &c\\
43& $\mathrm{O}+\mathrm{e^-} \to \mathrm{O^+}+\mathrm{2e^-} $ &
 $k_{43}=\Phi_\mathrm{col}(T_e,1.517\times10^{-9},360.1,1.329\times10^8,0.7574)$ &d\\
44& $\mathrm{Fe^+}+\mathrm{e^-} \to \mathrm{Fe}+\gamma $ &
 $k_{44}=\Phi_\mathrm{rec}(T,2.52\times10^{-7},0,7.9,0.701,0.25)$&c\\
45& $\mathrm{Fe}+\mathrm{e^-} \to \mathrm{Fe^+}+\mathrm{2e^-} $ &
 $k_{45}=\Phi_\mathrm{col}(T_e,2.735\times10^{-9},1314.0,4.659\times10^8,0.7568)$ &d\\
\hline

46& $\mathrm{O^+}+\mathrm{H} \to \mathrm{O}+\mathrm{H^+} $ &
 $k_{46}=4.99\times10^{-11}T^{0.405}+7.54\times10^{-10}T^{-0.458}$ &e\\
47& $\mathrm{O}+\mathrm{H^+} \to \mathrm{O^+}+\mathrm{H} $ &
 $k_{47}=\left[1.08\times10^{-11}T^{0.517}+4\times10^{-10}T^{0.00669}\right]\exp(-227/T)$ &e\\
48& $\mathrm{O}+\mathrm{He^+} \to \mathrm{O^+}+\mathrm{He} $ &
 $\mathrm{if}(T<6000\ \mathrm{K})\
    k_{48}=4.991\times10^{-15}\left(\frac{T}{10^4}\right)^{0.3794}
    \exp\left(-\frac{T}{1.121\times10^6}\right)$&e\\
 &&$\mathrm{if}(T\ge6000\ \mathrm{K})\
    k_{48}=2.780\times10^{-15}\left(\frac{T}{10^4}\right)^{-0.2163}
    \exp\left(\frac{T}{8.158\times10^5}\right)$&f\\
49& $\mathrm{C}+\mathrm{H^+} \to \mathrm{C^+}+\mathrm{H} $ &
 $k_{49}=3.9\times10^{-16}T^{0.213}$&e\\
50& $\mathrm{C^+}+\mathrm{H} \to \mathrm{C}+\mathrm{H^+} $ &
 $k_{50}=6.08\times10^{-14}\left(\frac{T}{10^4}^{1.96}\right) \exp\left(-\frac{1.7\times10^5}{T}\right)$&e\\
51& $\mathrm{C}+\mathrm{He^+} \to \mathrm{C^+}+\mathrm{He} $ &
 $\mathrm{if}(T\le200\ \mathrm{K})\ k_{51}=8.58\times10^{-17}T^{0.757}$&e\\
 &&$\mathrm{if}(200<T\le2000\ \mathrm{K})\ k_{51}=3.25\times10^{-17}T^{0.968}$&\\
 &&$\mathrm{if}(T>2000\ \mathrm{K})\ k_{51}=2.77\times10^{-19}T^{1.597}$&\\
52& $\mathrm{Si}+\mathrm{H^+} \to \mathrm{Si^+}+\mathrm{H} $ &
 $\mathrm{if}(T\le10^4\ \mathrm{K})\ k_{52}=5.88\times10^{-13}T^{0.848}$&e\\
 &&$\mathrm{if}(T>10^4\ \mathrm{K})\ k_{52}=1.45\times10^{-13}T$&\\
53& $\mathrm{Si}+\mathrm{He^+} \to \mathrm{Si^+}+\mathrm{He} $ &  $k_{53}=3.3\times10^{-9}$&e\\
54& $\mathrm{Si}+\mathrm{C^+} \to \mathrm{Si^+}+\mathrm{C} $ &  $k_{54}=2.1\times10^{-9}$&e\\

\hline
55& $\mathrm{H}+\mathrm{CR} \to \mathrm{H^+}+\mathrm{e^-} $ & $k_{56}=0.46\,\zeta_\mathrm{CR}$ &g\\
56& $\mathrm{H_2}+\mathrm{CR} \to \mathrm{2H}$ & $k_{57}=1.50\,\zeta_\mathrm{CR}$&g\\
57& $\mathrm{H_2}+\mathrm{CR} \to \mathrm{H_2^+}+\mathrm{e^-} $ &$k_{58}=0.96\,\zeta_\mathrm{CR}$ &g\\

\hline
\end{tabular*}
\end{center}
\end{table*}

\subsection{The chemical network for the ISM}\label{Gaschemistry}

The kernel of the numerical code modelling the gas
chemistry is the network of chemical reactions among different
elemental species (atoms and molecules, neutral or ionized dust
grains of different types) and free particles (such as the
electrons), the network of photochemical reactions between the
above elemental species, and the radiation field. In
addition to this, we must include all the processes creating and
destroying molecules and dust grains with particular attention to
those that are most efficient for cooling the gas.
For this reason our chemical network follows species like
$\mathrm H_2$, $\mathrm{HD}$, and metals (C, O, Si, Fe, and their ions).

The number of reactions depends on how many species are tracked and
how many details are included in their description. A number  of
codes study the behavior of the ISM
\citep{Cen92,Katz96,GalliPalla98,Anninos97,GloverSavin09}, each of
which has a different number of species to follow and a different
degree of sophistication for the physical processes taken into
consideration.

We keep track of the following 27 elemental species or molecules
plus the free electrons: \el{H}{}{}, \el{H}{}{+}, \el{H}{}{-},
\el{H}{2}{}, \el{H}{2}{+},
 \el{D}{}{}, \el{D}{}{+}, \el{D}{}{-}, \el{D}{2}{},
\el{HD}{}{}, \el{HD}{}{+},
 \el{He}{}{}, \el{He}{}{+}, \el{He}{}{++},
\el{C}{}{}, \el{C}{}{+}, \el{CH}{}{}, \el{CH}{2}{}, \el{CH}{2}{+},
\el{CH}{3}{+}, \el{CO}{}{}, \el{O}{}{}, \el{O}{}{+}, \el{Si}{}{},
\el{Si}{}{+}, \el{Fe}{}{}, \el{Fe}{}{+}, and \el{e}{}{-}.

These species are divided into four groups.  The first one contains
hydrogen-only based species. The second group is composed of
deuterium-based species, in which $\mathrm{HD}$ plays the key role
in the gas cooling. The third group lists helium and its ions.
Carbon, oxygen, silicon, and iron with their ions and compounds form
the fourth group. The free  electrons  link all the four groups
together. The species from \el{CH}{}{} through \el{CH}{3}{+} are
introduced to follow the formation and destruction of \el{CO}{}{} to
be described below.

The reactions in which all the above species are involved are
\begin{itemize}
  \item collisional ionization
    ($\mathrm A +\mathrm e^- \to \mathrm A^+ + 2\mathrm e^-$),
  \item photo-recombination
    ($\mathrm A^+ +\mathrm e^- \to \mathrm A + \gamma$),
  \item dissociative recombination
    ($\mathrm A^+_2 +\mathrm e^- \to 2\mathrm A$),
  \item charge transfer
    ($\mathrm A^+ +\mathrm B \to \mathrm A +\mathrm B^+$),
  \item radiative attachment
    ($\mathrm A +\mathrm e^- \to \mathrm A^- +\gamma$),
  \item dissociative attachment
    ($\mathrm A +\mathrm B^- \to \mathrm{AB} +\mathrm e^-$),
  \item collisional detachment
    ($\mathrm A^- +\mathrm e^- \to \mathrm A +2\mathrm e^-$),
  \item mutual neutralization
    ($\mathrm A^+ +\mathrm B^- \to \mathrm A +\mathrm B$),
  \item isotopic exchange
    ($\mathrm A^+_2 +\mathrm B \to \mathrm{AB}^+ +\mathrm A$)
  \item dissociations by cosmic rays
    ($\mathrm{AB} +\mathrm{CR}\to \mathrm A +\mathrm B$),
  \item neutral-neutral
    ($\mathrm{AB} +\mathrm{AB}\to \mathrm A_2+\mathrm B_2$),
  \item ion-neutral
    ($\mathrm{AB}^+ +\mathrm{AB}\to \mathrm{AB}_2^+ +\mathrm A$),
  \item collider
    ($\mathrm{AB} +\mathrm C \to \mathrm A +\mathrm B +\mathrm C$),
    \item
    ionizations by field photons ($\mathrm{A} +\gamma \to \mathrm{A^+} +\mathrm{e^-}$)
\end{itemize}

\noindent where $\mathrm A$ and $\mathrm B$ are two generic atoms.
To these reactions we must add those regulating the abundance of
\el{CO}{}{} to be described below.

The chemical network governing the ISM model is  a classical system
of differential equations in which each equation is a Cauchy problem
of the form

\begin{small}
\begin{equation}\label{cauchy_sys}
    \frac{\mathrm dn_i(t)}{\mathrm dt}=\sum_{lm}R_{lm}(T)n_l(t)n_m(t)-\sum_{j}R_{ij}(T)n_i(t)n_j(t)\,,
\end{equation}
\end{small}

\noindent where $n_i(t)$ is the number density  of the {\it i-th}
species with known initial value $n_i(0)$. In eqn.
(\ref{cauchy_sys}) $R_{lm}(T)$ is the rate of the reaction between
the {\it l-th} and the {\it m-th} species expressed in units of
$\mathrm{cm^3/s}$. Eqn. (\ref{cauchy_sys}) is written as the sum of
all the reactions forming the {\it i-th} species
($\sum_{lm}R_{lm}(T)n_l(t)n_m(t)$) minus the sum of all the
reactions destroying  the {\it i-th} species
($\sum_{j}R_{ij}(T)n_i(t)n_j(t)$). The indices $i, j, l$, and $m$
run from 1 to 28  as in the list of elemental species and free
electrons already mentioned. The one-to-one correspondence
between the indices in the eqn. (\ref{cauchy_sys}) and elemental
species is given in Table \ref{indices}.

To clarify the meaning of eqn. (\ref{cauchy_sys}) we show the case
of a two-reaction system

\begin{eqnarray}
 \mathrm A_0+\mathrm A_1 &\to& \mathrm A_2 + \mathrm A_3\nonumber\\
 \mathrm A_4+\mathrm A_5 &\to& \mathrm A_0 +\mathrm A_3\,,
\end{eqnarray}

\noindent where $A_i$ is the number density of the generic species
in units of $\mathrm cm^{-3}$. Looking at  the density variation of
the species $\mathrm A_0$ over the time step  $\mathrm dt$ the
density $n_{\mathrm A_0}$ changes as
\begin{eqnarray}
\mathrm dn_{\mathrm A_0}(t+\mathrm dt)=&-&k_{\mathrm{01}}(T)n_{\mathrm A_0}(t)n_{\mathrm A_1}(t)\mathrm dt\nonumber\\
&+&k_{\mathrm{45}}(T)n_{\mathrm A_4}(t)n_{\mathrm A_5}(t)\mathrm
dt\,.
\end{eqnarray}
In this equation the positive term is due to the second reaction
(between $\mathrm A_4$ and $\mathrm A_5$) that increases the total
quantity of $\mathrm A_0$ (and also $\mathrm A_3$), while the
negative term is due to the first reaction in which $\mathrm A_0$ is
destroyed because it reacts with $\mathrm A_1$ to form $\mathrm A_2$
and $\mathrm A_3$.

We adopt here a minimal description containing only the $28$ species
and/or the molecular compounds listed in Table \ref{indices}, the $64$
reactions listed in Tables \ref{network1} and \ref{network2} (where
reactions among hydrogen, deuterium, and helium are considered), in
Table \ref{network3} (where metals and cosmic rays are involved),
in Table \ref{TabCO} (where the reactions of the CO mini-network are
listed), and finally, the $12$ photochemical processes listed in
Table \ref{photocs}. In all the tables the gas temperature  $T$
is in Kelvin; the electron temperature $T_e$ is in $\mathrm{eV}$.
In Table \ref{network3} the cosmic ray
field is $\zeta_\mathrm{CR}$   $\mathrm s^{-1}$ and it is the rate
of  $\mathrm H_2$ ionization by cosmic rays. For Table \ref{TabCO} we have  $T_{300}=T/300$.
The expressions in Table \ref{photocs} are $\xi=E/E_{th}$,
$\Phi_\mathrm{ph}(a_f,y_w,x_f,y_a,P)=[(x_f-1)^2+y_w^2]y^{0.5P-5.5}[1+\sqrt{y/y_a}]^{-P}$,
$y=\sqrt{x_f^2+a_f^2}$,  $x_f(a,b)=E/a-b$, and
$x_{Si}=1.672\times10^{-5}$.

The various types of reaction are described
below in some detail. Tables \ref{network1}, \ref{network2},
\ref{network3}, and \ref{TabCO} also sample the chemical reactions
according to the same four groups in which we separated the
elemental species depending on the type of reaction and the
information at our disposal to derive the reactions rates. In doing
this,  the notation in use may appear rather complex. This is done
on purpose, and the main motivation is that we want to keep the same
formalism as adopted in the literature for easy comparison.
First of all,  there are 64 reactions among particles, whether atoms
or molecules or free electrons.  Each reaction is identified by a
progressive number from 1 to 64 and so the associated reaction
rate named $k_n$ (with $n$ from 1 to 64). Each reaction will be a
term in the righthand of the 28 differential equations of
system (\ref{cauchy_sys}) and the associated reaction rate will
coincide with one of the $R_{ij}$ terms of eqns. \ref{cauchy_sys}).
To establish the exact correspondence is a matter of patient work,
which is not the prime interest here.  Most of these reaction rates are
taken from \citet{Abel97} and \citet{GloverSavin09} to whom we refer
for all the details. Some of these reactions are discussed in this
section, and we pay attention to those rates that are vividly
debated in literature.

 \textbf{Minimal reaction network  for CO}. Since this molecule plays
 a key role in determining the properties of the ISM,
we have included  a simplified network of reactions taken from
\citet{NelsonLanger97} to follow the creation/destruction of the CO
in detail. These reactions  are listed in Table \ref{TabCO} together
with the corresponding rates of \citet{Woodall2007}. Note that
$T_{300}=T/300$, with $T$ in K.

\begin{table*}
\begin{center}
\caption{Reaction rates for the reactions belonging to the CO
network. References: (a)
\citet{Woodall2007}; (b) \citet{Petuchowski1989}.See the text for more details.} \label{TabCO}
\vspace{1mm}
\begin{tabular*}{148mm}{l l l l}
\hline
 &Reaction&Reaction Rate  &Ref \\
 & &(cm$^3$/s) &\\
\hline
58. & $\mathrm C^+ + \mathrm H_2$  $\rightarrow$  $\mathrm{CH}_2^+ + \gamma$ & $k_{58}$ = $4\times 10^{-16}\,T_{300}^{-0.2}$ & a\\
59. & $\mathrm{CH}_2^+ + \mathrm e^-$  $\rightarrow$  $\mathrm{CH} +\mathrm H$ & $k_{59}$ = $1.6\times 10^{-7}\,T_{300}^{-0.6}$ & a\\
60. & $\mathrm{CH}_2^+ + \mathrm H_2$  $\rightarrow$  $\mathrm{CH}_3^+ + \mathrm H$ & $k_{60}$ = $1.6\times 10^{-4}$& a\\
61. & $\mathrm{CH}_3^+ + \mathrm e^-$  $\rightarrow$  $\mathrm{CH}_2 + \mathrm H$ &$k_{61}$ = $6.6\times 10^{-11}$& a\\
62. & $\mathrm{CH} + \mathrm O$  $\rightarrow$  $\mathrm{CO} + \mathrm H$ &$k_{62}$ = $7.58\times 10^{-8}\,T_{300}^{-0.5}$& a\\
63. & $\mathrm{CH}_2 + \mathrm O$  $\rightarrow$  $\mathrm{CO} + 2\mathrm H$ &$k_{63}$ = $1.33\times 10^{-10}$& a\\
64. & $\mathrm{CO} + \mathrm e^-$  $\rightarrow$  $\mathrm{C} + \mathrm O +\mathrm e^-$  &$k_{64}$ = $2.86\times 10^{-3}\,T_{300}^{-3.52}\,\exp\left(-112700/T\right)$ & b\\

\hline
\end{tabular*}
\end{center}
\end{table*}

Table \ref{TabCO} shows how CO is formed from $\mathrm C^+$ and O
species, using CH, $\mathrm{CH}_2$, $\mathrm{CH}_2^+$ and
$\mathrm{CH}_3^+$ as intermediate products/reactants, and $\mathrm
H_2$ and $\mathrm e^-$ as main reactants.

Our model includes UV radiation, so we must consider the
photodissociation of the molecules included in the network. The
main effects of the radiation are on the carbon-based molecules and
the corresponding   photo-ionization rates are
\begin{equation}
    \Gamma_\mathrm{CO}=G_0\,10^{-10}\,\mathrm s^{-1}\,,
\end{equation}
and
\begin{equation}\label{photoCHx}
    \Gamma_\mathrm{CH_x^{(+)}}=G_0\,5\times10^{-10}\,\mathrm s^{-1}\,,
\end{equation}
where  $G_0$ is the ratio of the adopted UV flux  to that by Habing;
i.e. $I_\mathrm{Hab}=1.2\times10^{-4}\,\mathrm{erg/cm^2/s/sr}$. The
expression $\mathrm{CH_x^{(+)}}$ indicates that we are dealing with
the $\mathrm{CH}$, $\mathrm{CH}_2$, $\mathrm{CH}^+_2$, and
$\mathrm{CH}^+_3$ molecules.

Basing on eqn. (\ref{photoCHx}), we must consider the products of
all the $\mathrm{CH_x^{(+)}}$ molecules present in reactions like
$\mathrm{CH_x}^{(+)}+\gamma\rightarrow \mathrm{products}$. However,
at this stage we are not interested in the detail of these reactions;
therefore,  following \citet{NelsonLanger97}, we introduce the
parameter $\beta(G_0)$  controlling the efficiency of the reactions
from 58 through  61 in Table \ref{TabCO}.  This parameter is defined
as $\beta(G_0)=\exp[G_0\cdot\ln(\xi)]$ with $\xi=5\times 10^{-10}$
from eqn. (\ref{photoCHx}). In the absence of the UV radiation ($G_0=0$),
we get $\beta=1$: this means that the existing photons cannot affect
the formation of the various $\mathrm{CH_x^{(+)}}$. Instead, when
$G_0=1$, it follows $\beta=\xi$, which means that the formation
reactions have reduced their efficiency. Finally, when
$G_0\to\infty$, (unphysical) then $\beta\to0$, which means that the
$\mathrm{CH_x^{(+)}}$ molecules are completely destroyed before they
can interact. The coefficient of the $i$-th reaction now is
$k_i'=\beta\cdot k_i$, with $i\in[58,61]$ and eqn. (\ref{photoCHx})
is not included in the code.


\textbf{Photochemistry}. In addition to this, we consider the group
of photochemical processes listed in Table \ref{photocs} for which
we provide the cross section of the reaction  and the analytical
expression to derive the reaction rate as a function of the existing
radiation field.

In the present model of the ISM  we do not include  the chemical
reactions with lithium and its compounds (e.g.  $\mathrm{LiH}$ and
$\mathrm{LiH^+}$) because according to \citet{Prieto08} and
\citet{Mizusawa2005}, they are not important coolants. After we neglecting
lithium and compounds, the number of species to follow, including
metals and deuterium compounds, amounts to the 28 we have
considered.

There is some interest in the $\mathrm{CO}$ molecule because it is
an important coolant at very low temperatures. Among others, recipes
for the formation of the interstellar $\mathrm{CO}$  are presented
by \citet{Ruffle02} and \citet{Glover2009}. However, explicitly
following the formation of the $\mathrm{CO}$ molecule would be too
complicated for our aims.
See below for the cooling rate by  the CO molecule.

In our chemical network we also include the effect of  the cosmic
ray radiation field.  For convenience, we list in Table
\ref{network3} the ionization rates for cosmic rays given by
\citet{Walmsley04}. Cosmic rays are important for the cooling
because they can destroy molecules that contribute to it, thus
influencing the temperature of the interstellar medium (see the
reactions shown in Table \ref{network3}). In particular, cosmic rays
are able to destroy $\mathrm H_2$ and $\mathrm{HD}$ and to ionize
atoms. The role of the cosmic rays is still largely unknown because
the strength of their radiation field in various regions and epochs
of the Universe is not firmly determined. While their effect can be
explored in the vicinity of a strong source, for a random sample of
the Universe there are no exact measurements of the cosmic ray
radiation field. Furthermore, most NB-TSPH simulators like
\textsc{EvoL} do not contain the full description of cosmic rays,
but simply consider their presence as a parameter. Therefore, even
if these reactions have been considered and their rates presented in
Table \ref{network3}. they have not been included in practice in the
system of equations (\ref{cauchy_sys}).

\textsc{Photochemical reactions.} For the photochemical reactions
we adopt the model and rates of \citet{GloverJappsen2007} and
\citet{VernerFerland96}. In general the reaction rate is given by

\begin{equation}\label{ratephot}
 R_\mathrm{photo}=4\pi\int_{E_\mathrm{th}}^\infty \frac{\sigma(E)J(E)}{E}e^{-\tau(E)}[1+f(E)]\mathrm dE\,,
\end{equation}

\noindent where $J(E) = J(h\nu)$ is  the energy flux per unit
frequency and solid angle of the impinging radiation field,
$\sigma(E)$ is the cross-section, $\tau(E)$ the gas opacity at
the energy $E$, and $f(E)$ a numerical factor  accounting  for the
effects of the secondary ionization, which are negligible if the UV
radiation is not dominated by X-rays \citep{GloverJappsen2007}. The
rate is in \el{s}{}{-1}.

 The cross sections $\sigma(E)$ are
expressed in two different ways according to the reaction under
consideration. The reactions and associated rates are listed in
Table \ref{photocs}. For the reactions from 1 through 8 we note
that (i) the quantity $\epsilon = \sqrt{(E/E_\mathrm{th}) -1}$;
(ii) for $\mathrm H_2$ and HD photodissociation, we use the model
proposed by \citet{GloverJappsen2007}. We have

\begin{equation}
 \sigma_\mathrm{H_2}=1.38\times10^9J(h\bar\nu)\,\\
\end{equation}
 without considering self-shielding.
For the HD we have
\begin{equation}
 \sigma_\mathrm{HD}=1.5\times10^9 J(h\bar\nu)\,,\\
\end{equation}
 with $h\bar\nu=12.87\ \mathrm{eV}$.

For the remaining reactions involving metals (i.e. from 9 through
12), the rates contain the fits given by \citet{VernerFerland96}
using the expression

\begin{eqnarray}\label{phi_ph}
 \Phi_\mathrm{ph}(a_f,y_w,x_f,y_a,P)&=&\left[(x_f-1)^2+y_w^2\right]\nonumber\\
 &\times& y^{0.5P-5.5}\left[1+\sqrt{\frac{y}{y_a}}\right]^{-P}\,,
\end{eqnarray}
where $y=\sqrt{x_f^2+a_f^2}$. The meaning of the various symbols is
given in  Table \ref{photocs}. More details on the reactions and
companion quantities and symbols are given by
\citet{GloverJappsen2007} and \citet{VernerFerland96}. See also the
references therein.

The UV radiation field is calculated  as in \citet{Efstathiou92},
\citet{Vedel94}, and \citet{NavarroSteinmetz97}. In particular we
have

\begin{equation}\label{Jnu}
 J(\nu)=10^{-21}J_{21}(z)\left(\frac{\nu_\mathrm{H}}{\nu}\right)^{\alpha_\mathrm{UV}}\,,
\end{equation}
where $z$ is the redshift, $\nu_\mathrm{H}= E_\mathrm{H}/h$ is the
frequency corresponding to the hydrogen first-level energy
threshold, $\alpha_\mathrm{UV}=1$ and $J_{21}(z)$ is

\begin{equation}
 J_{21}(z)=\frac{J_{21}}{1+[5/(1+z)]^4}\,,
\end{equation}
where in our case $J_{21}=1$.
Finally, to derive the reaction rates, we integrate eqn.
(\ref{ratephot}) from $E=E_\mathrm{th}$ to $E=\infty$ over
the energy range of the radiation field photons.

The integrals must be calculated at each time step if the radiation
field changes with time or once for all at the beginning of the
simulation if the radiation field   remains constant. The integrals
are calculated  using Romberg's integration method.

\begin{table*}
\begin{center}
\caption{Cross sections of photochemical processes in
$\mathrm{cm}^2$. The energy $E$ is in $\mathrm{eV}$. The various
quantities in use are desctribed within the text. References: (a)
\citet{GloverJappsen2007}; (b) \citet{VernerFerland96}. See these
papers for further references.} \label{photocs} \vspace{1mm}
\begin{tabular*}{175mm}{r l l l l l}
\hline
 & &Reaction&Cross Section &Note&Ref \\
& &(cm$^3$/s) & \\
\hline
1 &&$\mathrm H + \gamma\to\mathrm H^+ + \mathrm e^-$ &
$\sigma_1=6.3\times 10^{-18}\xi^4
    \exp(4-4\epsilon^{-1}-\arctan \varepsilon) \left[1-\exp\left(\frac{-2\pi}{\epsilon}\right)\right]^{-1}$&
    $E_{th}=13.6$ & a\\
2 & &$\mathrm D + \gamma\to\mathrm D^+ + \mathrm e^-$ & $\sigma_2=\sigma_1$& $E_{th}=13.6$ &  a\\

3 & &$\mathrm{He} + \gamma\to\mathrm{He}^+ + \mathrm e^-$ &
$\sigma_3=3.1451\times 10^{-16} \xi^{7/2}
    \left[1.0 -4.7416\,\xi^{1/2} +14.82\,\xi \right.$ & $E_{th}=24.6$ &  a\\
&&&$\left.-30.8678\,\xi^{3/2}+37.3584\,\xi^2 -23.4585\,\xi^{5/2} +5.9133\,\xi^3\right]$&\\

4 & &$\mathrm{H^-} + \gamma\to\mathrm{H} + \mathrm e^-$ &
$\sigma_4=2.11\times 10^{-16}(E-E_{th})^{3/2}E^{-3}$&
    $E_{th}=0.755$ &   a\\

5 & &$\mathrm{H_2} + \gamma\to\mathrm{H} + \mathrm H$ & see text&&\\

6 & &$\mathrm{H_2^+} + \gamma\to\mathrm{H} + \mathrm H^+$ & if $(2.65<E<11.27)\,\sigma_5=\mathrm{dex}\left[-40.97 +15.9795\,\xi \right.$ &$E_{th}=2.65$ &a \\
&&&$\left.-3.53934\,\xi^2+0.2581155\,\xi^3\right]$&&\\
&&&if $(11.27<E<21)\,\sigma_5=\mathrm{dex}\left[-30.26 +7.3935\,\xi -1.29214\,\xi^2\right.$&\\
&&&$\left.+0.065785\,\xi^3\right]$&\\

7 & &$\mathrm{H_2} + \gamma\to\mathrm{H_2^+} + \mathrm e^-$ &if
$(15.4<E<16.5)\,\sigma_6=9.560\times10^{-17}\xi
-9.4\times10^{-17}$ & $E_{th}=15.4$ & a\\
&&&if $(16.5<E<17.7)\,\sigma_6=2.16\times10^{-17}\xi -1.48\times10^{-17}$&&\\
&&&if $(17.7<E<30.0)\,\sigma_6=1.51\times10^{-17}\xi -2.71$&& \\

8 & &$\mathrm{HD} + \gamma\to\mathrm{H} + \mathrm D$ & see text&&\\

9 & &$\mathrm{C} + \gamma\to\mathrm{C^+} + \mathrm e^-$ &$\sigma_7=5.027\times10^{-16}\Phi_\mathrm{ph}(1.607,0.09157,x_f,62.16,5.101)$&$E_{th}=11.26$ &a;b\\
&&&&$x_f(2.144,1.133)$&\\

10 & &$\mathrm{O} + \gamma\to\mathrm{O^+} + \mathrm e^-$ &$\sigma_8=1.745\times10^{-15}\Phi_\mathrm{ph}(0.1271,0.07589,x_f,3.784,17.64)$&$E_{th}=13.62$& a;b\\
&&&&$x_f(1.240,8.698)$&\\

11 & &$\mathrm{Si} + \gamma\to\mathrm{Si^+} + \mathrm e^-$ &$\sigma_9=2.506\times10^{-17}\Phi_\mathrm{ph}(0.4207,0.2837,x_f,20.57,3.546)$&$E_{th}=8.152$& a;b\\
&&&&$x_f(23.17,x_{Si})$&\\

12& &$\mathrm{Fe} + \gamma\to\mathrm{Fe^+} + \mathrm e^-$ &$\sigma_{10}=3.062\times10^{-19}\Phi_\mathrm{ph}(0.2481,20.69,x_f,2.671\times10^7,7.923)$&$E_{th}=7.902$& b\\
&&&&$x_f(0.05461,138.2)$&\\
\hline
\end{tabular*}
\end{center}
\end{table*}

\textsc{Heating by photodissociation.} Heating by
photodissociation of molecular hydrogen and UV pumping, H and He
photo-ionization, $\mathrm H_2$ formation in the gas and dust phase,
and finally ionization from cosmic rays are modeled as in
\citet{GloverJappsen2007}. All these processes are listed in Table
\ref{tabheat} where $n_\mathrm{cr}$ is the critical density
and R$_\mathrm{d}$ is the photodissociation rate. To describe the heating by the photoelectric effect
we adopt the model proposed by \citet{BakesTielens94} and discuss it
separately in the Section \ref{dustheat} below.

The H and He photo-ionizations increase the gas energy at the rate
\begin{equation}
 \Gamma=4\pi\int_{E_0}^\infty \frac{\sigma(E)J(E)}{E}e^{-\tau(E)}\left(E-E_0\right)\eta(E-E_0)\mathrm dE\,,
\end{equation}
where $\sigma(E)$ is the reaction cross-section, $J(E)$ is the photon flux, $e^{-\tau(E)}$ is
the optical depth for a photon of energy $E$, $\eta$ is the efficiency of the process (i.e. the
fraction of energy converted to heat), and $(E-E_{th})$ is the difference between the energy of
the photon and the energy of the atomic ground level. The rates $\Gamma$ are given in Table \ref{tabheat}.

\begin{table*}
\begin{center}
\caption{Heating processes.  References: (a)
\citet{GloverJappsen2007}; (b) \citet{VernerFerland96}. Details within the text.} 
\label{tabheat}                      
\begin{tabular*}{140mm}{ l l l l}
\hline
 &Process&Heating&Ref \\
\hline
 &$\mathrm H_2$ photodissociation & $\Gamma=6.4\times10^{-13} R_\mathrm{d}n_\mathrm{H_2}$ &a\\
 &UV pumping of $\mathrm H_2$ & $\Gamma=2.7\times10^{-11} R_\mathrm{d}n_\mathrm{H_2}
    \left(\frac{n}{n+n_\mathrm{cr}}\right)$&a\\
 &H photo-ionization & see text&a\\
 &He photo-ionization & see text&a\\
 &Gas-phase $\mathrm H_2$ formation &$\Gamma=\left(2.93\times10^{-12}k_{21}n_\mathrm{H^-}
    +5.65\times10^{-12}k_{11}n_\mathrm{H_2^+}\right) \left(\frac{n}{n+n_\mathrm{cr}}\right) $&a\\
 &Dust-phase $\mathrm H_2$ formation &$\Gamma=7.16\times10^{-12}k_\mathrm{dust}nn_\mathrm{H}
    \left(\frac{n}{n+n_\mathrm{cr}}\right)$&a\\
 &Cosmic-ray ionization& $\Gamma=3.2\times10^{-11}\zeta_\mathrm{tot}n$&a\\
 &Photoelectric effect&see Sect. \ref{dustheat}&b\\
 \hline
\end{tabular*}
\end{center}
\end{table*}

\subsection{Reaction by reaction: notes on a few cases}

In this  section we examine in some detail a few chemical reactions
that are widely debated and are affected by large
uncertainties. The reaction numbers are the same as those used in
Tables \ref{network1}, \ref{network2}, \ref{network3}, and
\ref{photocs} for the sake of an easy identification.

\begin{figure}
\begin{center}
\includegraphics[width=.45\textwidth]{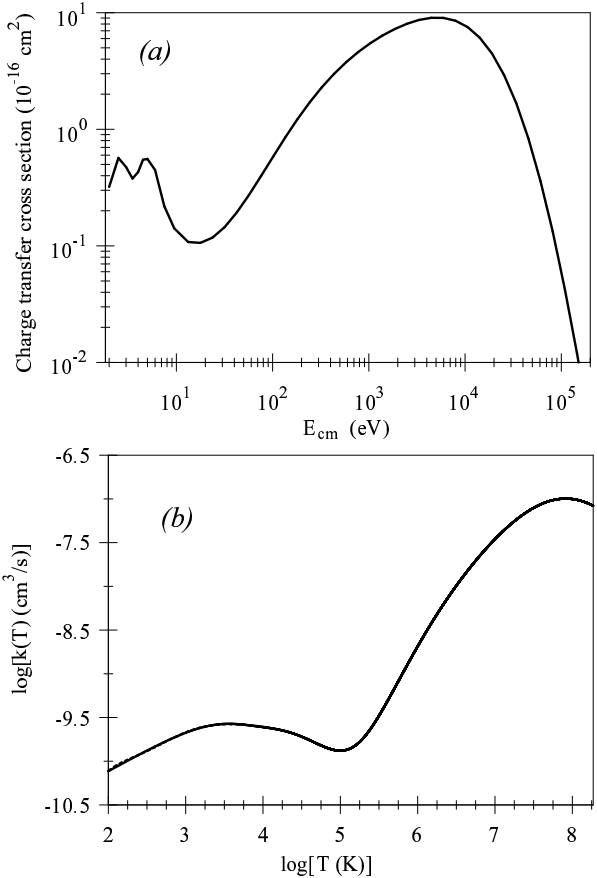}
\caption{Top panel: charge transfer cross section for the reaction
$\mathrm H_2+\mathrm H^+\to \mathrm H_2^++\mathrm H$. Bottom panel:
Rate coefficient $k(T)$ derived from the cross section in the top
panel. The fits of $k(T)$ (dashed line) given by eqn. (\ref{phi_ph})
and Table \ref{fitkrstic} cannot be visually distinguished from
$k(T)$.} \label{figkrstic}
\end{center}
\end{figure}

\textsc{$\mathrm H_2+\mathrm H^+\to \mathrm H_2^++\mathrm H$
(reaction 9)}. Each author has his own favored rate for this
reaction so that the overall uncertainty is large.  Our prescription
is as follows. Up to $3\times 10^4\ \mathrm K$ we adopt the data of
\citet{Savin04}, based on the latest and most accurate
quantum-mechanical calculations of the vibrationally resolved cross
sections for the charge transfer $\mathrm H_2+\mathrm H^+\to \mathrm
H_2^++\mathrm H$ at the center-of-mass collision energies from the
threshold ($\sim 1.8$ eV) up to 10 eV. For higher collision
energies, up to approximately $10^4$ eV, we derive the cross section
following  the suggestions by \citet{Barnett90} and \citet{Janev87},
which stand on the best known experimental \citep[see for
instance][]{Gealy87} and theoretical data, as well as on the recent
measurements by \citet{Kusakabe03}. These data were smoothly matched
to those of \citet{Krstic02} at low energies, thus yielding the most
updated  cross section for the charge transfer from $\mathrm H$ to
$\mathrm H_2^+$ in the vibrationally ground state. This cross
section is shown in the top panel of Fig. \ref{figkrstic}. Following
\citet{Savin04}, from this cross section we calculate the rate
coefficients for temperatures from the threshold up to $10^8$ K, as
shown in the bottom panel of Fig. \ref{figkrstic}.

We fit the reaction rate (in units of $\mathrm {cm}^3\mathrm s^{-1}$)
with the analytical expression

\begin{equation}\label{eqnkrstic}
    \log\left[k_9(T)\right]=\sum_{i=0}^7 a_i\log(T)^i\,,
\end{equation}
where $T$ is the
temperature in Kelvin. Two intervals are considered for the
temperature, i.e.  $T=[10^2,\,10^5]\ \mathrm K$ and
$T=[10^5,\,10^8]\ \mathrm K$ and two different fits are derived. The
coefficients $a_i$ for the two fits are given in Table
\ref{fitkrstic}. The rates  obtained from this analytical
fit are compared to those derived from  the numerical data. No
difference can be noticed as shown  in the bottom panel of Fig. \ref{figkrstic}.

\textsc{Associative detachments of $\mathrm H^-$ and $\mathrm D^-$
(reactions from 21 to 24).}  These processes  include the following
four reactions
\begin{eqnarray}
 \mathrm H^-+\mathrm H&\to& \mathrm H_2  +\mathrm e^-  \,,\nonumber\\
 \mathrm D^-+\mathrm H&\to& \mathrm {HD} +\mathrm e^-  \,,\nonumber\\
 \mathrm H^-+\mathrm D&\to& \mathrm {HD} +\mathrm e^-  \,,\nonumber\\
 \mathrm D^-+\mathrm D&\to& \mathrm {D_2}+\mathrm e^-  \,.\nonumber
\end{eqnarray}
Following \citet{Glover2006} and using their notation, the reaction
rate is expressed as $k=const \times \xi$ where  $\xi$ is a
parameter taking values in the interval  $[0.65,5.0]$. In our case
we obtain $k_{21}=k_{24}=10^{-9}\,\xi$ and
$k_{22}=k_{23}=10^{-9}\,\xi/2$. In \textsc{ROBO} the parameter $\xi$
can be varied in the above interval to investigate its effects on
the overall results. In the present study we adopt $\xi=0.65$.

\textsc{$\mathrm H^++\mathrm H^-\to \mathrm H+\mathrm H$ (reaction
29).} For the mutual neutralization of $\mathrm H^-$ and $\mathrm
H^+$ we adopt the cross section given by \citet{Croft99} and
\citet{GloverAbel08}. The mutual neutralization rate is related to
the rate of the associative detachment $\mathrm H^-+\mathrm H\to
\mathrm H_2+\mathrm e^-$, in competition for the available $\mathrm
H^-$ ions. Both rates are important for  the formation of  $\mathrm
H_2$ \citep[see][for more details]{GloverAbel08}. It is worth
recalling here that other estimates for  this reaction rate have
been reported by \citet{GloverAbel08}. \citet{Moseley70}
proposed the rate $
  k=5.7 \times 10^{-6}T^{-0.5}+6.3\times 10^{-8}-9.2\times 10^{-11}T^{0.5}\nonumber\\
  +4.4\times 10^{-13}T $ $\mathrm{cm^3\,s^{-1}}$,
whereas  \citet{DalgarnoLepp87} gave $
  k=7.0\times 10^{-7}T^{-0.5}$ $ \mathrm{cm^3\,s^{-1}}$.

\begin{table}
\begin{center}
\caption{Fit coefficients for the charge transfer reaction $\mathrm
H_2+\mathrm H^+\to \mathrm H_2^++\mathrm H$ as described in the eqn.
(\ref{eqnkrstic}). Coefficients are in the form $a(b)=a\times 10^b$.
All the temperature ranges are shown.} \label{fitkrstic}
\vspace{1mm}
\begin{tabular*}{.35\textwidth}{c |r@{.}l r@{.}l}
\hline
$a_i$ &\multicolumn{2}{c}{$10^2\le T<10^5\ \mathrm K$} &\multicolumn{2}{c}{$10^5\le T \le 10^8\ \mathrm K$}\\
\hline
$a_0$ & $-1$&$9153214(+2)$          & $-8$&$8755774(+3)$            \\
$a_1$ & $ 4$&$0129114(+2)$          & $ 1$&$0081246(+4)$            \\
$a_2$ & $-3$&$7446991(+2)$          & $-4$&$8606622(+3)$            \\
$a_3$ & $ 1$&$9078410(+2)$          & $ 1$&$2889659(+3)$            \\
$a_4$ & $-5$&$7263467(+1)$          & $-2$&$0319575(+2)$            \\
$a_5$ & $ 1$&$0133210(+1)$          & $ 1$&$9057493(+1)$            \\
$a_6$ & $-9$&$8012853(-1)$          & $-9$&$8530668(-1)$            \\
$a_7$ & $ 4$&$0023414(-2)$          & $ 2$&$1675387(-2)$            \\
\hline
\end{tabular*}
\end{center}
\end{table}

\textsc{$\mathrm{He}^+ +\mathrm e^-\to \mathrm{He}+\mathrm \gamma$
(reaction 35).} This process can occur either via direct radiative
recombination or di-electronic recombination, followed by radiative
relaxation. Therefore, the reaction rate  is the sum of two terms
\begin{equation}
 k_{35}=k_{35}^\mathrm{d}+k_{35}^\mathrm{e}\,,
\end{equation}
in units of $\mathrm {cm}^3\mathrm s^{-1}$. The first term is the
di-electronic recombination rate, described by
\begin{eqnarray}
 k_{35}^\mathrm{d}&=&1.544\times10^{-9}T_e^{-1.5}\exp(-48.596/T_e)\nonumber\\
 &\times&\left[0.3+\exp(8.1/T_e)\right]\,,
\end{eqnarray}
where $T_e$ is the temperature expressed in $\mathrm{eV}$. The
second term is the radiative recombination rate whose temperature
dependence is
\begin{equation}
 k_{35}^\mathrm{e}=3.925\times10^{-13}T_e^{-0.6353}\,.
\end{equation}
Both terms are taken from \citet{Abel97}.

\textsc{$\mathrm{Z}^++\mathrm e^-\to \mathrm{Z}+ \mathrm \gamma$
(reactions 38, 40, 42, 44).} An important process to include is the
metal recombination. The metals considered by our ISM model and
\textsc{ROBO} are $\mathrm{C},\mathrm{Si},\mathrm{O}$, and
$\mathrm{Fe}$. The recombination rates of the metals are calculated
according to the formalism proposed by \citet{VernerFerland96}

\begin{equation}
    \Phi_\mathrm{rec}(T,A,\tau_0,\tau_1,b)=\frac{A}{T_0\left(1+T_0\right)^{1-b}\left(1+T_1\right)^{1+b}}\,,
\end{equation}
in units of $\mathrm{cm}^{3}\mathrm{s}^{-1}$, where $T$ is the
temperature, $T_0=\sqrt{T/\tau_0}$ and $T_1=\sqrt{T/\tau_1}$. In
Table \ref{network3} we give the fit coefficients $A$, $b$, $\tau_0$
and $\tau_1$ for each metal.

\textsc{$\mathrm{Z}+\mathrm e^-\to \mathrm{Z}^+ + 2\,\mathrm e^-$
(reactions 39, 41, 43, 45).} For the collisions between electrons
and metals we use the fit proposed by \citet{Voronov97} expressed by

\begin{equation}
    \Phi_\mathrm{col}(T,A,P,\Delta E,X,K)=A\frac{1+P\sqrt{U}}{X+U}\,U^K e^{-U}\,,
\end{equation}
in units of $\mathrm{cm}^{3}\mathrm s^{-1}$, where $U=\Delta E/T$ in
which  $\Delta E$  is the energy difference between the two atomic
levels involved in the process, and $T$ is the temperature. Both
$\Delta E$ and $T$ are in $\mathrm{eV}$. The  parameters $A$, $P$,
$\Delta E$, $X$, and $K$ are given in Table \ref{network3}.


\subsection{The dust}\label{Dusts}

Dust grains take part to the process of molecule formation, e.g.
$\mathrm{HD}$ and $\mathrm H_2$ form on the surface of dust grains;
therefore, all the physical processes involving the dust must be
described and treated with the highest accuracy. To understand how
grains take part in the process of molecule formation, we need to
know the mechanisms  of grain formation and destruction, along with the
distribution of the grain temperature and size.

Given an initial set of dust composition and abundances, our dust
model follows the evolution of the dusty components during
the history of the ISM due to the creation of new grains and the
destruction of the existing ones. Creation of new dust grains is
governed by several processes that have different efficiencies
depending on the size of the dust particles. The same applies  for
the destruction that is mainly due to the thermal motion of the gas
particles and shocks from supernov{\ae}. Thermal
destruction is quite easy to model, because  the only parameter at
work is the gas temperature. Shock disruption  is more difficult to
evaluate. The main uncertainty comes from discrete
numerical hydrodynamical simulations only being able to follow shocks up to a
given (often too coarse) resolution, yet insufficient for the
microscopic description required here. To cope with this, we have
followed a ``statistical" approach. In brief, once identified the gas
particles of the NB-TSPH simulations with their turbulent
velocities (suggested by their velocity dispersion), we assume that
all the shocks  inside them follow the Kolmogorov law. See below for
more details.

Furthermore, grain destruction may depend on their size. This is the
case of the destruction by thermal motions, where the small
grains are destroyed before the large ones. As a consequence
of this, the size distribution function and the abundances
of dust grains of different type are continuously changing
with time. The ever-changing size distribution function plays an
important role in the formation mechanism of $\mathrm{HD}$ and
$\mathrm H_2$,  which are among the most efficient coolants. In brief,
changing the distribution function of the dust grains means
changing the quantity of key-role molecules produced by dust, as
shown in \citet{CazauxSpaans09}.

We analyze here the different aspects of the grain evolution and
their role in the formation of coolant molecules. First, we focus on
the distribution function of the dust grains, then we describe the
formation of coolants on grains. Finally, we analyze the destruction
and the formation of dust, as well as the effects of the grain
temperature.

\subsubsection{Size distribution function of dust grains}\label{distribution}

We adopt a simple power-law, MRN-like, grain size distribution
function \citep{MathisRumpl77,DraineLee84}. This is given by
$\mathrm dn(a)/\mathrm da\propto a^{-\lambda}$, with $a$ the grain
dimension, $n(a)$ the corresponding number density of grains with
dimension $a$,  and $\lambda=3.5$; this distribution covers the
range $5\ \mathrm{\AA}<a<2500\ \mathrm{\AA}$ and is extended to the
smallest grain dimensions in such a way as to include the typical
polycyclic aromatic hydrocarbons (PAHs) region
\citep{LiDraine2001b}. Even if more complicated distributions have
been proposed \citep{WeingartnerDraine01}, our simple choice is
suitable as an initial condition for the purposes of this work.

Indeed, it is worth pointing out here that the initial size
distribution function changes with time owing to dust destruction and
formation, which in turn depend on the grain size. Consequently, the
power law we started with does no longer apply. This will soon
affect the formation of $\mathrm H_2$ or $\mathrm{HD}$. For this
reason, the knowledge of the current size distribution function is
paramount.

First of all, we split the interstellar dust in two main components:
the carbon-based (thereinafter simply carbonaceous grains) and the
silicon-based grains (thereinafter the silicates). These play
different roles in the formation of molecular hydrogen, as well as in
different formation and destruction rates. In principle, there
should be an additional group to consider, namely the PAHs, because
the three types of dust have different efficiencies in the $\mathrm
H_2$ and $\mathrm{HD}$ formation. However, there is the lucky
circumstance that PAHs and graphite grains have similar efficiencies
to those shown by Fig. 2 of \citet{CazauxSpaans09}. Basing on this, we
lump together graphite grains and PAHs and treat separately the
silicates. Therefore, thereinafter we refer to the mixture graphite
grains plus PAHs as ``carbonaceous grains".

\subsubsection{Dust-driven $\mathrm{H_2}$ and $\mathrm{HD}$ formation}\label{h2bydust}

The model we adopt to describe   the so-called dust phase is the one
proposed by \citet{CazauxSpaans09}. In brief, the dust phase has its
own network of reactions that establish the relative abundances of
the various types of dust and  govern the formation of
$\mathrm{H_2}$ and $\mathrm{HD}$. We can neglect the presence of
other molecules like $\mathrm{CO}$. The newly formed $\mathrm{H_2}$
and $\mathrm{HD}$ are fed to the gas phase and enter the system of
equations governing the abundances of the ISM and vice versa. In
reality, the two phases should be treated simultaneously, governed
by a unique system of differential equations determining the number
densities of the elemental species, dusty compounds, and free
electrons at once. In practice this is hard to do because in general
the time scales of the various processes creating and destroying the
elemental species of the gas phase are much different from those
governing the formation/destruction of dust grains. To cope with
this difficulty, the \citet{CazauxSpaans09} model separates the two
phases, provides the results for the dust phase, and more important
provides a link between the two phases that allows determining the
formation rate of $\mathrm{H_2}$ and $\mathrm{HD}$ by dust as a
function of the gas temperature $T_g$ and dust temperature $T_d$.
This link is named the ``Sticking Function $S(T_g, T_d)$". The
sticking function somehow quantifies the probability that an
hydrogen atom striking a grain sticks to the surface rather than
simply bouncing off. The sticking function characterizes the
probability for an atom to remain attached to a grain. Here we
strictly follow this way of proceeding.

According to the \citet{CazauxSpaans09} model, the rate at which
$\mathrm H_2$  molecule  forms over the grain surface is

\begin{equation}\label{grainrate}
  R_d(\mathrm H_2)=\frac{1}{2}n(\mathrm H)v_H n_d\sigma
\epsilon_\mathrm{H_2}S(T_g,T_d)\ \ \mathrm{cm^{-3}\,s^{-1}}\,.
\end{equation}
where $n(\mathrm H)$ is the density of hydrogen in the gas phase,
$v_H=\sqrt{8kT_g/(\pi m_{\mathrm H})}$ is the gas thermal speed ($k$
is the Boltzmann constant and $m_{\mathrm H}$ is the hydrogen mass),
$n_d$ the dust number density, $\sigma$  the grain cross section
(i.e. $\pi a^2$), $\epsilon_\mathrm{H_2}$  the intrinsic efficiency
of the process,  and $S(T_g,T_d)$  the sticking function. This
latter is in turn a function of the dust and  grain temperatures

\begin{eqnarray}
  S(T_g,T_d)&=&\left[1+0.4\left(\frac{T_g+T_d}{100}\right)^{0.5}
+0.2\left(\frac{T_g}{100}\right)\right.\\ \nonumber
&+&\left.0.08\left(\frac{T_g}{100}\right)^2\right]^{-1}\,,
\end{eqnarray}
where  $T_g$ and $T_d$ are in Kelvin. With this function
the probability for a gas molecule to remain stuck on a dust grain
is higher in a cold gas than in a hot one. This probability is even
higher for the cold grains. For $\mathrm{HD}$   the expressions are
similar.

Equation (\ref{grainrate}) contains the intrinsic efficiency
$\epsilon_\mathrm{H_2}$ (or
 as $\epsilon_\mathrm{HD}$ for $\mathrm{HD}$), which is the
probability for the process to occur.  In our case it is the
probability that atoms, which are stuck to a grain, react to form
$\mathrm{H_2}$ or $\mathrm{HD}$.

\noindent  For the carbonaceous grains, the efficiencies
$\epsilon_\mathrm{H_2}$ and $\epsilon_\mathrm{HD}$ coincide and are
equal to

\begin{equation}\label{eC}
  \epsilon_\mathrm{C}=
  \frac{1-T_\mathrm{H}}{1+0.25
    \left(1+\sqrt{\frac{E_\mathrm{ch}-E_\mathrm{S}}{E_\mathrm{phy}-E_\mathrm{S}}}
    \right)^2e^{-\frac{E_\mathrm{S}}{T_\mathrm{d}}}}\,.
\end{equation}

\noindent The efficiency for the silicates $\epsilon_{\mathrm{Si}}$
is

\begin{equation}\label{eSi}
  \epsilon_{\mathrm{Si}}=\frac{1}{1+\frac{16T_\mathrm{d}}{E_\mathrm{ch}-E_\mathrm{S}}
    e^{-\frac{E_\mathrm{ph}}{T_\mathrm{d}}}
    e^{\beta_\mathrm{d}\mathrm{a_{pc}}\sqrt{E_\mathrm{phy}-E_\mathrm{S}}}}
  +\mathcal F\,,
\end{equation}

\noindent where $\beta_\mathrm{d}=4\times 10^9$ for $\mathrm H_2$
and $\beta_\mathrm{d}=5.6\times 10^9$ for $\mathrm{HD}$. The term
$\mathcal F$ is a function of the gas temperature and can be written
as
\begin{equation}
\mathcal
F(T)=2\frac{e^{-\frac{E_\mathrm{phy}-E_\mathrm{S}}{E_\mathrm{phy}+T_{g}}}}
         {\left(1+\sqrt{\frac{E_\mathrm{ch}-E_\mathrm{S}}{E_\mathrm{phy}-E_\mathrm{S}}}
           \right)^2}\,,
\end{equation}
where $T_\mathrm{H}$ is given by the expression
\begin{equation}
T_\mathrm{H}=4\left(1+\sqrt{\frac{E_\mathrm{ch}-E_\mathrm{S}}
{E_\mathrm{phy}-E_\mathrm{S}}}\right)^{-2}\exp\left(-\frac{E_\mathrm{ch}-E_\mathrm{S}}{E_\mathrm{ch}+T_g}\right)\,.
\end{equation}

The various quantities appearing in the above relationships are
listed in Table \ref{dustsdata}. For more details on these equations
see \citet{CazauxSpaans09}. Comparing eqns. (\ref{eC}) and
(\ref{eSi}), we see that the efficiency is high when the dust
temperature is low. For the silicates the efficiency window is
shorter than for the carbonaceous grains. For the silicates the
efficiency is high for temperatures up to $20\ \mathrm{K}$ and then
falls by two orders of magnitude. The carbonaceous grains are
efficient for temperatures up to $100\ \mathrm K$, where the
efficiency is still $0.1$ (instead of $0.01$ as for the silicates).
Finally, it is worth noticing that the efficiency profile is
smoother for the carbonaceous grains.

From all these considerations it follows that cold dust and
warm carbon-dominated dust in a cold gaseous environment
are the most efficient drivers for the formation of coolant
molecules like $\mathrm H_2$ and $\mathrm{HD}$.

\begin{table}
\begin{center}
\caption{The values used to compute the formation efficiency of
$\mathrm H_2$ and  $\mathrm{HD}$, with $E_\mathrm{phy}$, $E_\mathrm{ch}$,
and $E_S$ in $\mathrm K$, $a_\mathrm{pc}$ is in $\mathrm\AA$.
From \citet{CazauxSpaans09}.} \label{dustsdata} \vspace{1mm}
\begin{tabular*}{73mm}{l |l l l l}
\hline Surface &$E_\mathrm{phy}$
&$E_\mathrm{ch}$ &$E_\mathrm{S}$&$\mathrm{a}_\mathrm{pc}$\\
\hline
Carbons &$800$ K  &$7000$ K &$200$ K &$3$ \AA \\
Silicates &$700$ K  &$15000$ K &$-1000$ K &$1.7$ \AA \\
\hline
\end{tabular*}
\end{center}
\end{table}

\subsubsection{Grain formation}
Here we  briefly examine the formation of dust grains. We start with
an initial number density with the size distribution given in
Section \ref{distribution} and with a given ratio between the
silicates and the carbonaceous grains. The latter is a free
parameter varying in the range  $[0,1]$, where zero stands for dust
made of sole carbonaceous grains; one is for dust made of sole
silicates.

According to \citet{Dwek98}, the temporal variation in the size
distribution function of grains caused by accretion is given by

\begin{equation}\label{formation}
  \frac{\mathrm dn(a)}{\mathrm dt}=c_d\,\alpha(T_g,T_d)\pi a^2n_gn_d(a)v_g
\end{equation}
where all the quantities have the same meaning as in eqn.
(\ref{grainrate}), but  for $\alpha(T_g, T_d)$ and $c_d$.

The quantity  $\alpha(T_g, T_d)$  is a sort of sticking coefficient
depending  on the gas and dust temperature and the type of dust.
This coefficient is therefore forced to change in the course of the
evolution. Furthermore, $n_d$ (and consequentially $n$) are
functions of  $a$. Equation (\ref{formation}) is similar to eqn.
(\ref{grainrate}) because the process is similar, except that now
this mathematical description is applied to the carbon atoms that
remain stuck to the carbon lattice of the grain. Our expression
differs slightly from the original one of \citet{Dwek98} because the
parameter $c_d$ is introduced in eqn. (\ref{formation}) to take some
considerations made by \citet{Dwek98} himself into account.
Inserting $c_d=1$ we can obtain the time scale of the process. In a
standard cold gas this time scale is $\tau\approx 2\times 10^4\
\mathrm{yrs}$, which is significantly smaller than the observational
estimates. Normal evaporation, caused by cosmic rays or UV heating
and grain-grain collisions can halt the growing of the dust grains.
The factor $c_d$ somehow takes this phenomena into account. It is
estimated to be on the order of $c_d=10^{-3}$. We name $c_d$ the
delay factor.

\begin{figure}
\begin{center}
\includegraphics[width=.45\textwidth]{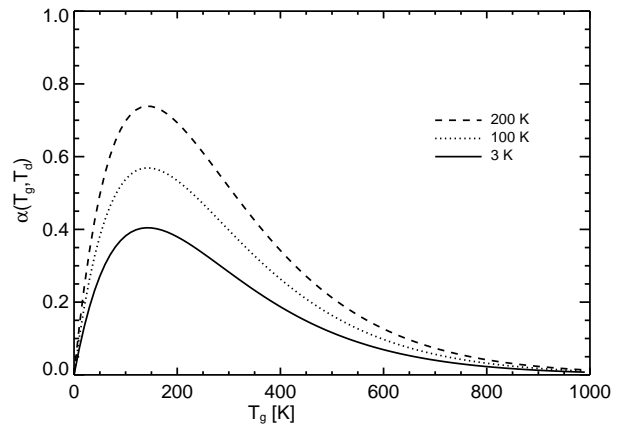}
\caption{The sticking coefficient for the carbon molecules over a
carbon lattice derived from \citet{LeitchWilliams85}. The different
lines indicate a different dust temperature: $3\ \mathrm K$ (solid),
$100\ \mathrm K$ (dotted), and $200\ \mathrm K$ (dashed). $T_g$ is
the gas temperature in Kelvin.} \label{stickLD}
\end{center}
\end{figure}

For the sticking coefficient $\alpha (T_g, T_d)$, we make use of the
data by \citet{LeitchWilliams85} and consider a carbon atom of the
gas phase as impinging on a carbon lattice. The  equation fitting
the data has the  form

\begin{eqnarray}
    \alpha(T_g,T_d)&=&0.0190\,T_g  \left(0.0017\,T_d+0.4000\right)\nonumber\\
    &\times&\exp\left(-0.0070\,T_g\right)\,,
\end{eqnarray}
where $T_g$  and $T_d$ are the gas and dust temperatures,
respectively (see Fig. \ref{stickLD}). The sticking coefficient
$\alpha(T_g,T_d)$ has no dimensions. The data of
\citet{LeitchWilliams85} provide a good dependence on the
gas temperature $T_g$, but a poor one on the dust
temperature $T_d$. Therefore an accurate fit is not possible. The
above relationship can be used in the temperature intervals  $10\
\mathrm K\le T_g\le 1000\ \mathrm K$ and $3\ \mathrm K\le T_d\le
300\ \mathrm K$. At present we also use the same model for silicate
grains even though this approximation might not be accurate. We plan
to improve upon this point in the future.

With this model the formation efficiency is higher when the
interstellar medium has a temperature $T \approx 100\
\mathrm K$ and when the dust grains have a temperature of
about $300\ \mathrm K$. For dust grains with temperature higher
than $300\ \mathrm K$ or lower than 3 $K$, the formation efficiency
is unknown. Though the fits from \citet{LeitchWilliams85}
are not very accurate, they are accurate enough for our purposes because we are
only interested in the shape and maxima of the curves in Fig.
\ref{stickLD}.

Finally, we note that we have already described the
formation of dust by means of a general process of accretion in the
ISM, leaving aside other sites of dust formation like the envelopes
of obscured AGB stars, Wolf-Rayet stars, and remnants of
supernov{\ae} \citep{Dwek98}. The reason behind this is that all of
these are external sources of dust that eventually enrich the ISM in  dust
content and therefore determine a different initial dust
content as input for ROBO. Indeed, for our purposes, only the
internal sources of dust in the unit volume are important. The
information about the initial conditions of the dust mixture should
be provided to ROBO by the companion NB-TSPH code \textsc{EvoL}, of
properly taking the dusty yields coming from the stars into account.

\subsubsection{Grain destruction}

In our model we assume that there are  three processes destroying
the grains of the ISM, i.e. shocks, in particular supernovae induced
shocks, vaporization by high-velocity shocks, and thermal
sputtering, in which dust is destroyed by the thermal motions of the
gas.  We need to describe the three processes in a way suited to the
NB-TSPH formalism.

 \textbf{Destruction by shocks.} This phenomenon is  difficult to
model. First of all shocks may easily induce turbulence in the ISM
and  to be suitably described one needs some assumptions about the
velocity fields of the fluid elements. The main difficulty arises
from the fact that within an SPH gas particle, because of the typical
mass resolution, unresolved shocks may take place at different
velocities. To avoid this difficulty, we treat the gas as a
turbulent fluid, assuming that even inside a gas particle the
turbulent nature of the fluid is preserved, and use the Gaussian
velocity distribution  $\phi(v)$ derived from the Kolmogorov law for
the power spectrum $E(k) \propto k^{-\alpha}\mathrm dk$ with
$\alpha$=5/3.  Second, grains of different masses and sizes move at
different velocities: the large, more massive grains moving slower
than the smaller less massive ones. This suggests adopting the
point of view in which the small fast grains are considered as
projectiles impinging on the massive slow grains considered as
targets. To summarize, the ISM we are dealing with is turbulent
because of the shocks crossing it and rich in dust grains of all
possible dimensions and masses interacting with each other and with the
shock fronts.

Given these premises, we model the destruction of dust grains by
shocks according to the picture by \citet{Hirashita2009}. The dust
grains are grouped in $N=100$ bins of number density according to
their mass. The mass of a grain is in turn expressed as the product
of a reference density, here assumed to be the density of the
graphite grains $\rho_\mathrm{gr}=2.3\, \mathrm{g/cm^3}$ times the
volume of the grain $V_{gr}=4/3\pi a^3$ where $a$ is the radius of
the grain (assumed to be spherical for simplicity). Therefore each
density bin contains grains whose mass goes from $m_{low}$ to
$m_{up}$, where both mass limits vary with the bin. The
corresponding radii are given by $m_{low} = 4/3\pi a^3_{low} $ and
$m_{up} = 4/3\pi a^3_{up} $. The grain radii follow the distribution
law given by $dn(a)= C_{nor} \times a^{-3.5}da$, where $C_{nor}$ is
a suitable normalization factor to be determined. Therefore, the
number density of the grains in each mass interval is
\begin{equation}
   n_\mathrm{i}=  \int_{a_\mathrm{low}}^{a_\mathrm{up}}n(a) \,   da
   = C_{nor}\int_{a_\mathrm{low}}^{a_\mathrm{up}}a^{-3.5} \,   da
\end{equation}
 The total number
density of dust grains of any size and hence mass is

\begin{equation}
 n_\mathrm{dust}=\sum_{i=0}^{N-1} n_i\,.
\end{equation}

When a projectile of mass $m_j$ hits a target of mass $m_i$, the
target loses a fraction of mass $f_\mathrm{sh}m_i$ if $m_j<m_i/2$,
or it is totally destroyed if $m_j \ge m_i/2$ (i.e.
$f_\mathrm{sh}=1$). In both cases the projectile is destroyed. The
lost mass, i.e. a  fragment of lower mass,  so the dimension  is
assigned to the mass (size) bin  according to the adopted power law.
The remaining part of the target (remnant) of mass
$(1-f_\mathrm{sh})m_i$  is added to the appropriate mass, hence size
bin, so the number density variation of the {\it i-th} bin is
\begin{equation}
  \frac{\mathrm dn_i}{\mathrm dt}= \left(\frac{\mathrm dn_i}{\mathrm dt}\right)_\mathrm{L}+
   \left(\frac{\mathrm d n_i}{\mathrm dt}\right)_\mathrm{F}+
    \left(\frac{\mathrm dn_i}{\mathrm dt}\right)_\mathrm{R}\,,
\end{equation}
where the subscript L indicates the term representing the mass lost
in fragments, F the mass gained from the fragmentation, and finally R
the mass of the remnants  moving from other bins to the
\textit{i-th} one. The first term must be negative.

To estimate the various contributions to the total number density
variation, we   must consider the probability of an impact. This
will be proportional to the number density of the targets $n_i$, the
density of the projectiles $n_j$, their relative speed $v_{ij}$, and
their sizes ($a_i$ and $a_j$). The assumptions on the speed will be
discussed later in this section. For the targets belonging to the
{\it i-th} bin, we find that the mass lost in fragments corresponds
to the density change of
\begin{equation} \label{dnidt}
   \left(\frac{\mathrm dn_i}{\mathrm dt}\right)_\mathrm{L}=-\sum_j^{j\le i}\epsilon_{ij}\alpha_{ij}n_i n_j\,,
\end{equation}

\noindent where
\begin{equation}
   \epsilon_{ij} = \left\{
   \begin{array}{cl}
      \frac{f_\mathrm{sh} n_i+ n_j}{n_i+ n_j}& \mathrm{if}\,m_j\le m_i/2\,,\\
       1  & \mathrm{if}\,m_j>m_i/2\,
   \end{array} \right.
\end{equation}

\noindent and the impact coefficient $\alpha_{ij}$is
\begin{equation} \label{alphaij}
   \alpha_{ij}=v_{ij}\pi(a_i+a_j)^2\,.
\end{equation}

\noindent Consequently, the total density variation of newly created
fragments is
\begin{equation}
   \frac{dn_\mathrm{frag}}{\mathrm dt}=\sum_i \frac{\mathrm dn_i}{\mathrm dt}\,.
\end{equation}
They  will be distributed among the bins with a power law to obtain
the $(\mathrm d n_i/\mathrm dt)_\mathrm{F}$ of each bin.

The variation in the remnant number density is  given by
\begin{equation}
 \left(\frac{\mathrm dn_k}{\mathrm dt}\right)_\mathrm{R}=(1-\epsilon_{ij})\frac{\mathrm dn_i}{\mathrm dt}\,,
\end{equation}
where k is the index of the bin receiving the remnant given by
\begin{equation}
 k=\mathrm{int}\left[\frac{m_i(1-f_\mathrm{sh})-m_\mathrm{low}}{\Delta m}\right]\,,
\end{equation}
with $i$ the index of the initial grain that suffered fragmentation
and $\Delta m$ the mass range of each  bin,
 $(m_\mathrm{up}-m_\mathrm{low})/N$.

The shock velocities obey a Gaussian distribution
$\phi\left(v\right)$ normalized to
\begin{equation}\label{kolmo}
  \int_{v_\mathrm{low}}^{v_\mathrm{up}}\phi(v)\,\mathrm dv=1\,,
\end{equation}
where $v_\mathrm{low}$ and $v_\mathrm{up}$ are the limits of the
shock velocities over which the normalization of $\phi(v)$ applies,
thus fixing the normalization constant. The relative speed $v_{ij}$
of the grains is obtained from the velocity distribution of eqn.
(\ref{kolmo}) with $v_\mathrm{low}=1$ km/s and $v_\mathrm{up}=200$
km/s and a total of 30 velocity bins. The  velocity  $\phi(v)$ is a
Gaussian centered at $v=100$ km/s with a dispersion of $30$ km/s.
The integral
\begin{equation}
  \xi(v)=\int_{v_\mathrm{i}}^{v_\mathrm{i+1}}\phi(v)\mathrm dv\,,
\end{equation}
yields the relative weight of each velocity bin. The relative
velocities of the projectiles are distributed according to these
weights and the total density of targets in the i-th bin changes as
described by eqns. (\ref{dnidt}) and (\ref{alphaij}).

\textbf{Vaporization.} Since we also deal with high-velocity shocks,
vaporization of dust grains into the gas phase may become important.
We use the formalism of \citet{Tielens1994} for the impact of carbon
grains. The fraction of vaporized material is
\begin{equation}
  F_v=\frac{f_\mathrm{vap}}{1+f_\mathrm{vap}}\,
\end{equation}
with $f_{vap}$ given by
\begin{equation}\label{frac_vap}
  f_\mathrm{vap}=\left[f_{v1}+f_{v2}\sqrt{1-\frac{v_t}{v}}\right]\frac{m_j}{m_i}\,,
\end{equation}
where $v_t=23$ km/s is the shock threshold velocity, $m_i$ and $m_j$
are the masses of target and projectile grains, respectively.  Equation
(\ref{frac_vap}) must satisfy the condition $v\ge v_t$. The quantities
$f_{v1}$ and $f_{v2}$ are taken from \citet{Tielens1994} and are
given by the relations
\begin{equation}
  f_{v1}=2.59\frac{a_i}{a_i+a_j}\,,
\end{equation}
and
\begin{equation}
  f_{v2}=\frac{2.11}{\sigma_1^{8/9}}\,,
\end{equation}
where
\begin{equation}
  \sigma_1=\frac{0.3\,\left(s+\mathcal{M}^{-1}-0.11\right)^{1.3}}{\left(s+\mathcal{M}^{-1}-1\right)}\,,
\end{equation}
with $\mathcal{M}$ the Mach's number and $s=1.9$.


\textbf{Destruction by thermal sputtering.} To evaluate the fraction
of grains destroyed by thermal sputtering, we adopt  the
approximation by \citet{DraineSalpeter79a}. A grain of dust in a
medium with temperature  $T\leq 10^6\ \mathrm K$ has a destruction
time (i.e. its lifetime)

$$\tau_\mathrm{dist}\approx 10^3a /n_\mathrm{H}\,\,\,\,  \mathrm{yr}.$$

\noindent with $a$ in nm. From this we can obtain the destruction
rate per second. The above relation  is only valid for high
temperatures.

To include  the temperature dependence of the destruction rate, we
refer to \citet{Tielens94}, with the aid of which we model
the dependence on the gas temperature of the lifetime of the dust.
As expected, dust grains with lower temperature have a longer
lifetime. Finally, according to \citet{DraineSalpeter79a}, the small
grains have a shorter lifetime than the large ones in the same
environment.

The grain temperature depends on the radiation field generated by
all the stellar sources. However, for the sake of simplicity, in
this study we used a fixed value for the grain temperature.
Therefore, a big improvement  would be given by   including a
description of the photon diffusion
\citep[e.g.][]{Mathis83}. Our choice is partially justified by the
fact that the companion NB-TSPH code \textsc{EvoL} does not
yet include photon diffusion. This leads us to
postpone the implementation of different grain temperatures to when
photon diffusion is included in \textsc{EvoL}.

\subsection{Heating and cooling}\label{Cooling}

At this stage, it is worth discussing in some detail the role played
by the heating and cooling during the history of star formation in a
galaxy or a cosmological simulation. We have already emphasized that
coolants are the key elements for the formation scenario. As the gas
cools down, more and more spatial structures and stars (with a
certain mass function, hence mass-luminosity law) are born.
Without coolants neither structures nor stars would form. In this
context, the  grains and their temperature in turn play the key role
in the formation of  efficient coolant molecules like $\mathrm H_2$
and $\mathrm{HD}$.  Cold grains form more molecules than the warm
ones. Since the dust temperature depends on the surrounding photon
flux, it means that stars in the neighborhood are crucial for
heating the dust particles. All  this forms a closed loop of
interwoven physical processes: dust forms coolants - coolants induce
star formation - stars heat the dust.
Understanding the details of this mutual
interaction will allow us to get clues on the  star formation
history in general.

In this scene a starring actor is the gas cooling. Chemical
reactions are sensitive to the gas temperature, hence to the gas
cooling; indeed, the formation of coolants depends on the gas
temperature, which in turn depends on the cooling process. We split
the cooling of the gas in several sources characterized  by the
physics of the dominant process. Above $10^4\,\mathrm K$, there are
two very popular descriptions of the cooling, namely \citet{Cen92}
and \citet{SutherlandDopita93}. At lower temperatures we have the
\emph{metal} cooling \citep{Maio07}, the \emph{molecular hydrogen}
cooling \citep{GloverAbel08,GalliPalla98}, and finally, the
\emph{$\mathrm{HD}$} cooling \citep{Lipovka05}.

\subsubsection{Heating by photoelectric ejection of electrons from dust grains}\label{dustheat}

The photoelectric ejection of electrons by dust grains is  an
important source of heating. The model proposed here is based on
\citet[][]{BakesTielens94} and \citet{WeingartnerDraine01b}. The
photoelectric heating is given by
\begin{eqnarray}
 H(N_C,\mathrm Z)&=&W\pi\int_{\nu_\mathrm{Z}}^{\nu_\mathrm{H}}
 \sigma_\mathrm{abs}(N_C)Y_\mathrm{ion}(N_C,IP_\mathrm{Z})\nonumber\\
    &\times&F(\nu)g(N_C,IP_\mathrm{Z})\mathrm d\nu\,,
\end{eqnarray}
where W is the FUV dilution factor, $\sigma_\mathrm{abs}$  the
photon absorption cross section, $Y_\mathrm{ion}$ the photoelectric
ionization yield, $F(\nu)$  the UV radiation flux, and
$g(N_C,IP_\mathrm{Z})$ the kinetic energy partition function. Here, $N_C$
is the number of carbon atoms that form the dust ($N_C \sim
0.5\,a^3$ with $a$ the radius of the grain in \AA, according to
\citet{LiDraine2001b}) and $IP_\mathrm{Z}$ the ionization potential
of a grain of charge Z, and $\nu_H$ is the Lyman frequency and
$\nu_\mathrm{Z}$  the frequency corresponding to $IP_\mathrm{Z}$. In
our model the range of grain sizes goes from $5$\AA \, to $100$\AA,
since the total heating is mostly due to small grains \citep[see][\,
for details]{BakesTielens94}.

The total photoelectric heating is
\begin{equation}\label{dustH}
 \Gamma=\int_{N_-}^{N^+}\sum_\mathrm{Z}H(N_C,Z)f(N_C,Z)n(N_C)\mathrm dN_C\,,
\end{equation}
where $f(N_C,Z)$ is the probability to find a grain composed of
$N_C$ carbon atoms at a certain charge Z. This can be computed
by considering the collisions with electrons and ions. The detailed
balance yields
\begin{equation}
 f(Z)\left[J_\mathrm{pe}(Z)+J_\mathrm{ion}(Z)\right]=f(Z+1)J_e(Z+1)\,,
\end{equation}
with $J_\mathrm{pe}$ the rate of photoelectron emission,
$J_\mathrm{ion}$, and $J_e$ the accretion rates of ions and
electrons. For the detailed calculation of $f(Z)$ see
\citet[][]{BakesTielens94}, while for all the details about
$J_\mathrm{pe}$, $J_\mathrm{ion}$ and $J_e$ see
\citet{WeingartnerDraine01b}.

As the FUV absorption cross section $\sigma_\mathrm{abs}$ depends on
the absorption efficiency of FUV photons by the grains, we use the
data of \citet{DraineLee84} and \citet{LaorDraine93} for the
graphite grains and
\citet{LiDraine2001a,LiDraine2001b,LiDraine2001c} for the PAHs. We
also use the model proposed by
\citet{LiDraine2001a,LiDraine2001b,LiDraine2001c} to create a mixed
population of PAHs and graphite grains. Following their paper the
total absorption efficiency is
\begin{equation}
 Q_\mathrm{abs}(a,\lambda)=\xi(a)Q_\mathrm{abs}^\mathrm{PAH}(a,\lambda)+
 \left[1-\xi(a)\right]Q_\mathrm{abs}^\mathrm{gra}(a,\lambda)\,,
\end{equation}
defining
\begin{equation}
 \xi(a)=(1-q)\mathrm{min}\left[1,\left(\frac{a_0}{a}\right)^3\right]\,
\end{equation}
by $q=0.01$ and $a_0=50\,\AA$ \citep[see][for more
details]{LiDraine2001a,LiDraine2001b,LiDraine2001c}.

\begin{figure}   
\begin{center}   
\includegraphics[width=.45\textwidth]{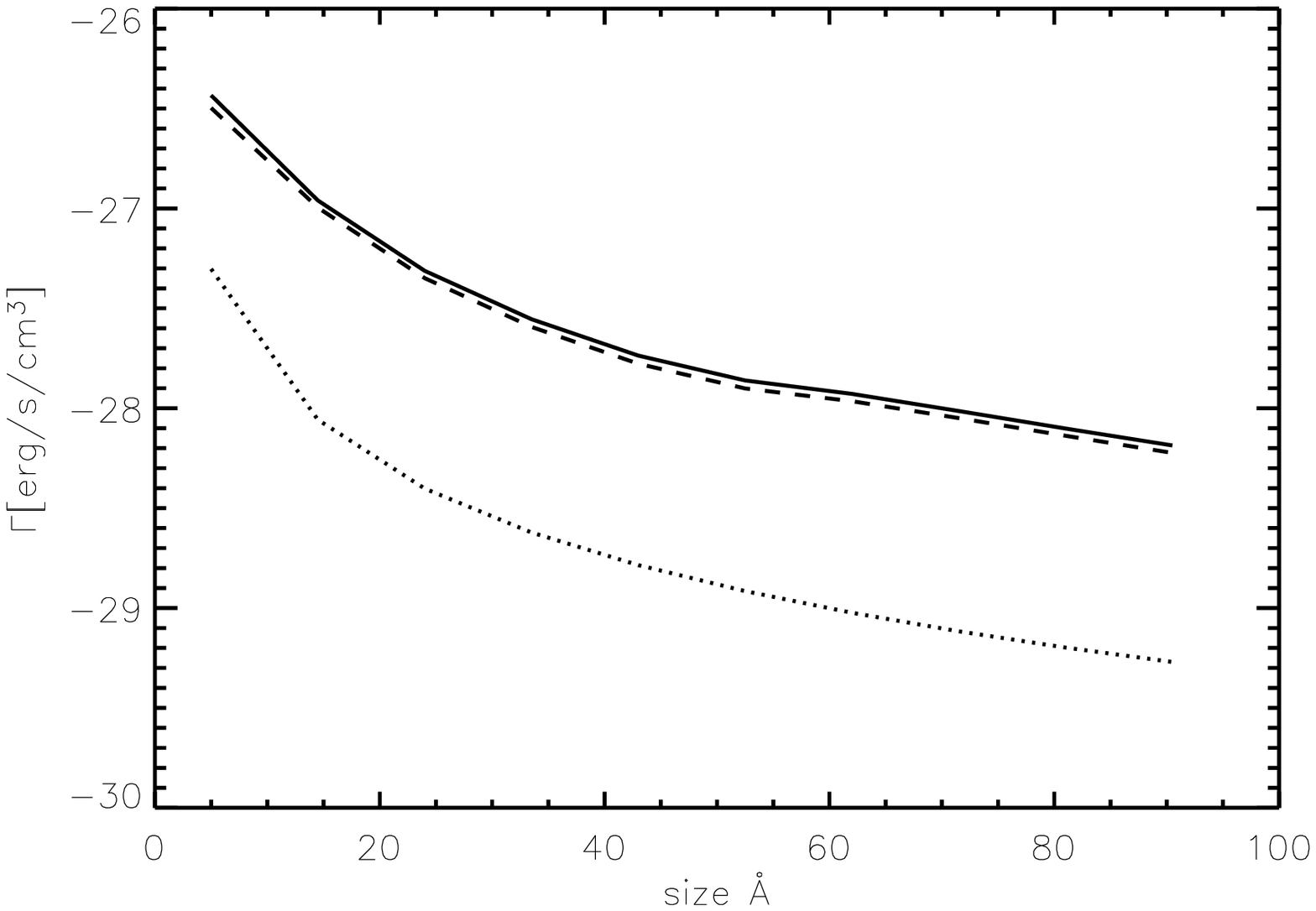}   
\includegraphics[width=.45\textwidth]{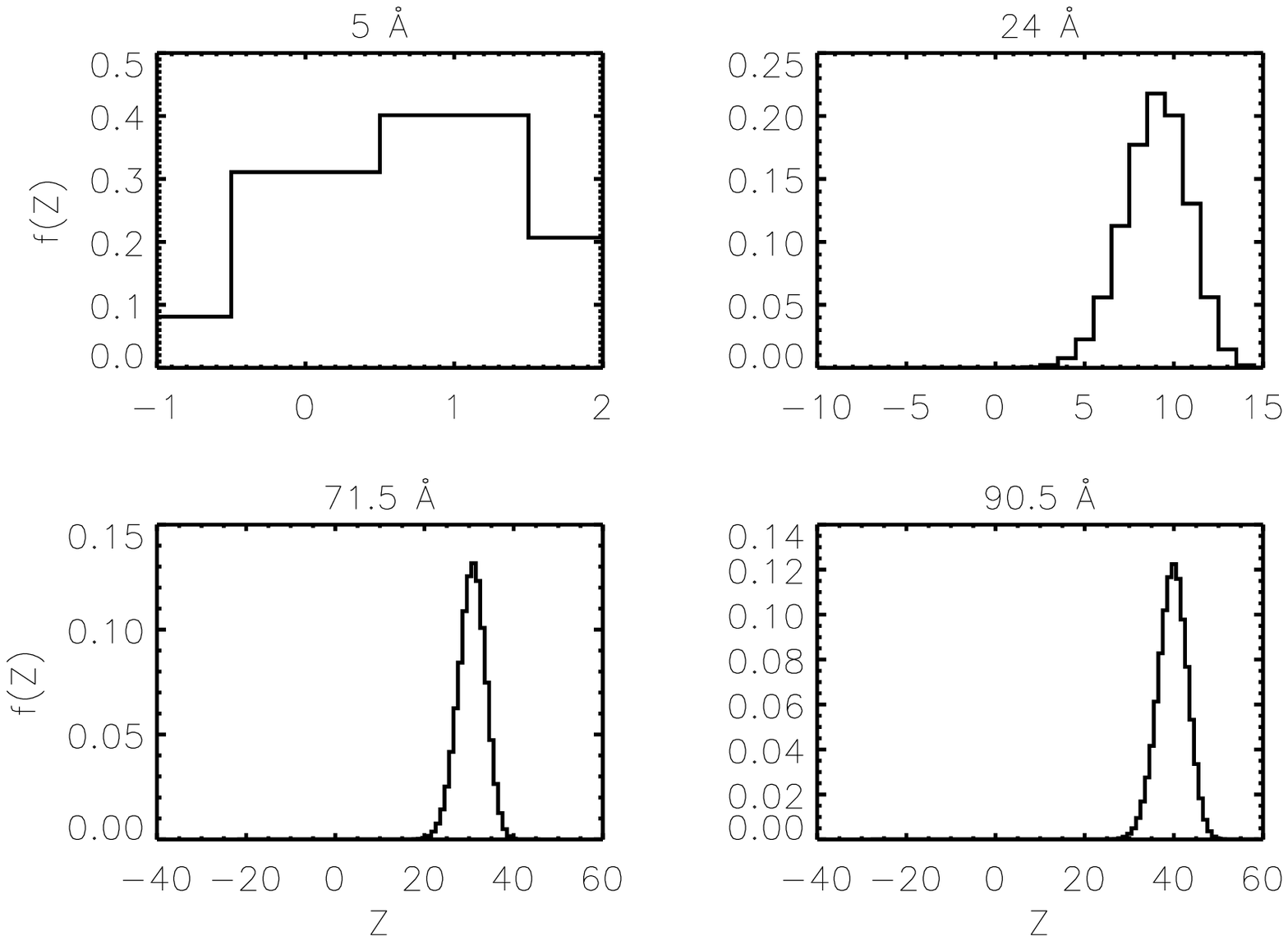}
\caption{Top panel: heating by dust (solid line), cooling (dotted
line) and the difference between heating and cooling (dashed line)
for different diameters of the dust particles. Bottom panel: the
probability of finding a grain with a certain charge Z as a function
of the grain dimensions. }\label{dustheatfig}\end{center}
\end{figure}

In the same way we can also obtain the cooling associated with the
electron recombination. This is given by
\begin{equation}
 C(N_C,Z)=n_i s_i \sqrt{\frac{8kT}{\pi m_i}}\pi a^2\hat\Lambda(\tau,\nu)\,,
\end{equation}
where $n_i$ is the density of the charged particle with sticking
coefficient $s_i=1$ and mass $m_i$ and $a$ is the grain size. The
term  $\hat\Lambda(\tau,\nu)$ is described in detail in
\citet[][]{BakesTielens94}. The total cooling is obtained as in eqn.
(\ref{dustH}).
In Fig. \ref{dustheatfig} we show the results from a simple model
with a grain distribution with $n_\mathrm{dust}=10^{-5}\
\mathrm{cm}^{-3}$ and a gas with a temperature of $10^2$ K.

\subsubsection{Which are the sources of cooling to consider?}

\textbf{Cooling at high temperatures ($T \geq 10^4$)}. Two main
sources of cooling can be found in the literature: (i) The formulation
proposed by \citet[][]{Cen92} which includes the following
mechanisms:

\begin{itemize}
  \item collisional ionization - \el{H}{}{}, \el{He}{}{}, \el{He}{}{+},
    \el{He(2^3S)}{}{},
  \item recombination - \el{H}{}{+}, \el{He}{}{+}, \el{He}{}{++},
  \item dielectronic recombination - \el{He}{}{},
  \item collisional excitation - \el{H}{}{}(all $n$), \el{He}{}{+}($n=2$),
    \el{He}{}{}($n=2,3,4$),
  \item bremsstrahlung - all ions,
\end{itemize}
which is  particularly suited in presence of non equilibrium
reactions for the H and He chemistry. (ii) The  description of
\citet[][thereinafter SD93]{SutherlandDopita93}, which is
particularly suited to treat cooling in presence of metals and to
describe it as function of the gas metallicity. Their case for $Z=0$
can be left aside.  Both cooling models are  used for temperature
higher than $10^4\,\mathrm{K}$. We refer to them by the acronyms C92
or SD93. In our models of the ISM we adopt both sources as
appropriated to the current physical conditions.

A difference between C92 and SD93 is that the cooling
rate is described in the former  by numerical fits (no further interpolations are
needed), whereas in the latter one has to interpolate huge tabulations of data in
temperature and metallicity. To this aim we
used a surface fit (routine SFIT in IDL) where the plane surface
passes through four points that correspond to two discrete
values in temperature and two discrete values in metallicity. The
analytical expression  of this plane yields  the cooling rate. Given
the two pairs of interpolating points, $(T_0,Z_0)$, $(T_0,Z_1)$,
$(T_1,Z_0)$, and $(T_1,Z_1)$, the  analytical function that describes
the surface can be written in the form

\begin{equation}\label{surface}
 \Lambda_\mathrm{SD93}(T,Z)=a_0+a_1 Z+a_2 T+a_3T Z\,,
\end{equation}

\noindent with $a_i$ the fit coefficients. So, for a generic point
$(T,Z)$, for which $T_0\le T\le T_1$ and $Z_0\le Z\le Z_1$, the rate
cooling is given by eqn. (\ref{surface}).

\textbf{$\mathrm H_2$ cooling.} The $\mathrm H_2$ cooling requires a
different description. \citet{HollenbachMcKee79} first evaluated the
molecular hydrogen cooling. A modern, widely used source of $\mathrm
H_2$ cooling is by \citet{GalliPalla98}. Although this cooling
function is quite accurate,  we prefer to add some more details by
including the function found by \citet{GloverAbel08}, which takes
not only the $\mathrm H_2-\mathrm H$
interaction into account, but also the collisions with $\mathrm{He}$, $\mathrm
H^+$, $\mathrm H_2$, and free electrons. It is described by
\begin{equation}
  \Lambda_{\mathrm H_2}=\sum_k\sum_i\mathrm{dex}\left[a_{ik}\log(T_3)^i\right]
    n_\mathrm{H_2}n_k\ \mathrm{erg\,cm^{-3}\,s^{-1}}\,,
\end{equation}
where $T_3=T/10^3\ \mathrm K$ with $T$ the gas temperature, $a_{ik}$
is the {\it i-th} fit coefficient of the {\it k-th} species
($k=\{\mathrm{H,H^+,H_2,e^-,He}\}$), and $n$ are the number
densities. The orto-para ratio is assumed to be $3:1$. Outside the
temperature range of the fits we use the molecular hydrogen cooling
by \citet{GalliPalla98}.

\textbf{$\mathrm{HD}$ cooling.} To describe the cooling by the
deuterated molecular hydrogen we use the model proposed by
\citet{Lipovka05}. It includes  $\mathrm{HD}$ roto-vibrational
structures, radiative and collisional transitions for $J\le 8$
rotational levels, and  the vibrational levels $v=0,1,2,3$. It has
been found that including the roto-vibrational transitions increases
the cooling efficiency of the $\mathrm{HD}$. The fit provided by the
authors depends on the gas density and temperature. It can be
parameterized as
\begin{equation}\label{fitHD}
  \Lambda(T,n)_\mathrm{HD}=\sum_{i=0}^4\sum_{j=0}^4c_{ij}\log_{10}(n)^i\log_{10}(T)^j\,,
\end{equation}
where $c_{ij}$ is the matrix whose  elements are  given in Table
\ref{HDtab}.
Using eqn. (\ref{fitHD}) is particularly convenient from a numerical
point of view as it provides fast evaluations of the cooling by the
$\mathrm{HD}$ molecule.

\begin{table*}
\begin{center}
\caption{Coefficients $c_{ij}$ used  in the eqn. (\ref{fitHD}) and
taken from \citet{Lipovka05}.} \label{HDtab} \vspace{1mm}
\begin{tabular*}{106mm}{r r r r r r}
  \hline
   & $i=0$ & $i=1$ & $i=2$ & $i=3$ & $i=4$\\
  \hline
  $j=0$ & $-42.56788$ &$0.92433$ &$0.54962$ &$-0.07676$&$0.00275$\\
  $j=1$ & $21.93385$ &$0.77952$ &$-1.06447$ &$0.11864$&$-0.00366$\\
  $j=2$ & $-10.19097$ &$-0.54263$ &$0.62343$ &$-0.07366$&$0.002514$\\
  $j=3$ & $2.19906$ &$0.11711$ &$-0.13768$ &$0.01759$&$-0.00066631$\\
  $j=4$ & $-0.17334$ &$-0.00835$ &$0.0106$ &$-0.001482$&$0.000061926$\\
\hline
\end{tabular*}
\end{center}
\end{table*}

\textbf{Cooling by the CO molecule}. Owing to founding hypotheses of
our model of the ISM presented in the introduction, the ISM is
optically thin so that it is worth considering the straight cooling
of the CO molecules by radiative  transitions among rotational and
vibrational  energy levels.

To this aim,  we may adopt the steady-state equations of
\citet{MSGW82} and   the rates  calculated by \citet{Schinke1985}
by applying the $L\rightarrow0$ coefficients to the
\citet{Goldflam1977} method and including the \citet{DePristo1979}
correction (see these studies for more details).

To calculate the total rotational cooling, we use the equation of
statistical equilibrium from \citet{MSGW82}:
\begin{eqnarray}
n_j[A_j&+&n_\mathrm{H_2}\sum_i\gamma_{ji}] \nonumber \\
&-&\left[n_\mathrm{H_2}\sum_i\gamma_{ij}n_i+n_{j+1}A_{j+1}\right]=0\,,
\end{eqnarray}
where $n_j$ is the number density of the CO molecules in the state
$j=J$, $n_\mathrm{H_2}$ the number density of the molecular
hydrogen, $\gamma_{ij}$ the rate coefficient of the transition,
and $A_j$ the $A$-coefficient from the state $j$ to $j-1$. The
system of equations is completed  by the condition
$\sum_jn_j=n_\mathrm{CO}$. The solution of this linear system is
straightforward. The emission due to  the transition from $J$ to
$J-1$ is
\begin{equation}
I_J=\frac{h\nu}{4\,\pi}\frac{A_Jn_J}{n_\mathrm{CO}}\ \mathrm{erg/s/molecule/sr}\,,
\end{equation}
from which we get the total cooling rate
$\Lambda_\mathrm{rot}=n_\mathrm{CO}\sum_JI_J$.

The vibrational component is taken from \citet{HollenbachMcKee79},
considering collisions with $\mathrm H_2$ whose rates, are

\begin{eqnarray}
\gamma_{01}&=&4.3\times10^{-14}T\exp\left(-68/T_3\right)\exp\left(-E_{10}/kT\right)\,\nonumber\\
\gamma_{02}&=&2.7\times10^{-13}T\exp\left(-172/T_3\right)\exp\left(-E_{20}/kT\right)\,,
\end{eqnarray}
where $E_{10}/k=0.5E_{20}/k=3080\,\mathrm{K}$, $T_3=T^{1/3}$, and $k$
is the Boltzmann's constant.


\textbf{Cooling by the metals.}\label{metalcooling} The cooling
process by the metals is included using the list of the metals
adopted by \citet{Maio07}, namely  CII, SiII, OI and FeII \citep[see
the Appendix B in][]{Maio07}. This source of cooling is particularly
important for temperatures lower than $10^2\,\mathrm{K}$. The
cooling is due to the fine structure level transitions of the
ionized carbon $2\mathrm P(3/2) \to 2\mathrm P(1/2)$, and similarly
for the ionized silicates. There are two other species that take
part in this process: neutral oxygen (nine transitions between
levels ${}^1\mathrm S_0$, ${}^1\mathrm D_2$, ${}^3\mathrm P_0$,
${}^3\mathrm P_1$, and ${}^3\mathrm P_2$) and ionized iron (five
transitions between levels ${}^6\mathrm D_i$ with $i$ odd in the
range $[1,9]\in\mathbb N$). The cooling from each transition is
additive and is given in erg/cm$^{3}$/s by

\begin{small}
\begin{equation}
  \Lambda_i=\frac{\gamma_{ji}^{\mathrm H}+\gamma_{ji}^{\mathrm e}n_{\mathrm e}/n_{\mathrm H}}
       {\gamma_{ji}^{\mathrm H}+\gamma_{ij}^{\mathrm H}
         +\left(\gamma_{ji}^{\mathrm e}+\gamma_{ij}^{\mathrm e}\right)n_{\mathrm e}/n_{\mathrm H}
       +A_{ij}/n_{\mathrm H}}n_{tot}A_{ij}\Delta E_{ij}\,
\end{equation}
\end{small}

\noindent where $\gamma_{ij}^{\mathrm H}$ is the reaction rate for
the hydrogen, and $\gamma_{ij}^{\mathrm e}$ is the same but for the
electrons, $A_{ij}$ is the Einstein's coefficient between the {\it
i-th} and {\it j-th} levels, $\Delta E_{ij}$ is the energy
difference between the levels, $n_{tot}$ is the total number density
of the species considered, and finally, $n_{\mathrm H}$ and
$n_{\mathrm e}$ are the hydrogen and electron number density,
respectively. To complete the equation we need to know
\begin{equation}
  \gamma_{ji}=\frac{g_i}{g_j}\gamma_{ij}\exp\left[-\Delta E_{ij}/\left(k_BT_{\mathrm{gas}}\right)\right]\,,
\end{equation}
both for hydrogen and electrons; $g_i$ and $g_j$ are the level
statistical weights, $k_B$ is the Boltzmann's constant, and
$T_{\mathrm{gas}}$ is the gas temperature.

If the model is calculated at high redshift, the CMB
pumping must be considered. This process is described by including
the stimulated emission coefficient in the treatment of the cooling.
The black-body radiation field of the CMB has a temperature that
depends on the redshift as $T_\mathrm{CMB}(z)=(1+z)
T_\mathrm{CMB}(0)$ K. However, as the models of the ISM refer to the
current age, the temperature is kept constant and equal to the
present day value (see below). In any case, owing to the low value
of the present-day CMB temperature, this effect is neglected.
To obtain the total cooling by the metals we use
$\Lambda_{\mathrm{metals}}=\sum_i\Lambda_i$ where $i$ indicates a
generic level transition for any of the four metals we have
included.

The cooling proposed by \citet{Maio07} is  improved implementing the
cooling rates  by  \citet{GloverJappsen2007}; in particular, we use
the de-excitation rates calculated for the collisions with ionized
and molecular hydrogen when available. We also include the cooling
from neutral C, Si (as  in \citet{GloverJappsen2007}), Fe, and
ionized O (as in \citet{HollenbachMcKee89}).

\textbf{Total cooling.} The cooling rate by all the above processes
is additive and can be described by

\begin{eqnarray}
\Lambda_{\mathrm{tot}}& =& \Lambda_{\mathrm{X}}+\Lambda_{\mathrm
H_2}+\Lambda_{\mathrm{HD}}  + \Lambda_\mathrm{CO} +  \nonumber\\
                      && + \left(\Lambda_{\mathrm
C^+}+\Lambda_{\mathrm
O}+\Lambda_{\mathrm{Si^+}}+\Lambda_{\mathrm{Fe^+}}\right)\,.
\end{eqnarray}
where $\Lambda_\mathrm{X}$ is either $\Lambda_{\mathrm{C}92}$ or
$\Lambda_{\mathrm{SD}93}$ as appropriate, and all the $\Lambda$s are
functions of the temperature. Both  $\Lambda_{\mathrm H_2}$ and
$\Lambda_{\mathrm{HD}}$ are the cooling functions of the two
molecular species, and the terms in brackets are the cooling of the
metals.

The  overall rate of  temperature change due to the heating and
cooling is given by the following equation
\begin{equation}\label{cooling}
  \frac{\mathrm dT}{\mathrm dt}=\frac{\gamma-1}{k_B\sum_in_i}(\Gamma-\Lambda)
  \ \mathrm{K/s}\,,
\end{equation}
where $\Gamma$ is the heating source (if present), $\Lambda$ the
cooling source, $n_{i}$ the number density (the sum is over
all the elements), $k_B$ the Boltzmann constant, and $\gamma$ 
the adiabatic index defined in \citet{Merlin11} as
\begin{equation}\label{adiab_index}
  \gamma=\frac{5\,x_\mathrm{H} +5\,x_\mathrm{He} + 5\,x_\mathrm{e^-} + 7\,x_\mathrm{H_2}}
    {3\,x_\mathrm{H} +3\,x_\mathrm{He} + 3\,x_\mathrm{e^-} + 5\,x_\mathrm{H_2}}\,,
\end{equation}
where $x$ is the  fraction of the element indicated by  the
subscript. For an ideal gas of pure hydrogen this value is $5/3$. If
the gas is made only of molecular hydrogen we have $\gamma=7/5$. In
eqn. (\ref{adiab_index}) we use a linear fit of the data proposed by
\citet{Boley2007} for the  adiabatic index of $\mathrm
H_2$ under $\log(T)\le2.6$ (see also erratum), considering a 3:1 ortho/para
ratio mix.

The generic heating term $\Gamma$ can be used to introduce
heating phenomena like SN{\ae} feedback, cosmic rays, and others. In
the current version of \textsc{ROBO}, $\Gamma$ is used as a free
parameter that is kept constant during a single run. In any case,
the $\Gamma$ term can be specified by the user according to his
scientific aims.

\section{Code description}\label{Code}

The code has been developed with IDL\footnote{IDL is a product of
ITT Visual Information Solutions, http://ittvis.com/}. Its user
friendly features help the development of applications that are
otherwise difficult to build in FORTRAN. The code is divided in
self-explanatory procedures (routines) that are grouped in four
classes (gas chemistry, dust, cooling, and general code behavior).
The main routine calls all other subroutines that are needed to
calculate the gas evolution. The first group of routines calculates
the reaction rates  and updates the density of each species.

\textsc{Mass conservation.}  The total mass per unit volume of the
ISM at time $t$ is $M(t)=\sum_in_i(t)\,m_i$, where $n_i(t)$ is the
current number density of the {\it i-th} species, $m_i$  its
molecular or atomic mass, and the sum is over all the species. While
the number densities can vary with time, the total mass must be
conserved. In the numerical integration of the system of
differential eqns. (\ref{cauchy_sys}), the conservation of the total
mass is not always guaranteed, because it depends on the physical
processes, the choice of initial parameters and the time step.
Particular care is paid in our test computations to securing and 
checking run-by-run the
 conservation of the total mass, i.e.
$\left|M(t_0)-M(t)\right|/M(t_0)\ll\mathcal O(10^{-10})$ where
$M(t_0)$ is the initial total mass of the system per unit volume.

\textsc{Time steps.}  Even with short time steps, it may happen that
some species reach negative values in a single time step. This is
because the differential of a species could be negative, and its
absolute value higher than the relevant number density. This problem
could be solved by forcing the species to be positive. This would
cause difficulties with the mass conservation, and then in the
subsequent time step the solver would again change the sign of the
species abundance, and finally, the cooling would produce some nonphysical 
effects. In brief, a negative value of the abundance of the
species would artificially turn the cooling into heating, which is
clearly impossible. The obvious way out is to suitably choose the
time step.

\textsc{Numerical Solvers.} In our system of differential equations
(\ref{cauchy_sys}) there are  $76$ reactions to deal with ($64$
reactions plus $12$ photochemical processes), so a fast and accurate
solver is needed. We use the routine LSODE of IDL. LSODE uses
adaptive numerical methods to advance the solution of a system of
ordinary differential equations by one time step \citep{Hindmarsh83,
Petzold83}, optimizing the process for user-defined time steps. In
our case LSODE absolute tolerance is $10^{-40}$, and the relative one
is $10^{-12}$.

 \textsc{Integrators.}
The number densities of all the particles (atomic species,
molecules, dusty components, and free electrons) are governed by the
balance between creation and destruction processes and, even more
important, span very wide ranges of values.

\textsc{Ranges of applicability.} \textsc{ROBO} can be safely used
in the following intervals for temperature, density, and
metallicity: $10 \leq T \leq 10^7$ K, $10^{-12} \leq n \leq 10^3$,
and $10^{-12} \leq Z \leq 10$. The density range is somewhat limited
towards the high value end: observational estimates of the density
in molecular clouds can indeed reach higher values. Work is in
progress to extend the density range.  Furthermore, let us remind
the reader that $Z=1$ means $\mathrm{[Fe/H]=0}$, the solar case, and
$Z=10$ means $\mathrm{[Fe/H]=1}$; i.e., the ratio of the number
densities (Fe/H) is equal to ten times solar. Finally, as not all
the reaction rates cover the whole range of values, some
extrapolations are required.

The aforementioned ranges of applicability are chosen in such a way as
to guarantee the numerical stability of the system. Indeed, the core
of the model is the system of differential equations that describe
the chemical evolution of the gas according to the reactions listed
in Tables \ref{network1} through \ref{photocs}. The  stability of
the system is measured by the conservation of the mass of the ISM
elements. Therefore, we carefully checked whether \textsc{ROBO}
satisfied this conditions for values of the parameters spanning over
large volumes of the parameter space. The results  show that this is
always the case.

\subsection{Free parameters}\label{freeparam}
\textsc{ROBO} contains 47 parameters governing  the physics, the
mathematics, and the numerical procedure. Here we briefly comment on
the most important ones.

\begin{itemize}
    \item \emph{Ionized fraction of metals}: it fixes the fraction of
    $\mathrm C$, $\mathrm O$, $\mathrm {Si}$, and $\mathrm{Fe}$ at the first ionization
    level.

    \item \emph{Metallicity}: the metallicity of the gas is defined as  $Z=10^\mathrm{[Fe/H]}$. It is worth
    recalling here that $Z=1$ is the solar case corresponding to $\mathrm{[Fe/H]=0}$ (see below for more details).

    \item \emph{Number densities}: the number densities of the 28 elemental species and/or molecules,
      free electrons, and dusty components are all in  $\mathrm{cm}^{-3}$.

    \item \emph{Fraction of carbon dust}: the percentage of carbonaceous grains. If it is equal to $1$,
    all the dust is made by graphite grains and PAHs; if this parameter is equal
      to $0$, all the dust is made by silicates. Intermediate values are also possible.

    \item \emph{Gas temperature}: the temperature of the gas depends on the
     cooling and heating processes and changes during the gas evolution.

    \item \emph{Dust temperature}: the temperature of the dust is kept constant during each run.
    It controls the formation of the molecules
    on the surface of the grains and also the accretion of the dust grains themselves.

    \item \emph{CMB temperature}: the gas temperature cannot go below
    the temperature of the cosmic microwave background (CMB).
    At present the CMB temperature is kept constant during each run.

    \item \emph{Cosmic ray field}: the field formed by the cosmic rays that populate the gas. It destroys
    or ionizes molecules like $\mathrm H_2$. This field is expressed in
    $\mathrm s^{-1}$, thus corresponding to the rate of
    ionization of the $\mathrm H_2$ by the cosmic rays.

    \item \emph{Total integration time of a model}: the total time  of each run in s.

    \item \emph{Time step}: the time step used in the models.
    A typical value is $10^3$ years. However,
    it can be changed by the routine LSODE.

    \item Finally, there is a number of flags activating or switching off
    the different sources of heating and cooling and the mechanisms of dust
    formation and destruction we have described in the previous sections.

\end{itemize}

\section{Models  and discussion}\label{Results}
To validate  \textsc{ROBO} we calculated two groups of models
of gas evolution: the first one  consists of $600$ dust-free cases.
Each case corresponds to a different combination of the parameters.
The second group consists of  $32$ models of gas evolution, but the 
presence of dust grains is taken into account in them.

The aim here is to investigate the evolution of the same elemental
species in the presence or absence of the dust. In addition to this, the
large number of cases per group allow us to investigate the model
response at varying the key parameters. Indeed, owing to the large
number of parameters at our disposal, it is not possible to explore
the whole parameter space.

Finally, it is worth noticing that the choice of the parameters for
the two groups of models is somehow guided by 
\textsc{ROBO} mainly being designed to become a sort of ancillary tool
for  NB-TSPH codes. Therefore, in view of this, the two groups of
models use the same parameter space adopted  by \textsc{EvoL}.

\subsection{Dust-free models}
Each model is a simulation of a unitary volume of gas in the absence of
dust. The results consist of $600$ (see below) evolutionary models
of the gas temperature and number densities of the 28 species under
consideration plus the free electrons. As already emphasized, in
each model special care is paid to secure the mass conservation and
to avoid unphysical negative number densities. The models of this
set are calculated neglecting the presence of any type of dust, and
they are meant to check the internal consistency of \textsc{ROBO}.

\subsubsection{Set up of the parameters}\label{testsetup}

Each model gives the thermal and density history of the gas for an
assigned set of parameters. The  number of parameters and the values
assigned to each of them determine the total number of models to
calculate. This is given by $N_m=\prod_{i=1}^Np_i$, where $N_m$ is
the total number of models, $p_i$ the number of different values
for the {\it i-th} parameter, and $N$ the total number of the
parameters, which in our case amounts to $47$ (see Sect.
\ref{freeparam}).

For all the models, we must specify the initial value of the
metallicity, the number densities $n_{\mathrm H_2}$, $n_{\mathrm
H^+}$, $n_{\mathrm e^-}$, and the gas temperature. We adopt four
values for the metallicity $Z=10^{\left[\mathrm{Fe/H}\right]}$:
$\{0,10^{-6},10^{-3},1\}$; five values for the $\mathrm H_2$ number
density: $\{10^{-10},10^{-6},10^{-2},10^{-1},1\}$
$\mathrm{cm}^{-3}$; three values for the $\mathrm H^+$ number
density: $\{10^{-10},10^{-1},1\}$ $\mathrm{cm}^{-3}$; two values for
the electron number density: $\{10^{-10},10^{-1}\}$
$\mathrm{cm}^{-3}$; five values for the gas temperature:
$\{10,10^{3},10^{4},10^{6},10^{8}\}$ K. More details on the
metallicity are given below when we discuss the number densities of
the metals. All the other parameters of the models have a constant
value.

In all the models, the following quantities have a fixed initial
value. They are $n_{\mathrm H}=1.0$ for the neutral hydrogen,
$n_{\mathrm H^-}=10^{-9}$ for the hydride, $n_{\mathrm
H_2^+}=10^{-11}$ for the molecular hydrogen and $n_\mathrm{He}=0.08$
for the helium. No helium ions are present at the beginning,  so we
have $n_\mathrm{ He^+}=n_\mathrm{ He^{++}}=0$. The initial values of
all the deuteroids are set to $10^{-25}$: these are the deuterium
atom, its ions ($\mathrm D^+$ and $\mathrm D^-$), the molecules
$\mathrm{HD}$ and $\mathrm D_2$, and the ion $\mathrm{HD}^+$.

Metals (i.e. C, O, Si, Fe and their ions $\mathrm C^+$, $\mathrm
O^+$, $\mathrm{Si}^+$, and $\mathrm{Fe}^+$) are computed from the
metallicity as follows\footnote{\textsc{ROBO} also lets us insert the
value for single metallic species without using the total
metallicity.}. We start from the general
\begin{equation}
 \left[\mathrm{Fe/H}\right]=\log_{10}\left(\frac{n_\mathrm{Fe}}{n_\mathrm{H}}\right)-
    \log_{10}\left(\frac{n_\mathrm{Fe}}{n_\mathrm{H}}\right)_\odot\,,
\end{equation}
where $n_\mathrm{Fe}$ and $n_\mathrm{H}$ are the number densities of
iron and hydrogen, and the definition of metallicity we have
adopted, namely  $Z=10^{\left[\mathrm{Fe/H}\right]}$. It must be
emphasized  that this notation for $Z$ is different from the
commonly used definition of metallicity, that is, 
$\mathrm{Z}=1-\mathrm{X}-\mathrm{Y}$ with X and Y indicating the
abundances  by mass of hydrogen and helium. In the usual meaning, $Z$
is therefore the mass fraction of all the species heavier than
helium. In our definition $Z$ is simply related to the iron content
$\left[\mathrm{Fe/H}\right]$ by the expression
$\mathrm{Z=10^{[Fe/H]}}$. With this notation, $Z$ can be higher than
one: $Z=1$ corresponds to the solar iron abundance. Considering now
a generic metal indicated by X we can write
\begin{equation}
 \frac{n_\mathrm{Fe}}{n_\mathrm{X}}=\left(\frac{n_\mathrm{Fe}}{n_\mathrm{X}}\right)_\odot\,,
\end{equation}
 and
\begin{equation}
 n_\mathrm{X}=Z\,n_\mathrm{H}\left(\frac{n_\mathrm{X}}{n_\mathrm{H}}\right)_\odot\,.
\end{equation}
Therefore if we know the ratio
$\left(n_\mathrm{X}/n_\mathrm{H}\right)_\odot$, the total
metallicity $Z$, and the number density $n_H$ of hydrogen in the
model, we can derive the number density of the generic metal X. 
By doing this we assume that for any metallicity the relative
abundances of the elements follow the solar partition. In other
words we do not consider here the possibility that the gas may have
a different distribution of the heavy elements 
 from that in the Sun with respect to the hydrogen ($\alpha$-enhancement problem).

We use the same method  to derive the  number density of metal ions
given by
\begin{equation}
 n_\mathrm{X^+}=\kappa\,Z\,n_\mathrm{H}\left(\frac{n_\mathrm{X}}{n_\mathrm{H}}\right)_\odot\,,
\end{equation}
where $\kappa$ represents the ionized fraction
$n_\mathrm{X^+}/n_\mathrm{X}$ of the metal X. In the present models
we set $\kappa=0$, so that all the metals start as neutral species.
If the initial temperature is high enough, the hot gas can ionize
the metals  very early on  in the course of the ISM evolution.

The temperature of the gas is a free parameter, but cannot be
lower than the temperature of the CMB. For the present models we
set $T_\mathrm{CMB}=2.73\ \mathrm K$, the present day measured value
\citep{Boggess1992,Fixsen2009}.

Each model of the ISM is followed during a total time of $3.15\times
10^{14}\ \mathrm s$, which approximately corresponds to  $10^7\
\mathrm{yrs}$. This choice stems from the following considerations.
Each model of the ISM is meant to represent the thermal-chemical
history of a unit volume of ISM whose initial conditions have been
established at a given arbitrary time and whose thermal-chemical
evolution is followed over a time scale long enough for secular
effects to develop but short enough to closely correspond to a sort
of  instantaneous picture of large-scale evolution of the whole
system hosting the ISM unit volume. The initial physical conditions
are fixed by a given set of parameters each of which can vary over
wide ranges. One has to solve the network of equations for a time
scale that is long enough to reveal the variations due to important
phenomena such as star formation, cooling, and heating, but not too
long to let the system  depart from the instantaneous situation one
is looking at. The value of $10^7$ yr resulted in a good
compromise.

The initial value of the time step  is $3.15\times 10^{10}\ \mathrm
s\approx 10^3\ \mathrm{yrs}$. This time step determines the minimum
number of steps required to cover the time spanned by a model.  It
means that each simulation needs at least $10^4$ iterations to be
completed. The LSODE integrator may introduce shorter time steps
depending on the complexity of the problem, so $10^4$ is the minimum
number of required steps. This value for the time step seems to keep
the system stable during the numerical integration.

While integrating, all the sources of cooling are kept active: metal
cooling, $\mathrm H_2$ cooling according to the prescriptions by
\citet{GalliPalla98} and \citet{GloverAbel08}, and finally, the
cooling from deuterated hydrogen.

In general, the initial values of the number densities fall
into three groups:  (i) the elemental species with constant initial
values, the same for all the models (namely $\mathrm{H^-}$,
$\mathrm{H_{2}^+}$, $\mathrm{He^+}$ and $\mathrm{He^{++}}$); (ii)
the elemental species whose initial values are derived from other
parameters (namely He, all the deuteroids and the metals, such as 
$\mathrm{C}$ and $\mathrm{O}$, which depend on the choice
the total metallicity $Z$), and finally, (iii) the elemental species
with free initial conditions (namely $\mathrm{H}$, $\mathrm{H_2}$,
$\mathrm{H^+}$, and $e^{-}$).

\noindent Hydrogen group: $\mathrm{H}$, $\mathrm{H^+}$,
$\mathrm{H^-}$, $\mathrm{H_2}$, and \textsc{$\mathrm{H_2^+}$}. The
initial values of the species $\mathrm{H^-}$ and $\mathrm{H_2^+}$
are $n_{\mathrm{H}^{-}}=10^{-9}$cm$^{-3}$,
$n_{\mathrm{H}_2^{+}}=10^{-11}$cm$^{-3}$ according to
\citet{Prieto08}, while the three other hydrogen species have free
initial values.

\noindent Deuterium group: $\mathrm{D}$, $\mathrm {D^+}$, $\mathrm
{D^-}$, $\mathrm{D_2}$, $\mathrm{HD}$, and $\mathrm{HD^+}$. The
number densities of the deuteroids are calculated from their
hydrogenoid counterparts. For the single atom species we have
$n_\mathrm{D}=f_\mathrm{D}\,n_\mathrm{H}$,
$n_\mathrm{D^+}=f_\mathrm{D}\,n_\mathrm{H^+}$, and
$n_\mathrm{D^-}=f_\mathrm{D}\,n_\mathrm{H^-}$, where
$f_\mathrm{D}=n_\mathrm{D}/n_\mathrm{H}$. For the molecules, we can
consider the ratio $f_\mathrm{D}$ as the probability of finding an
atom of deuterium in a population of hydrogen-deuterium atoms. This
assumption allows us to calculate the $\mathrm{HD}$, $\mathrm D_2$,
and $\mathrm{HD^+}$ number densities as a joint probability. For
$\mathrm{HD}$ and $\mathrm{HD}^+$ we have
$n_\mathrm{HD}=f_\mathrm{D}\,n_\mathrm{H_2}$ and
$n_\mathrm{HD^+}=f_\mathrm{D}\,n_\mathrm{H_2^+}$, but for $\mathrm
D_2$ is $n_\mathrm{D_2}=f_\mathrm{D}^2\,n_\mathrm{H_2}$ as the
probability of finding two deuterium atoms is $f_\mathrm{D}^2$. This
is valid as long as $f_\mathrm{D}\ll1$.

\noindent  Helium group:  $\mathrm{He}$, $\mathrm{He}^+$,
$\mathrm{He}^{++}$. The ratio $\mathrm{n_{He}/n_H}$ is set to 0.08,
thus allowing the initial value of $\mathrm{n_{He}}$ to vary
according to the initial value for $\mathrm{n_{H}}$. The initial
number densities of the species $\mathrm{He}^+$, $\mathrm{He^{++}}$
are both set equal to zero in all the models.

\noindent Metals group:  $\mathrm{C}$,  $\mathrm C^+$, O,
$\mathrm O^+$, $\mathrm{Si}$, $\mathrm{Si}^+$, and
$\mathrm{Fe}$, $\mathrm{Fe}^+$.  The $\mathrm{Fe}$ number
density of the ISM is
\begin{equation}
n_\mathrm{Fe}=n_\mathrm{H}\cdot \mathrm{dex}\left\{[\mathrm{Fe/H}]+
\log\left(\frac{n_\mathrm{Fe}}{n_\mathrm{H}}\right)_\odot\right\}\,,
\end{equation}
where $\left(n_\mathrm{Fe}/n_\mathrm{H}\right)_\odot$ is the
iron-hydrogen ratio for the Sun. To retrieve the number density of a
given metal X we use $n_\mathrm{X}=n_\mathrm{Fe}\cdot f_\mathrm{X}$,
where $f_\mathrm{X}$ is the metal-iron number density ratio in the
Sun.

The list of the species whose initial number densities are kept
constant in all the models of the ISM is given in Table
\ref{fixparam}. The values listed here are either fixed
to a constant value or based upon the number density of one of the
\textit{free} hydrogen species $n_{\textrm{H}}$,
$n_{\textrm{H}_{\textrm{2}}}$ and $n_{\textrm{H}^{+}}$ via the
$f_{\textrm{D}}$ factor. Since $\mathrm{H^-}$ and $\mathrm{H_2^+}$
are constant, then also $\mathrm{D^-}$ and $\mathrm{HD^+}$ are
fixed. Values are indicated as $a(b)=a\times 10^b$. 

The chemical composition of the ISM is typically
primordial with a mass abundances of hydrogen $X= 0.76$, and
helium $Y=0.24$ and all metals $Z \simeq 0$ . The helium-to-hydrogen
number density ratio corresponding to this primordial by mass
abundances is $n_{\textrm{He}}/n_{\textrm{H}} \simeq 0.08$. The
adopted cosmological ratio for the deuterium is $f_\textrm{D} =
n_\textrm{D}/n_\textrm{H} \simeq 10^{-5}$. With the aid of these
numbers and the above prescriptions we get the number density ratios
listed in Table \ref{fixparm} and the initial values of the number
density in turn.

\renewcommand{\arraystretch}{1.1}
\begin{table} \label{fixparm}
\begin{center}
\caption{Initial values for the number densities of the hydrogen and
helium elemental species and the deuteroids. See the
text for details. } \label{fixparam} \vspace{1mm}
\begin{tabular*}{.32\textwidth}{l l| l l}
\hline
$\mathrm{H^-}$     &$1.0 (-9)$   &$\mathrm{D}$    &$f_{\textrm{D}}$n$_{\mathrm{H}}$\\
$\mathrm{H_2^+}$   &$1.0 (-11)$  &$\mathrm{D^+}$  &$f_{\textrm{D}}$n$_{\mathrm{H}^{+}}$\\
$\mathrm{He}$      &$0.8 (-1) $ n$_{\mathrm{H}}$  &$\mathrm{D^-}$  &$f_{\textrm{D}}$n$_{\mathrm{H^{-}}}$\\
$\mathrm{He^+}$    &$0.0 (+0) $   &$\mathrm{HD}$   &$f_{\textrm{D}}$n$_{\mathrm{H_2}}$\\
$\mathrm{He^{++}}$ &$0.0 (+0) $   &$\mathrm{HD^+}$ &$f_{\textrm{D}}$n$_{\mathrm{H_{2}^{+}}}$\\
                    &               &$\mathrm{D_2}$&$f_{\textrm{D}}^{2}$n$_{\mathrm{H_2}}$\\
\hline
\end{tabular*}
\end{center}
\end{table}
\renewcommand{\arraystretch}{1}


\subsubsection{Results for dust-free models}

In this section we discuss the results we have obtained for the
dust-free models. Together with this we also consider a side group
of models calculated with some specific assumptions (e.g. without
metal cooling and others to be explained in the course of the
presentation)  to underline some effects and to better understand
the physical processes taking place in the ISM.

Special attention is paid to check whether there are some unexpected
effects or drawbacks in the models and in \textsc{ROBO}.  We also
describe how different physical quantities affect the overall
behaviour of the gas. This is achieved by varying a single parameter
at a time and keeping constant all the others.

\begin{figure}
\begin{center}
\includegraphics[width=.45\textwidth]{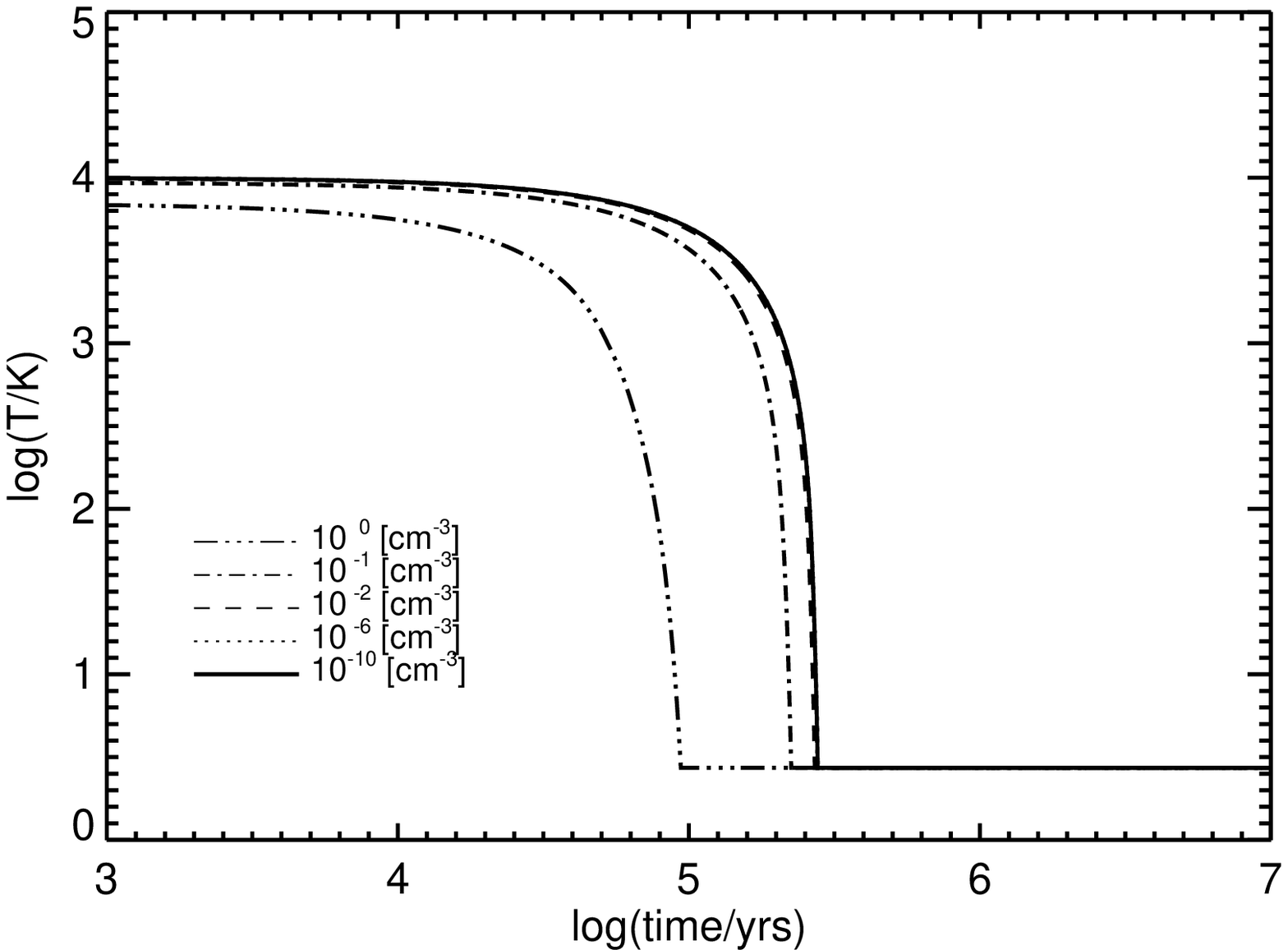}
\includegraphics[width=.45\textwidth]{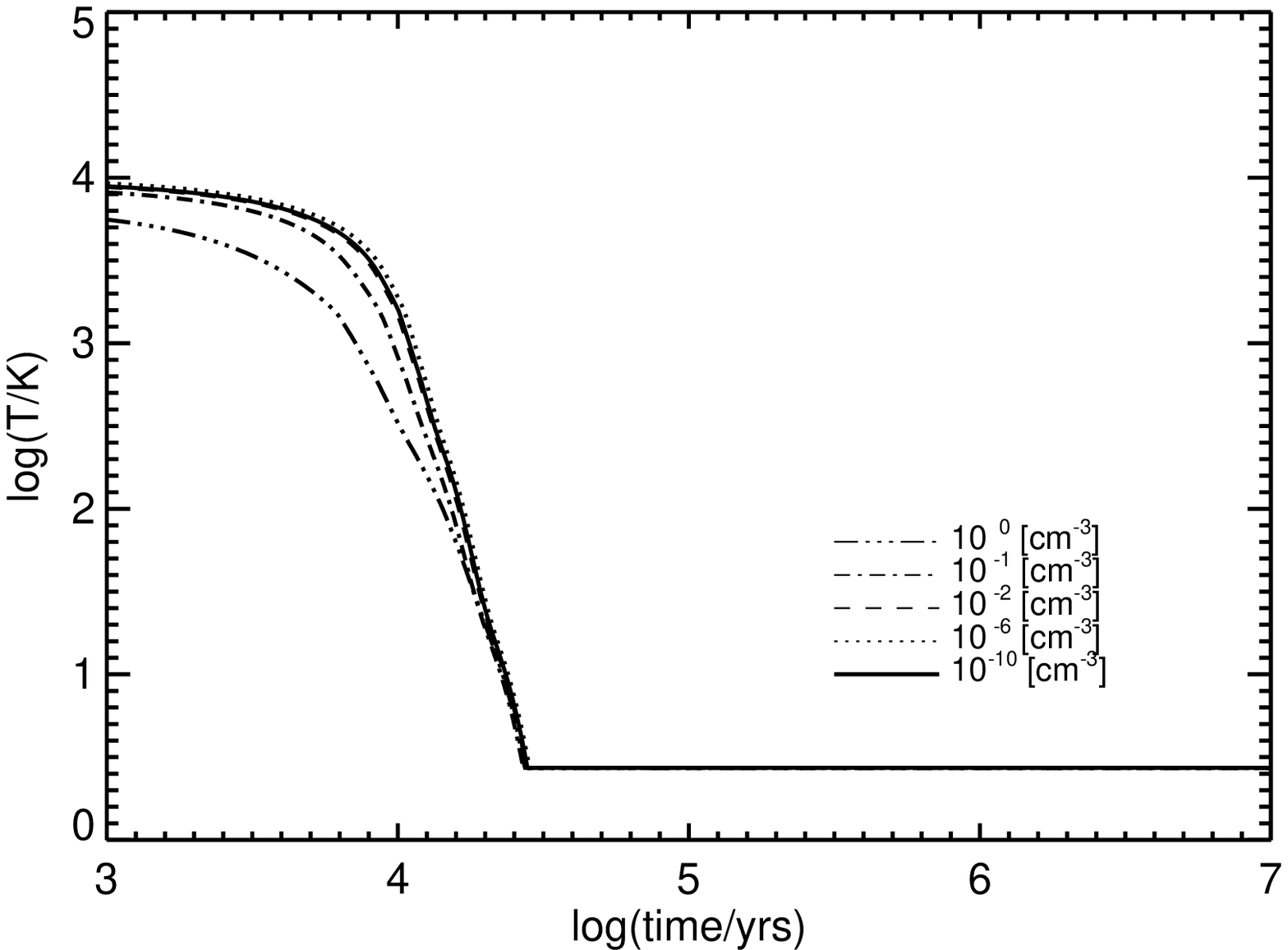}
\caption{Temperature evolution for different values of the initial
$\mathrm H_2$ densities. The lines represent the time evolution in
the cases $\mathrm H_2=10^{-10}$ (solid line), $\mathrm H_2=10^{-6}$
(dotted line), $\mathrm H_2=10^{-2}$ (dashed line), $\mathrm
H_2=10^{-1}$ (dashed-dotted line), and finally, $\mathrm H_2=1$
(three dots-dashed line). All the densities are in $\mathrm
{cm}^{-3}$. Top panel: the models with $Z=10^{-10}$. Bottom panel:
the same as in the top panel but for $Z=1$. See the text for more
details.} \label{pap10e11_T}
\end{center}
\end{figure}

{\it Temperature and $\mathrm{H_2}$ density}. In Fig.
\ref{pap10e11_T} we compare the temperature evolution for different
values of the initial molecular hydrogen densities in two cases with
very different metallicity, i.e. $\mathrm Z=10^{-10}$ and $\mathrm
Z=1$. The important role played by the metal cooling is soon visible
comparing the two panels. In the bottom panel (high-metallicity
$Z=1$), cooling is so strong that all the models simultaneously
reach in a short time interval  nearly the same final temperature of
about $10\ \mathrm K$. In the top panel (low-metallicity case) the
cooling time scale is much longer than in the bottom panel because
the  contribution to  cooling by the metals is negligible. Comparing
each curve with the others in both panels, the effect of the
molecular hydrogen is evident: higher $\mathrm H_2$ densities imply
steeper cooling curves and a shift toward lower temperatures. In the
metal-free models, the temperature systematically shifts to lower
values at increasing the density of the molecular hydrogen (top
panel of Fig. \ref{pap10e11_T}). The effect is similar but enhanced
in the case of metal-rich models (bottom panel of Fig.
\ref{pap10e11_T}).

\begin{figure}
\begin{center}
\includegraphics[width=.45\textwidth]{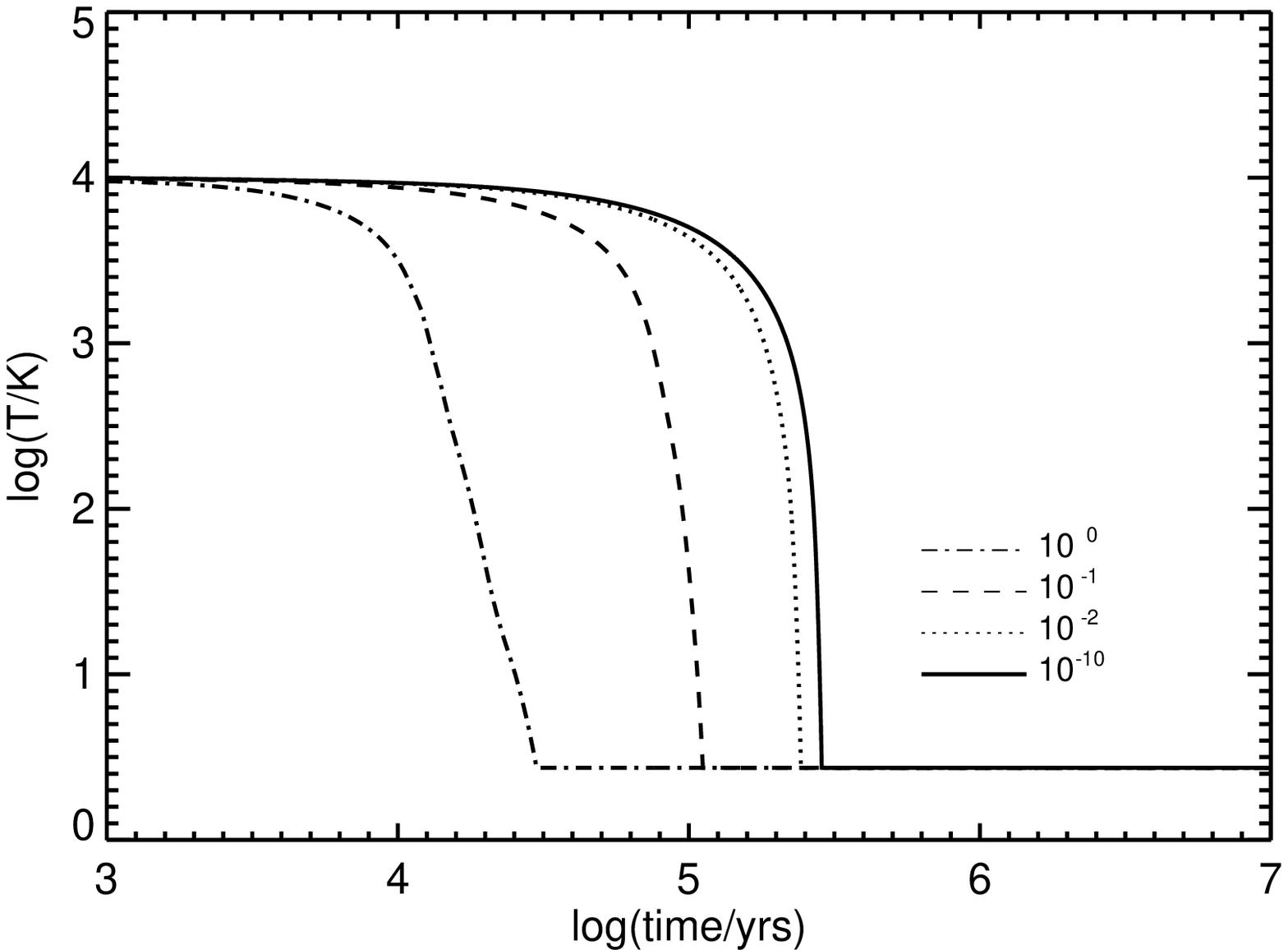}
\includegraphics[width=.45\textwidth]{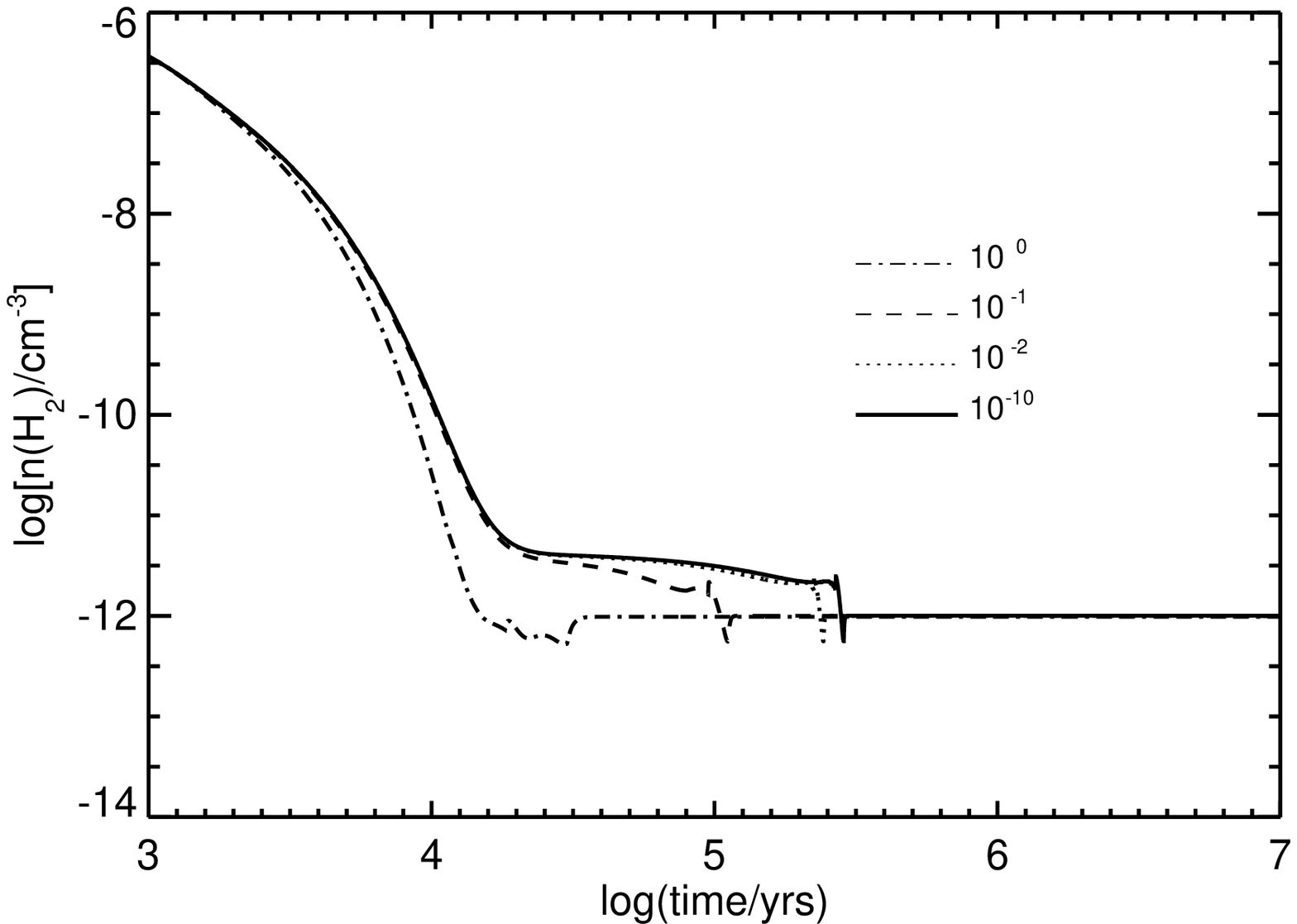}
\caption{Top panel: Temperature evolution for different
metallicities $Z$ and in presence of the UV radiation field. Bottom
panel: evolution of the number density of molecular hydrogen for the
same set of parameters. We represent the cases $\mathrm Z=1$ (dot-dashed
line), $\mathrm Z=10^{-1}$ (dashed line), $\mathrm Z=10^{-2}$
(dotted line), and finally $\mathrm Z=10^{-10}$ (solid line).
See the text for more details.}\label{pap1_T_H2}\end{center}
\end{figure}

\begin{figure}
\begin{center}
\includegraphics[width=.45\textwidth]{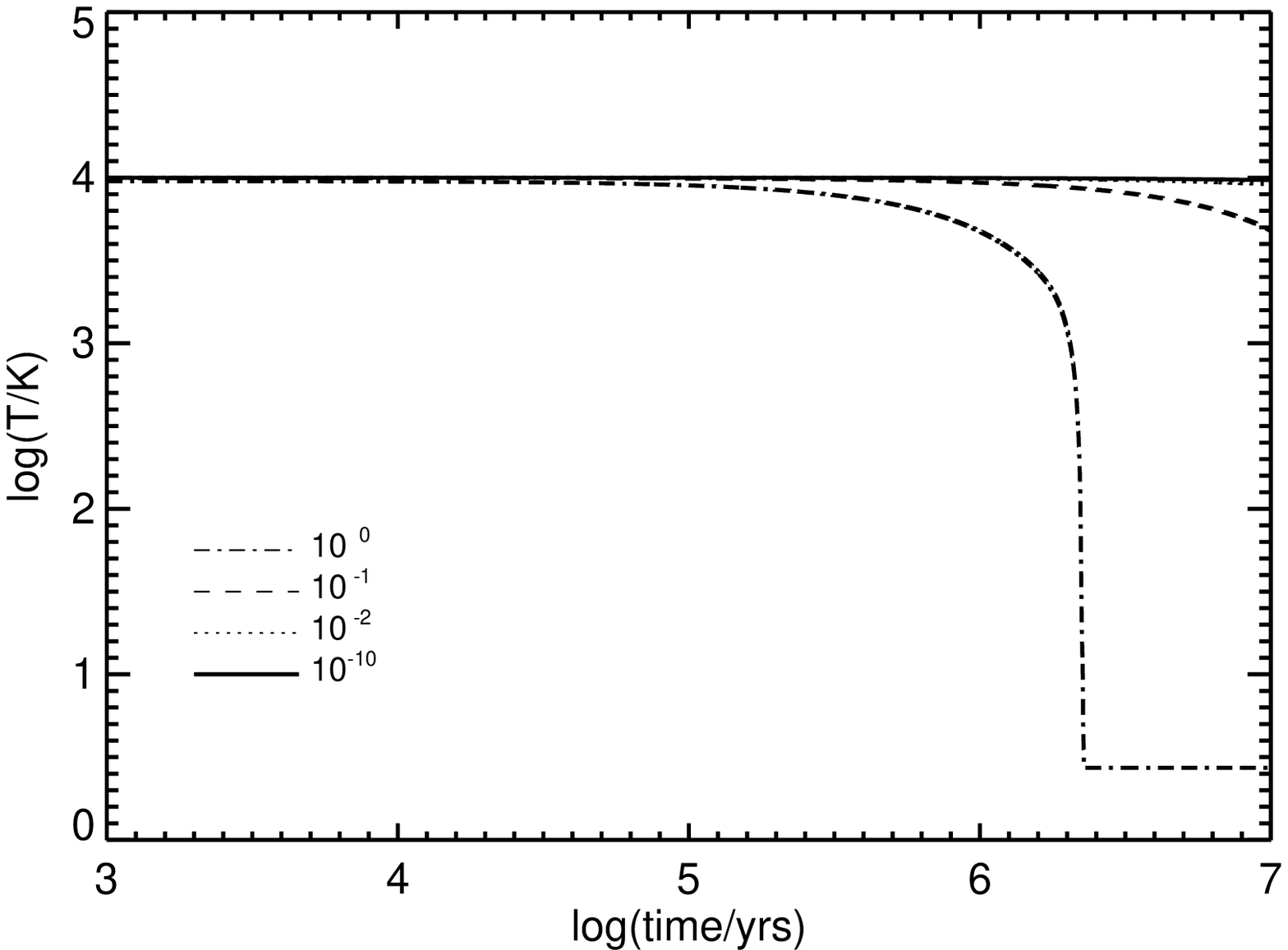}
\includegraphics[width=.45\textwidth]{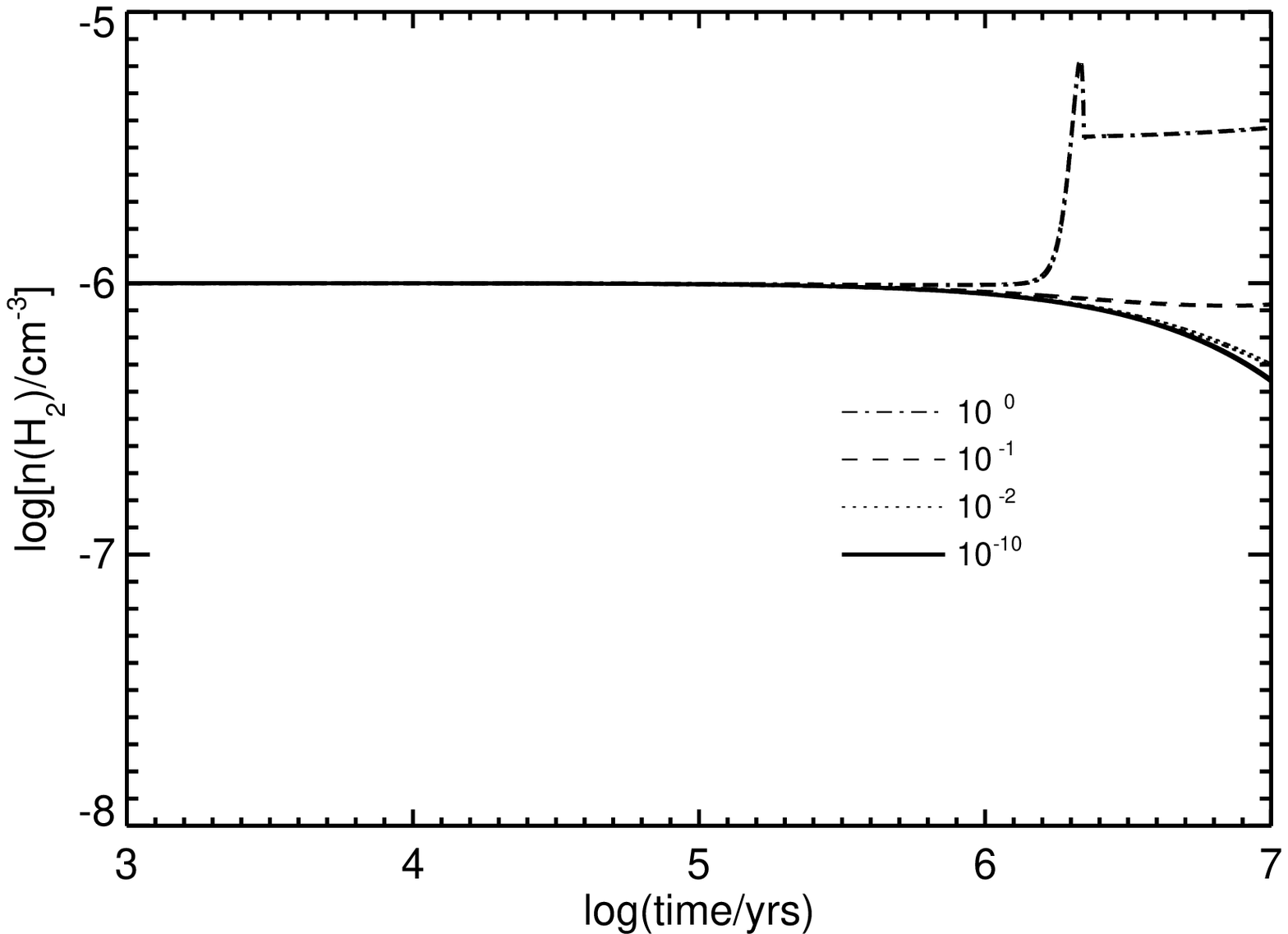}
\caption{Top panel: Temperature evolution for different
metallicities $Z$ but in the absence of the UV radiation field. Bottom
panel: evolution of the molecular hydrogen  for the same set of
parameters. We represent the following cases: $\mathrm Z=1$ (dot-dashed
line), $\mathrm Z=10^{-1}$ (dashed line), $\mathrm Z=10^{-2}$
(dotted line), and finally, $\mathrm Z=10^{-10}$ (solid line).
See the text for more details. }\label{pap5_T_H2}\end{center}
\end{figure}

\begin{figure}
\begin{center}
\includegraphics[width=.45\textwidth]{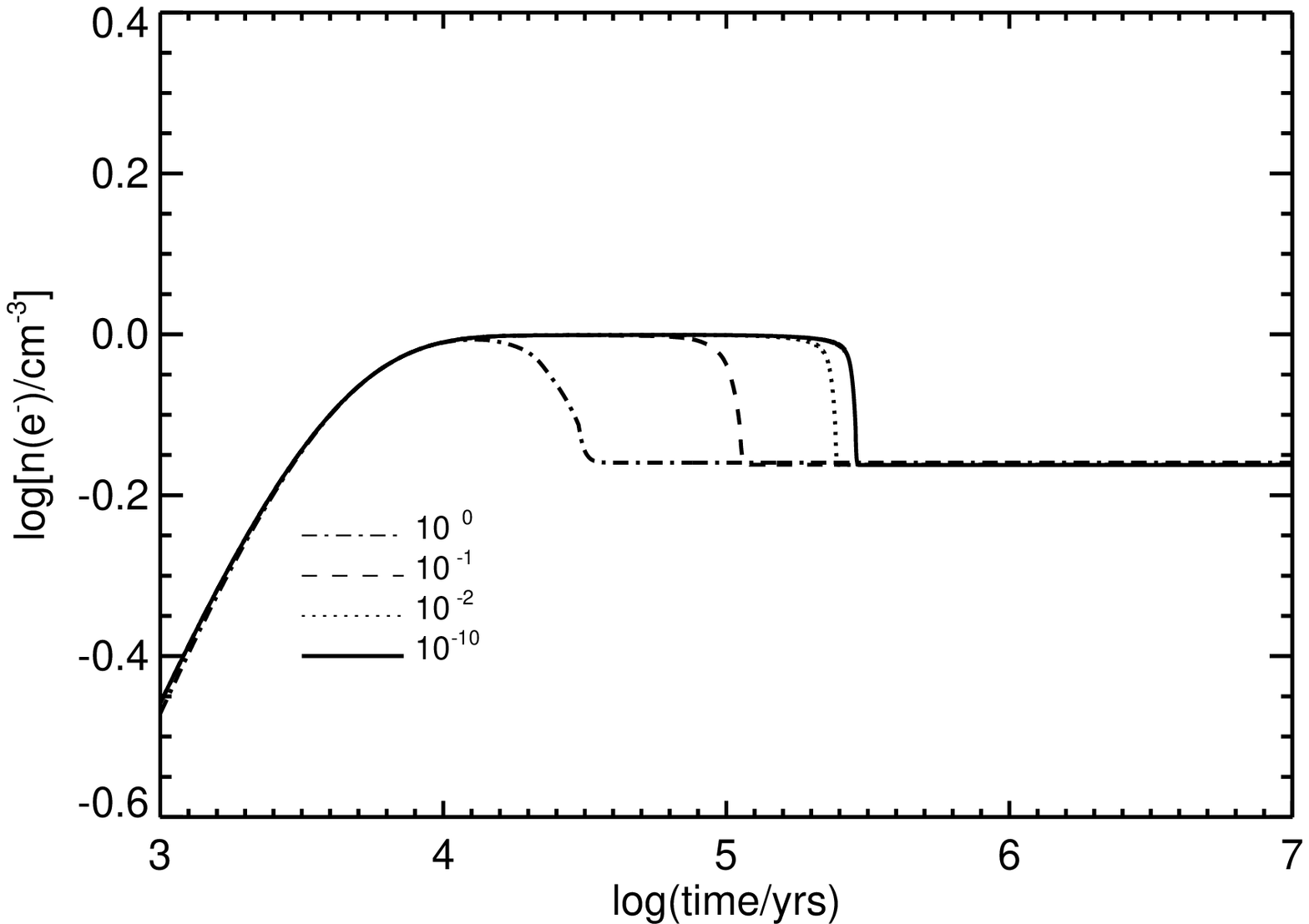}
\includegraphics[width=.45\textwidth]{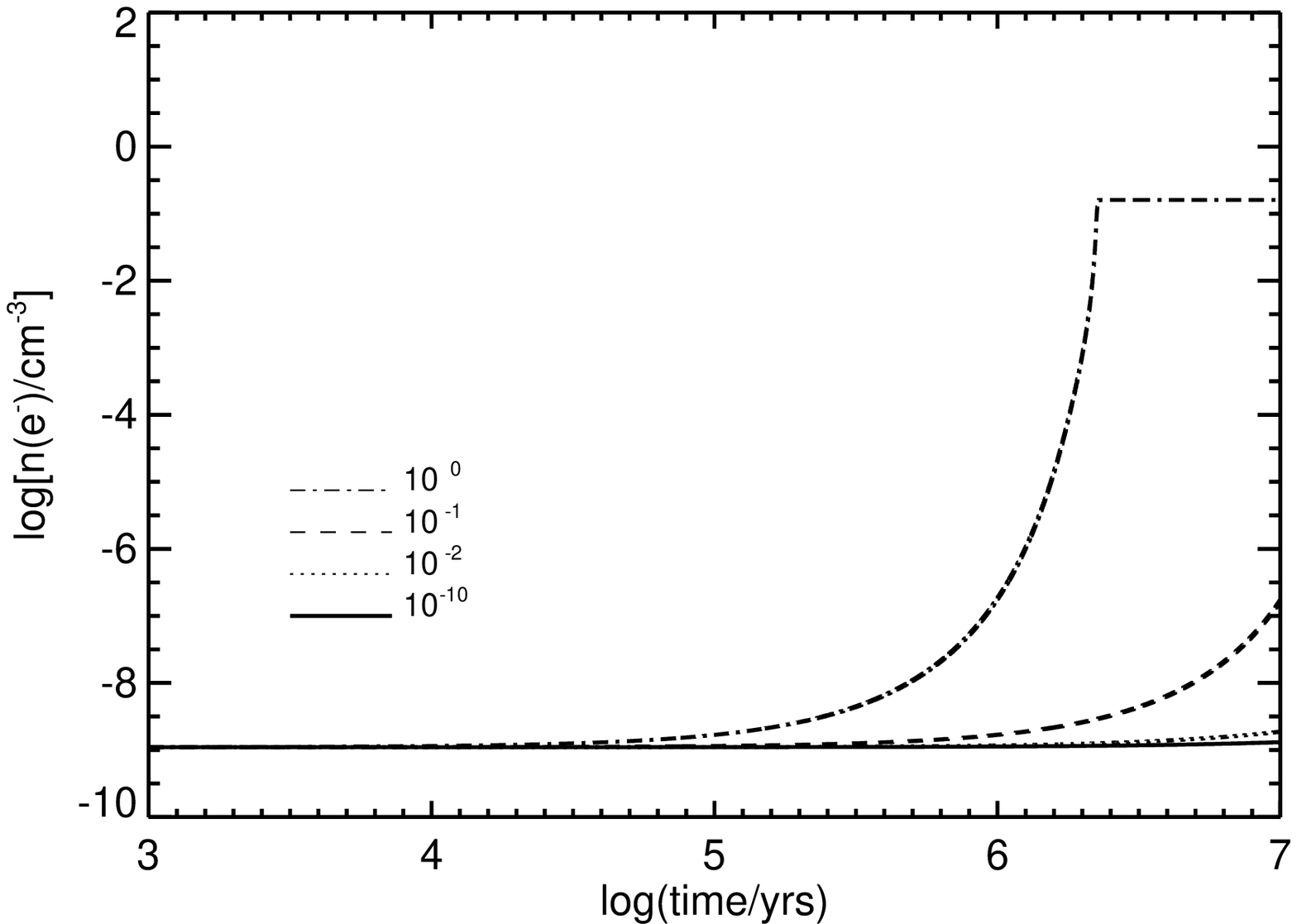}
\caption{ Temporal evolution of the electrons number density for
different metallicities $Z$ and in presence of the UV radiation
field (top panel, related to Fig. \ref{pap1_T_H2}) or without the UV
field (bottom panel, same model of Fig. \ref{pap5_T_H2}). The
metallicities are  $\mathrm Z=1$ (dot-dashed line), $\mathrm
Z=10^{-1}$ (dashed line), $\mathrm Z=10^{-2}$ (dotted line), and
finally $\mathrm Z=10^{-10}$ (solid line). See the text for more
details.   }\label{eUV}\end{center}
\end{figure}

{\it Metallicity}. Figure \ref{pap1_T_H2} shows the evolution of the
temperature (top panel) and $\mathrm{H_2}$ number density (bottom
panel) of models with different metallicity $Z$. Each curve is
characterized by a different initial metallicity. 
The sources of UV radiation are at work in these models. 
As in the previous figure,
the cooling strongly depends on the metallicity. The marked knee in
the temperature-age relationship is caused by the cooling via the
metallicity. Looking at the $\mathrm{H_2}$-age relationship, we note
that the formation of $\mathrm{H_2}$ is favored when more free
electrons are present (an effect that becomes clear when considering
Figs. \ref{pap1_T_H2}, \ref{pap5_T_H2}, and \ref{eUV}).

{\it UV radiation}. The UV radiation plays a key role in the
$\mathrm{H}_2$ density evolution. To cast light on the issue, we
calculated models with no UV sources. They are displayed in Fig.
\ref{pap5_T_H2}. The top panel shows the temperature\,-\,age
relationship, whereas the bottom  panel shows the $\mathrm{H}_2$
number density\,-\,age relationship. Comparing the bottom panel of
Fig. \ref{pap1_T_H2} to that of Fig. \ref{pap5_T_H2}, it is clear the
effect of the UV radiation. When the UV radiation is present (Fig.
\ref{pap1_T_H2}), $\mathrm{H}_2$ is destroyed very early on, whereas
when the UV radiation is absent (Fig. \ref{pap5_T_H2}) the number
density of $\mathrm{H}_2$ remains nearly constant during the whole
evolution. We also note that the effect of the
cooling is stronger in the first case. This happens because the UV flux increases the
density of free electrons, and since they act as colliders, the
eventual higher number of collisions determines a marked improvement
in the cooling process. This effect of the UV flux on the number
density of free electrons is displayed in Fig. \ref{eUV}, where we
show the time evolution of this quantity at varying the metallicity
and switching on (top panel) and off (bottom panel) the UV
radiation.

\begin{figure}
\begin{center}
\includegraphics[width=.45\textwidth]{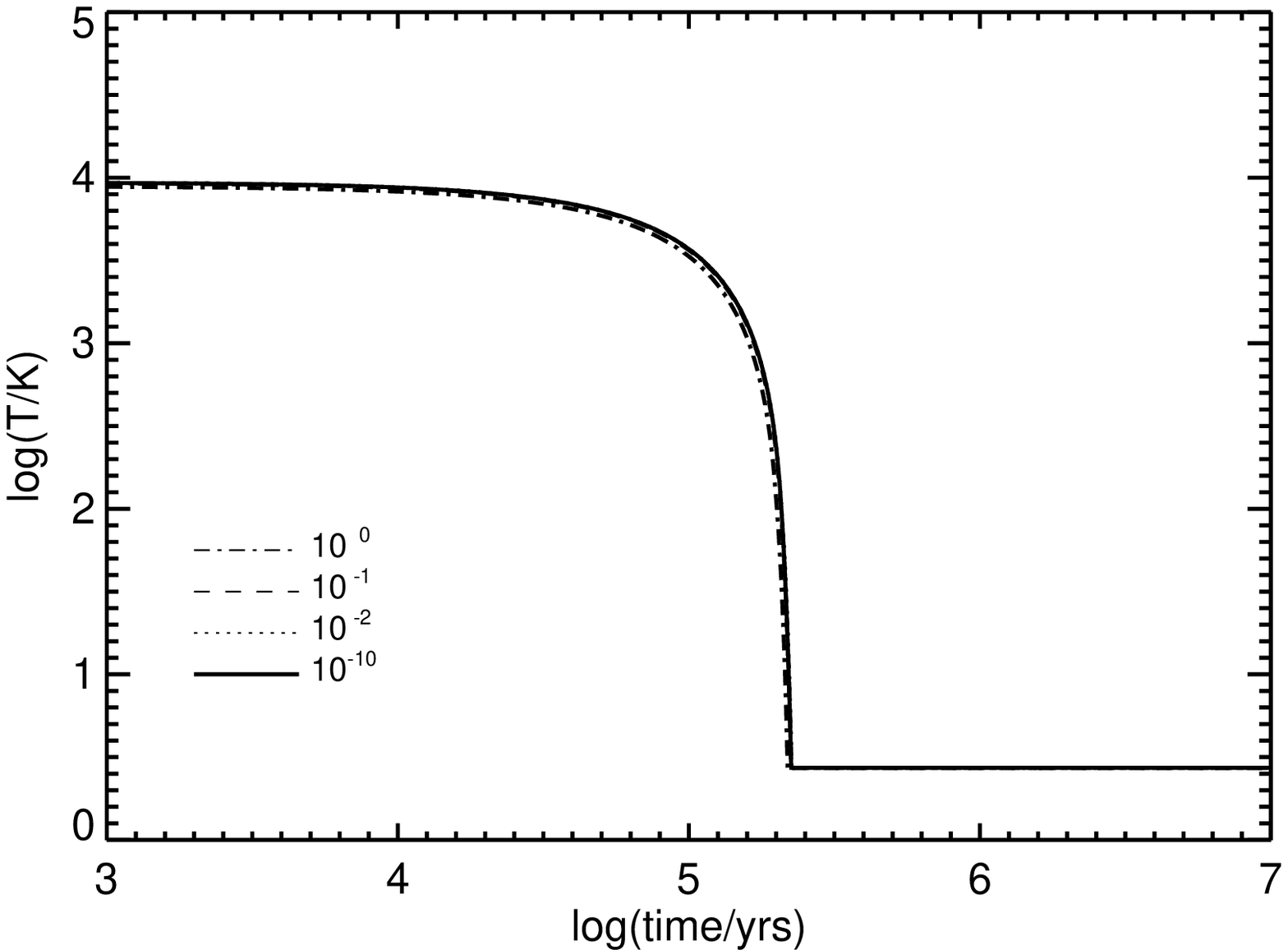}
\includegraphics[width=.45\textwidth]{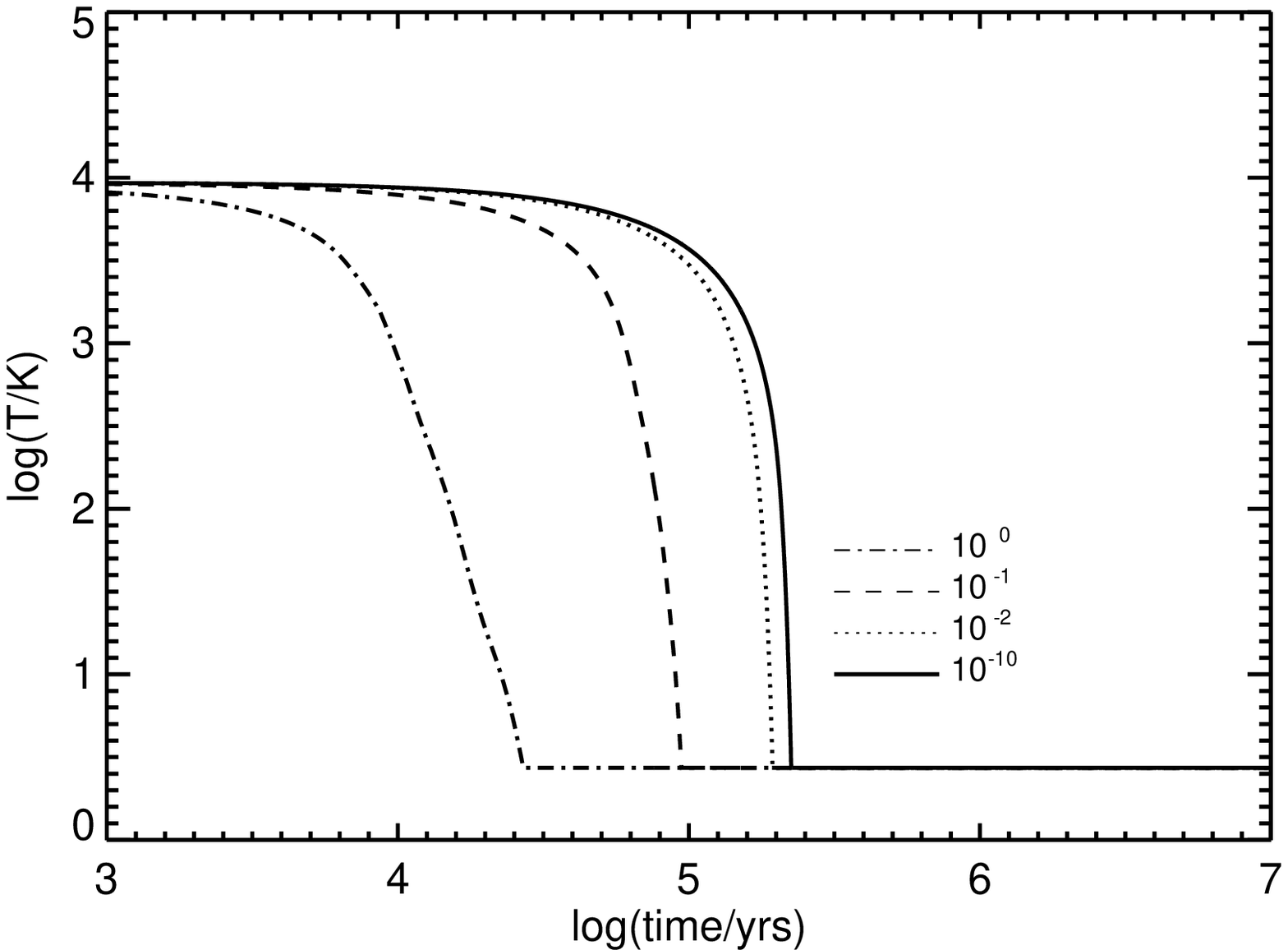}
\caption{Temperature evolution for different metallicities $Z$ and
with the cooling contribution by ionized or neutral metals
alternatively switched off. The \textit{UV flux is switched on}. Top
panel: no cooling by ionized metals. Bottom panel: no cooling by
neutral metals. In both panels the meaning of the lines is as
follows: the dot-dashed line is for $\mathrm Z=1$, the dashed line
for $\mathrm Z=10^{-1}$, the dotted line for $\mathrm Z=10^{-2}$,
and the solid line for $\mathrm Z=10^{-10}$. See the text for more
details.}\label{pap13e14_T}\end{center}
\end{figure}

{\it Ionized versus neutral metals}. The difference in the cooling
between Fig. \ref{pap1_T_H2} (top panel) and Fig. \ref{pap5_T_H2}
(top panel) could be attributed to the different efficiencies of the
cooling by ionized and neutral metals. Indeed, it is worth noticing
that, in order to make the effects of the metals evident, the
initial number density of H$_2$ in Figs. \ref{pap1_T_H2} and
\ref{pap5_T_H2} is set to $10^{-6}$. The UV radiation affects the
ionization state of atoms and molecules, thus varying the partition
between ionized and neutral metals. To clarify this issue, we first
compare models in the presence of UV flux in Fig. \ref{pap13e14_T},
but  the metal cooling by ionized species (top panel)
or neutral species (bottom panel) are alternatively switched off. The
initial number density of H$_2$ is always set to $10^{-6}$. For each
metallicity, the values shown  in Fig. \ref{pap1_T_H2} (top panel),
where both neutral and ionized metals are included, are roughly the
sum of the values presented in Fig. \ref{pap13e14_T} (both panels).
In the top panel of Fig. \ref{pap13e14_T}, the cooling by ionized
metals is switched off so that we would expect  a behavior similar
to that  of Fig. \ref{pap5_T_H2} (top panel), where the UV radiation
is switched off and  the neutral metals are favored. It must be
pointed out, however, that when the UV is switched off (Fig.
\ref{pap5_T_H2}) most metals are neutral. Even
if the ionized metals are switched off (Fig. \ref{pap13e14_T}), the
fraction of neutral metals will be much lower. Indeed, the shape of
the curves displayed in Fig. \ref{pap13e14_T} (top panel) remains
similar to those shown in Fig. \ref{pap1_T_H2} (top panel), and  the
long cooling time scales of Fig. \ref{pap5_T_H2} cannot be
reproduced.

For the same effect of the UV flux on metals, in the top panel of
Fig. \ref{pap13e14_T} (where the UV flux is  on and the ionized
metals are off)  cooling does not depend on the metallicity. The UV
radiation  easily ionizes metals like silicon, carbon, and iron: UV
radiation below 13.6 eV from various astrophysical sources generates a UV background
that ionizes atoms with a first ionization potential lower than 13.6
eV \citep{Maio07}. Consequently, with the UV flux switched on, the
total amount of neutral metals becomes negligible. If the cooling by
ionized metals is switched off and the contribution to the total
cooling by neutral metals is negligible, it follows that the metals
almost do not contribute to  cooling. When the ionized metals
are put back and  the negligible amount of neutral metals is dropped
(Fig. \ref{pap13e14_T} - bottom panel),  we recover the
situation of Fig. \ref{pap1_T_H2} as expected.

\begin{figure}
\begin{center}
\includegraphics[width=.45\textwidth]{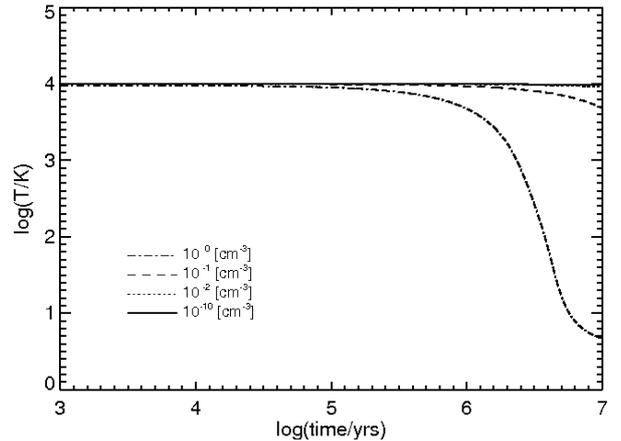}
\caption{Temperature evolution for different metallicities $Z$ and
with the cooling contribution by ionized metals switched off. Also,
the \textit{UV flux is switched off}. The meaning of the lines is as
follows: the dot-dashed line is for $\mathrm Z=1$, the dashed line
for $\mathrm Z=10^{-1}$, the dotted line for $\mathrm Z=10^{-2}$,
and the solid line for $\mathrm Z=10^{-10}$. See the text for more
details.}\label{pap8b_T}\end{center}
\end{figure}

Now  the UV flux is switched off as in Fig. \ref{pap5_T_H2} and the
metal cooling by ionized species (top panel) or neutral species
(bottom panel) is alternatively removed. As we can see in Fig.
\ref{pap8b_T}, with no cooling by ionized metals and only with the
contribution by neutral metals, we are able to reproduce the
behavior observed in Fig. \ref{pap5_T_H2} (top panel). The case
without neutral metals and  only with  cooling by ionized metals is
not shown because it would lead to an almost constant temperature in
all the cases, consistent with the picture we have just described.
In brief, in this case, both H$_{2}$ (the input number
density of H$_{2}$ is set to a lower value, thus 
contributing less to the cooling) and the ionized metals (without UV
flux, as expected, most of the metals are neutral) little contribute
to the cooling process.  We can finally conclude that the partition
between ionized and neutral metals fully explains the difference
between Figs. \ref{pap1_T_H2} and \ref{pap5_T_H2}.

\begin{figure}
\begin{center}
\includegraphics[width=.45\textwidth]{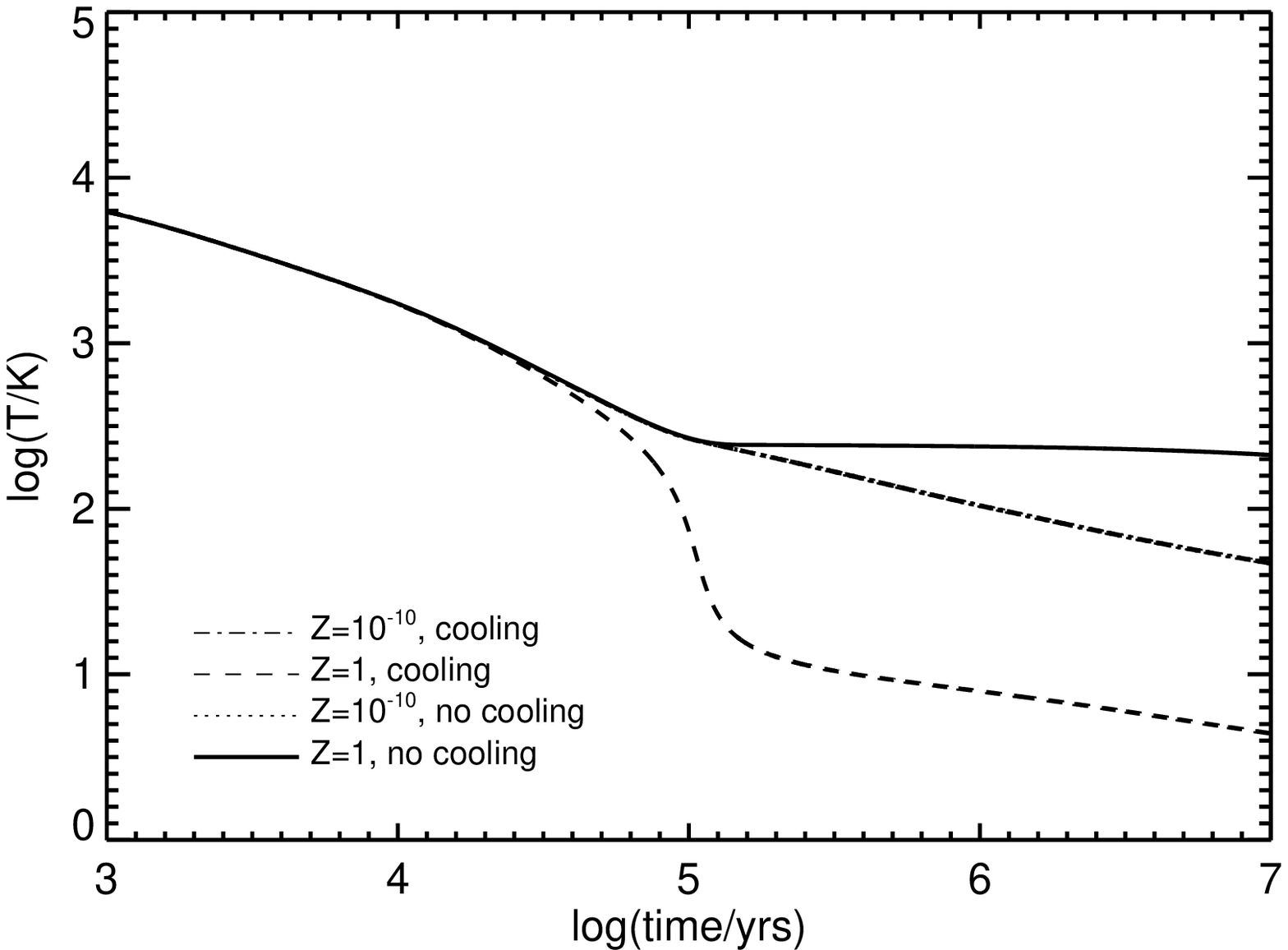}
\includegraphics[width=.45\textwidth]{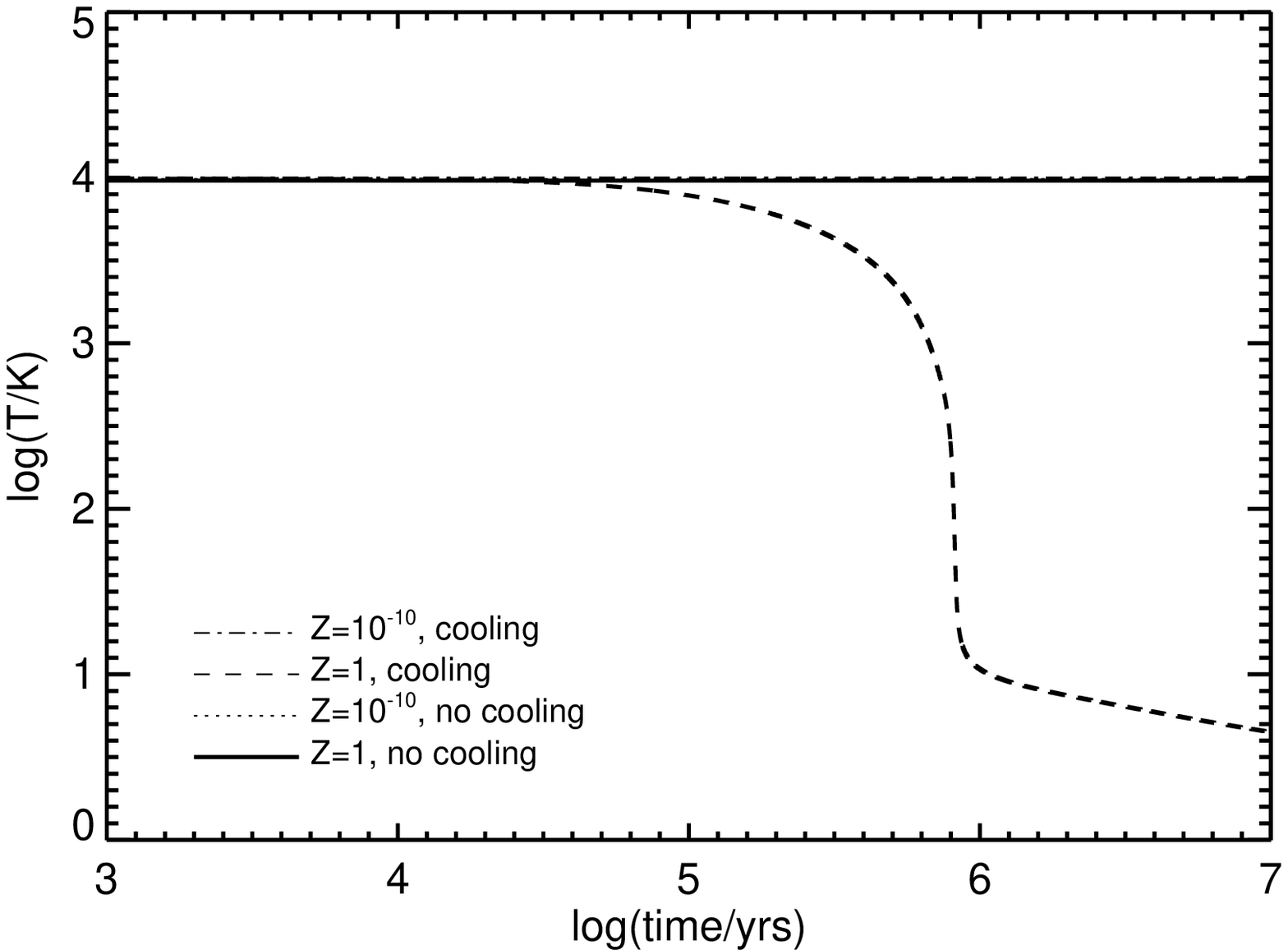}
\caption{Temperature evolution with different metallicities $Z$ and
cooling options by metals. The meaning of the lines is as follows:
high $Z$ and no cooling by metals (solid line), low $Z$ and no
cooling by metals (dotted line), high $Z$ and cooling by metals
included (dashed line), low $Z$, and cooling by metals included
(dashed-dotted). The dashed-dotted and dotted lines overlap. Top
panel: the $\mathrm H_2$ cooling is enabled. Bottom panel: the
$\mathrm H_2$ cooling is switched off. The cooling by \citet{Cen92}
is switched off in both panels.} \label{T_cooling}\end{center}
\end{figure}

\begin{figure}
\begin{center}
\includegraphics[width=.45\textwidth]{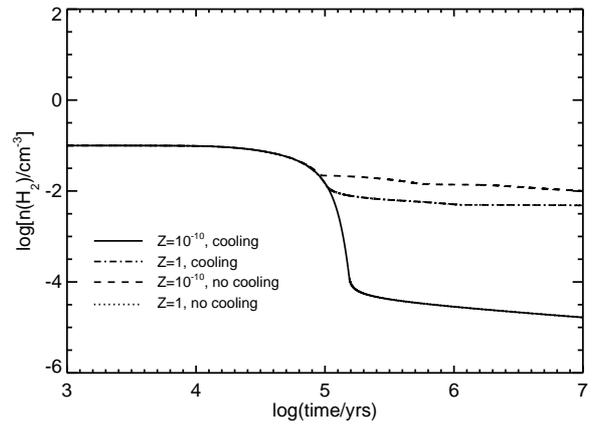}
\caption{Evolution of the number density of molecular hydrogen
H$_{2}$ for different initial metallicity $Z$ and options for the
cooling by metals. The meaning of the lines is as follows: high $Z$
and no cooling by metals (solid line), low $Z$ and no cooling by
metals (dotted line), high $Z$ and cooling by metals enabled (dashed
line), low $Z$ and cooling by metals switched off (dashed-dotted
line). The dashed-dotted and dotted lines overlap. The cooling by
\citet{Cen92} is switched off in both panels.}
\label{H2_cooling}\end{center}
\end{figure}

 {\it \textbf{Metallicity and H$_{2}$ vs. Cooling.}} In Fig.
\ref{T_cooling} are plotted four different models with different
combinations of the metallicity (very high or very low) and cooling
by metals (included or switched off)\footnote{In these models we
switch off the C92 cooling to better highlight the sole effects due
to different metallicities.}. Furthermore,  to  examine the effect
of the H$_{2}$ cooling, we  set the input number density of H$_2$
equal to $10^{-1}$ cm$^{-3}$ (see Sect. \ref{testsetup} for more details on
the selected input values for H$_2$). This plot highlights how the
cooling by metals and H$_2$ affects the temperature and thus the
physical status in the gas. The steep decrease of the temperature
from the beginning of the simulations and common to all four
models is due to the strong H$_2$ cooling. As expected, the model
with high-metallicity and cooling by metals (top panel of Fig. \ref{T_cooling}, dashed line)
is the one experiencing the
strongest cooling. The two models with very low-metallicity have a
similar behavior independent of the presence or absence of the
cooling by metals. The last case, with high-metallicity and no
cooling by metals, is the one with the lowest temperature decrease.
The reason for it is explained in Fig. \ref{H2_cooling}, where the
evolution of $\mathrm{H}_2$ for the four cases under consideration
is shown.

The model with high-metallicity and no cooling by metals has the
lowest $\mathrm{H}_2$ density compared to the other three and
consequently has low cooling. From Fig. \ref{H2_cooling} we also see
that the model with high-metallicity and cooling by metals (dashed
line) has the highest $\mathrm{H}_2$ density. This can explain the
results in the top panel of Fig. \ref{T_cooling}, in the sense that
the cooling process here could be mainly due to the contribution
coming from the molecular hydrogen. The bottom panel of Fig.
\ref{T_cooling} shows the same models with no cooling by $
\mathrm{H}_2$. In this case only the model with cooling by metals
included and high-metallicity undergo a significant cooling. Indeed,
if the effect of $ \mathrm{H}_2$ is neglected, we need metals in
significant amounts to have strong cooling.

\subsection{Models with dust: parameter set up}

In this section we consider models in which the effect of the dust
grains is taken into account. We adopt here the same values for the
parameters as in the dust-free models.

The minimum initial metallicity  is $Z=10^{-6}$, and the initial
number density of the molecular hydrogen is $n_\mathrm{H_2}=10^{-6}\
\mathrm{cm}^{-3}$.  The number densities of electrons and $\mathrm
H^+$ are free parameters; they are
$n_\mathrm{H^+}=\{10^{-10},10^{-1}\}$ and
$n_\mathrm{e^-}=\{10^{-10},10^{-1}\}$, both in units of
$\mathrm{cm}^{-3}$ as usual. The initial temperature is set to
$T=10^4\ \mathrm K$.

The only  difference with respect to the previous models is the
presence of dust. We adopt four values for the number density of
dust grains, namely $n_\mathrm{dust}=\{0,10^{-3},10^{-2},10^{-1}\}$
$\mathrm{cm}^{-3}$. In these models the composition of the dust
mixture is 50\% carbonaceous grains and 50\% silicates.

We also calculate models  with or without dust sputtering by shocks,
in order to describe the behavior of  a turbulent gas particle with
and without the grain depletion due to the shocks.

All the other parameters remain the same as in the dust-free models,
such as the number densities of different elements and the cooling
processes. In all these models thermal sputtering and dust formation
are active, so the dust properties are let change during the
evolution of the interstellar medium.

\subsubsection{Results for models with dust}

\begin{figure}
\begin{center}
\includegraphics[width=.45\textwidth]{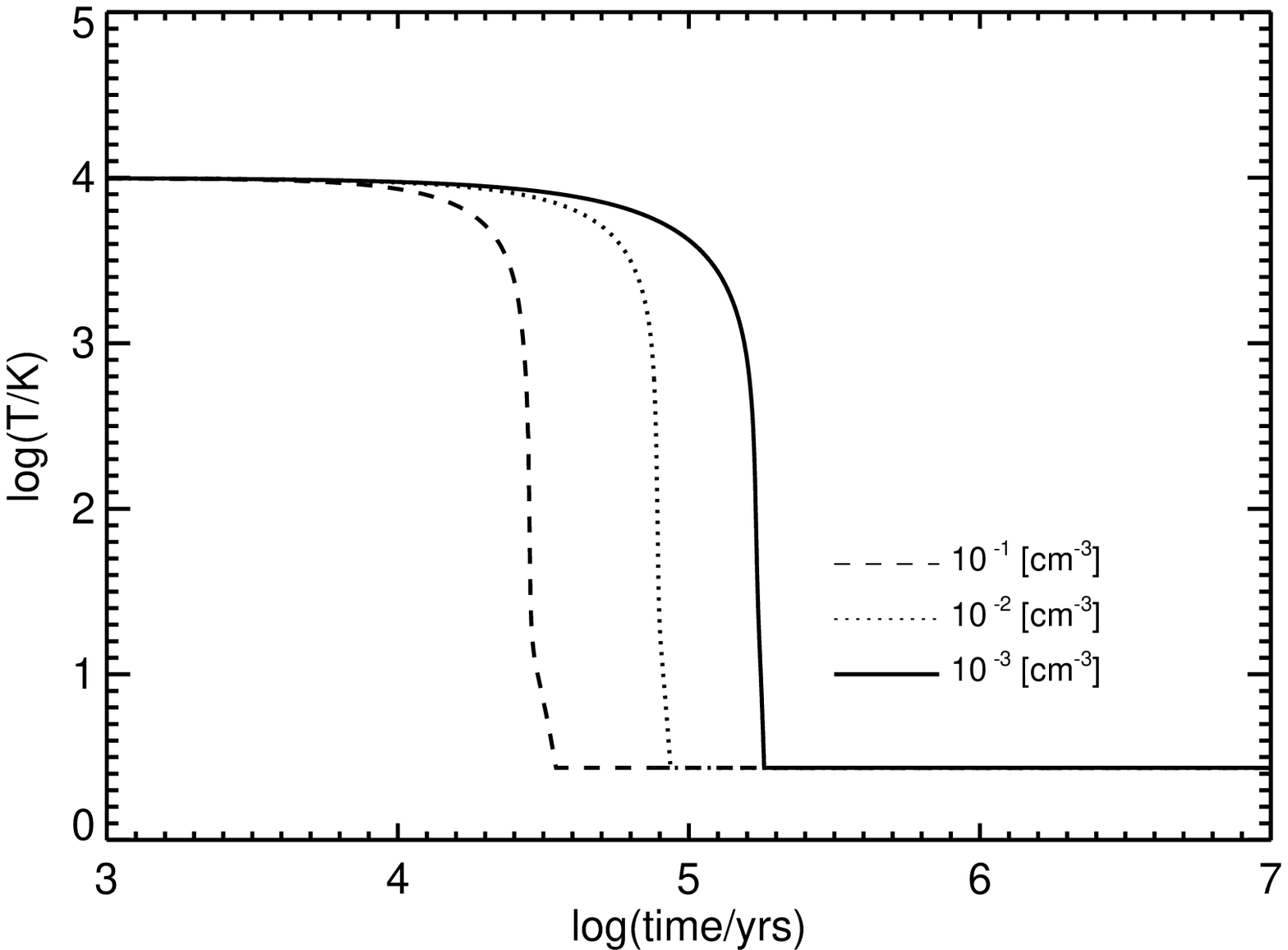}
\includegraphics[width=.45\textwidth]{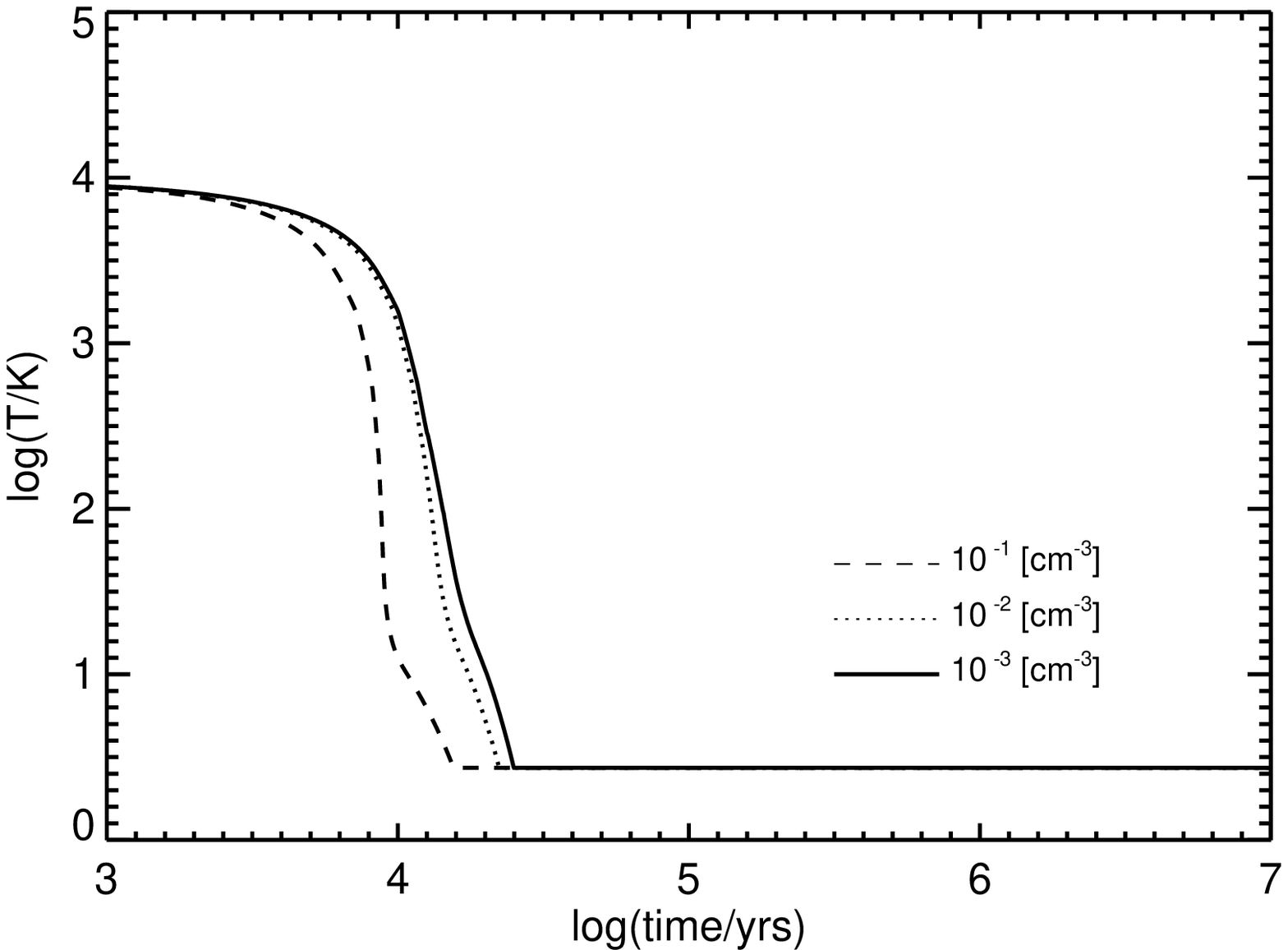}
\caption{Temperature evolution for different amounts of initial
dust. Top panel: low-metallicity cases. Bottom panel: high
metallicity cases. In both panels the meaning of the lines is as
follows: the solid line is for $n_\mathrm{dust}=10^{-3}$, dotted
line is for $n_\mathrm{dust}=10^{-2}$, and finally, the dashed line
is for $n_\mathrm{dust}=10^{-1}$. See the text for more details.}
\label{pap7_T}\end{center}
\end{figure}

\begin{figure}
\begin{center}
\includegraphics[width=.45\textwidth]{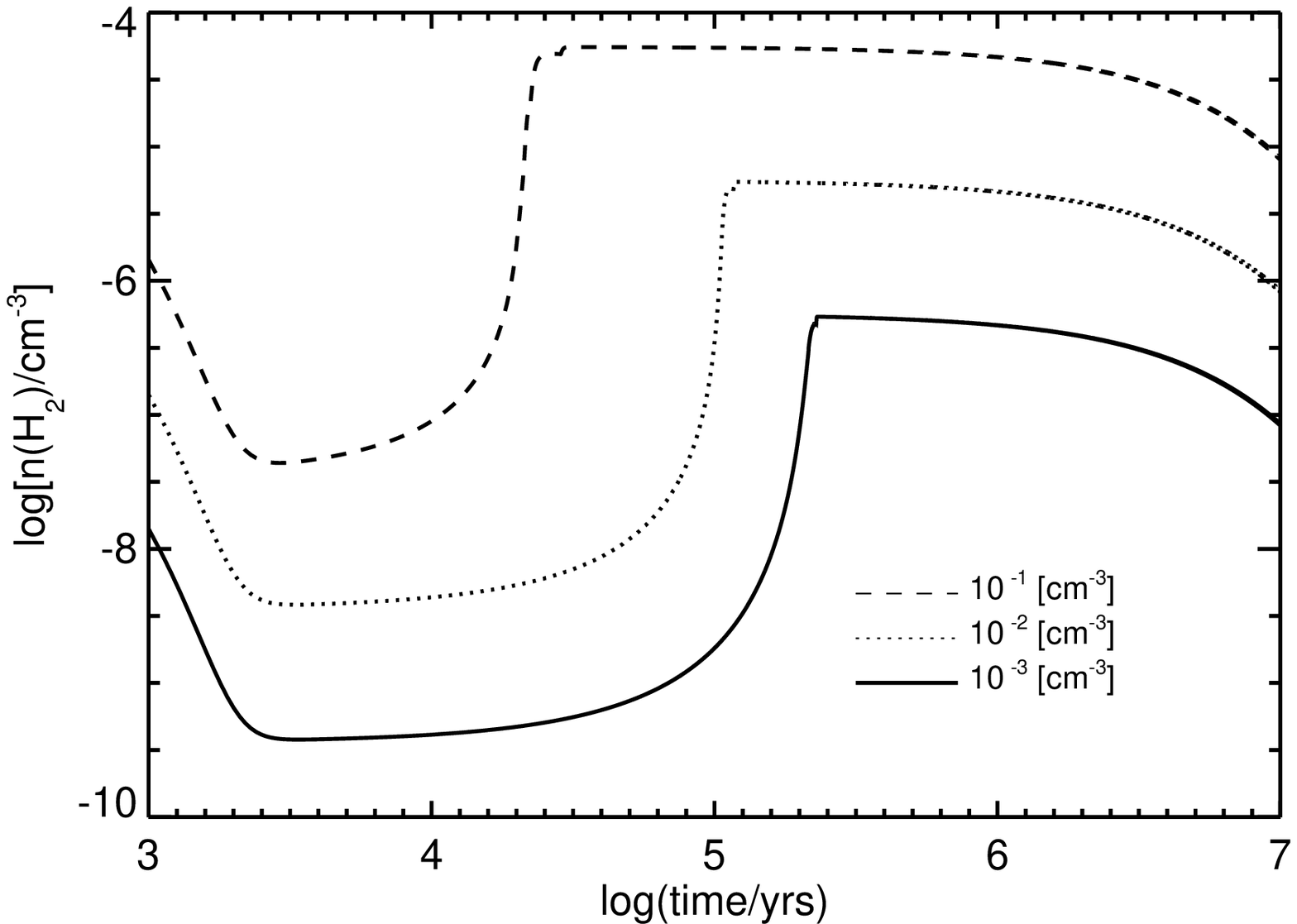}
\includegraphics[width=.45\textwidth]{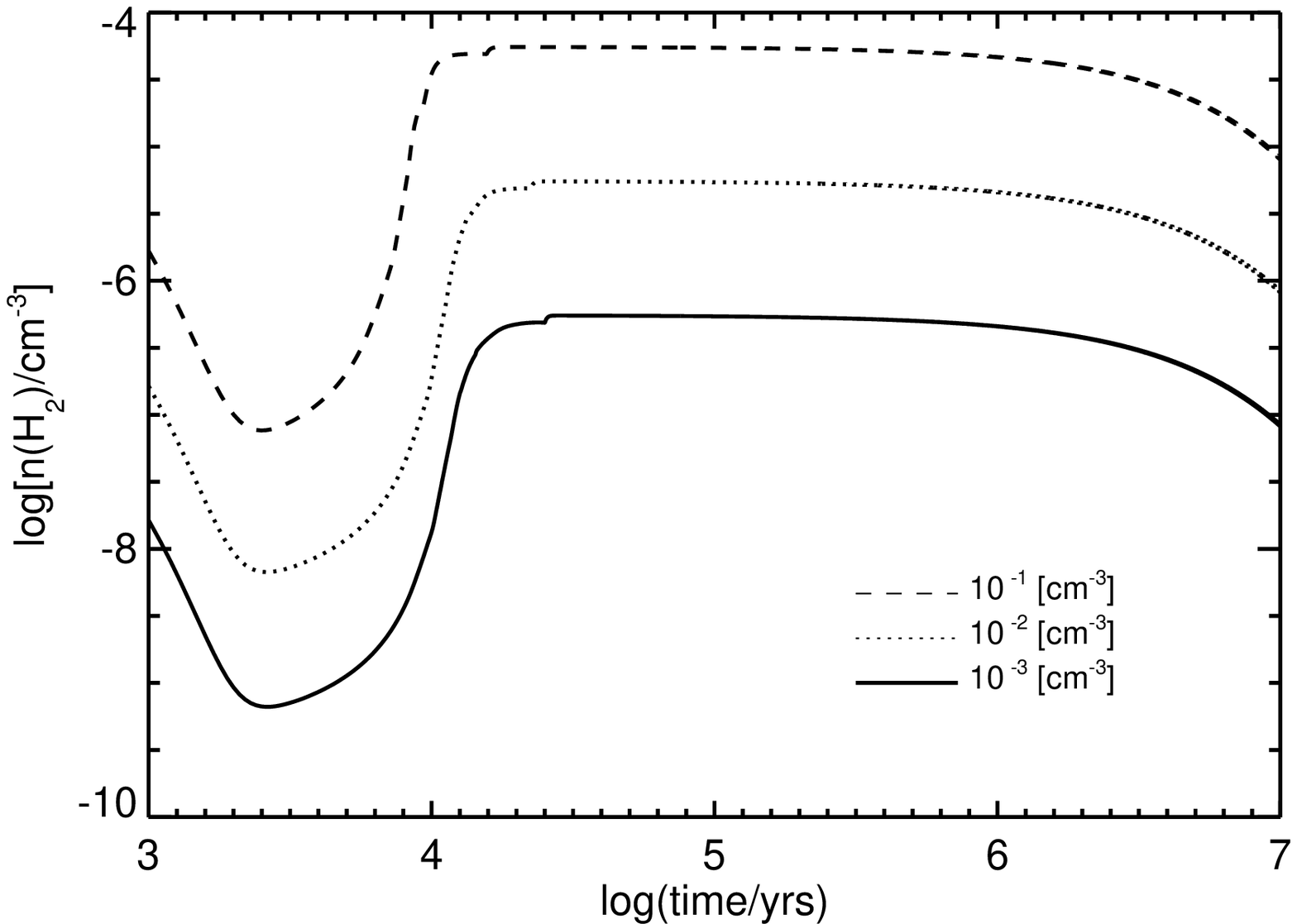}
\caption{Evolution of molecular hydrogen  for different amounts of
initial dust. Top panel: low-metallicity cases. Bottom panel: high
metallicity cases. In both panels the meaning of the lines is the
same as in Fig. \ref{pap7_T}. See the text for more details.}
\label{pap7_H2}\end{center}
\end{figure}

\begin{figure}
\begin{center}
\includegraphics[width=.45\textwidth]{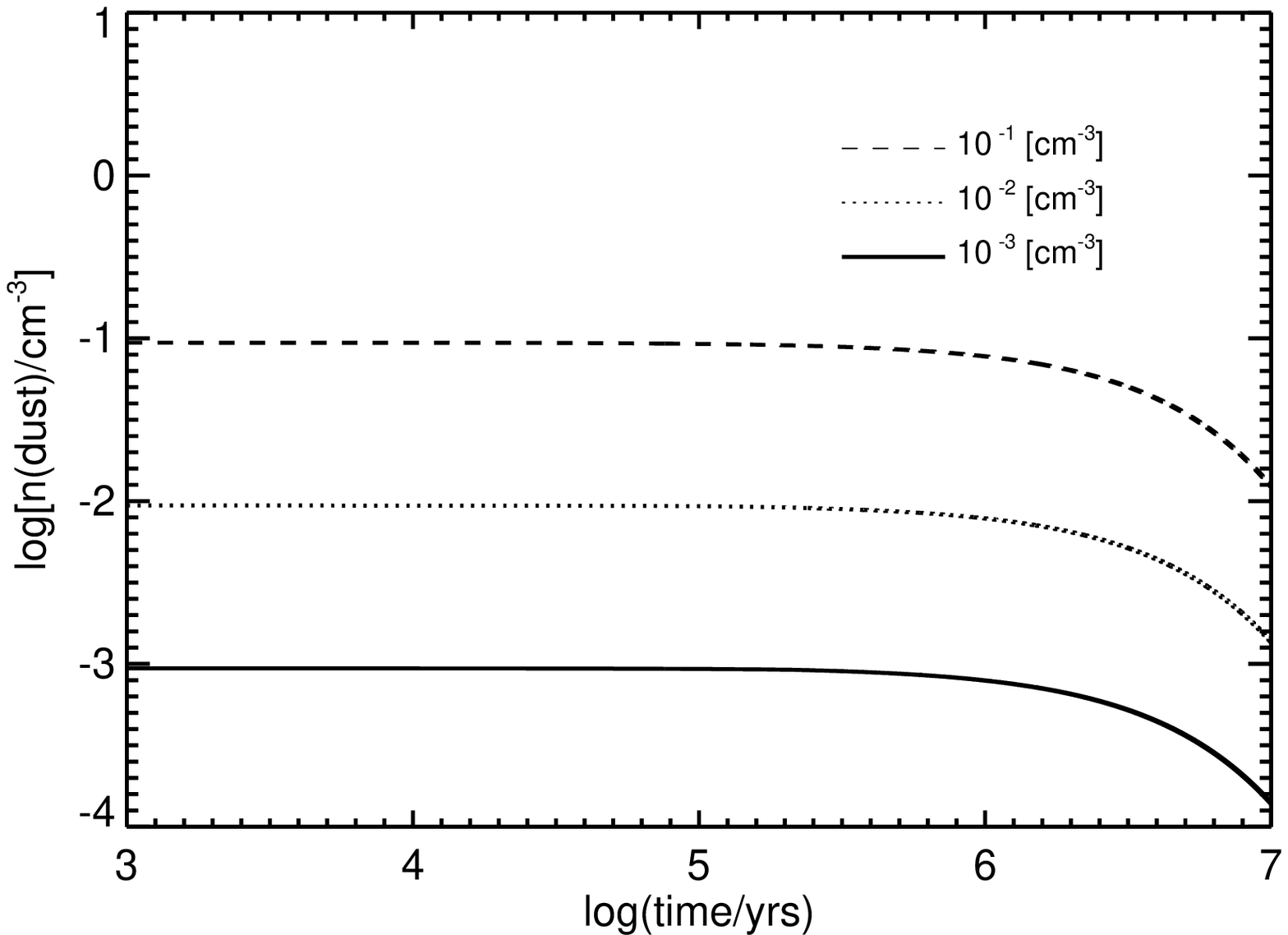}
\includegraphics[width=.45\textwidth]{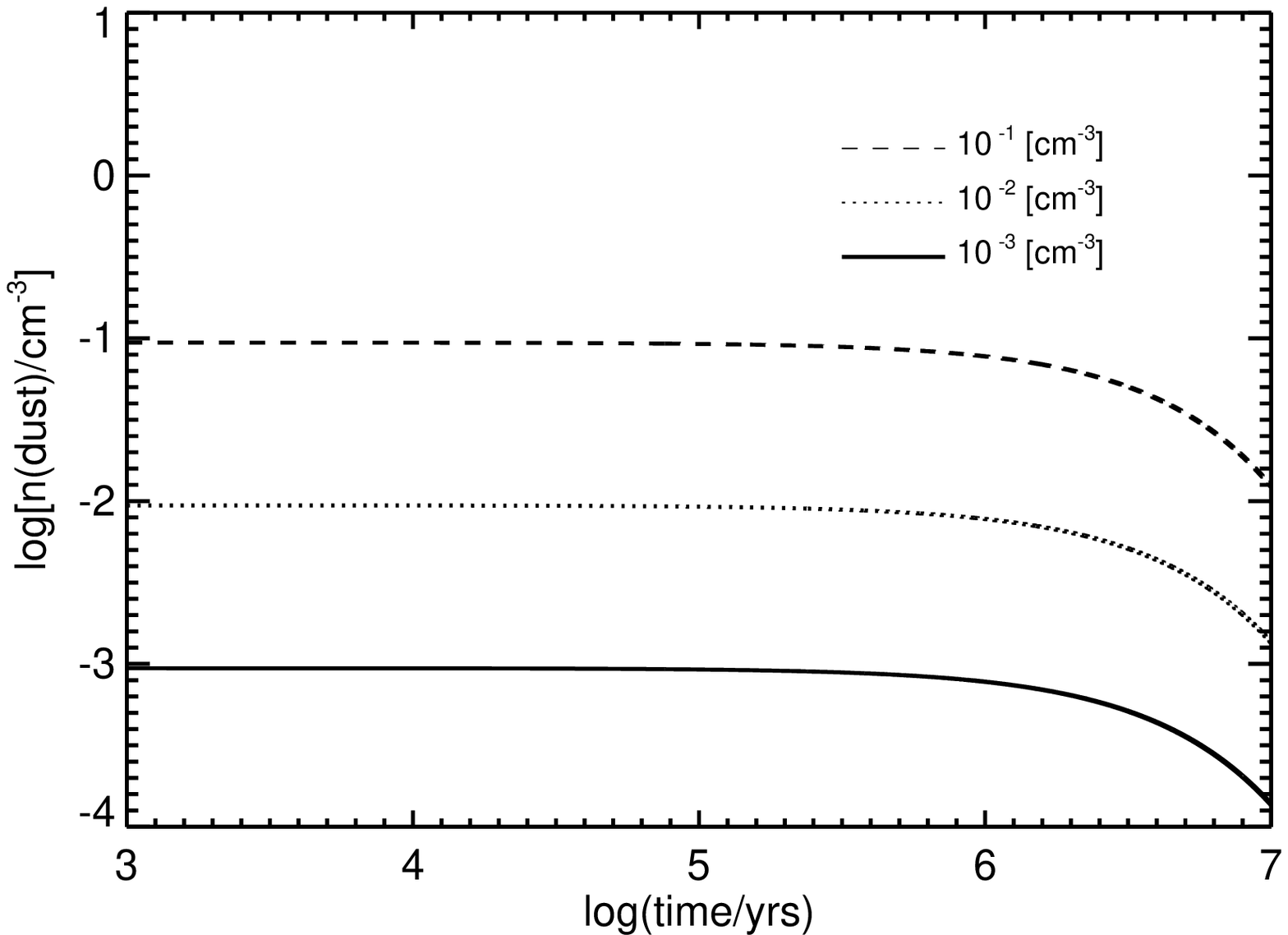}
\caption{Dust density evolution for different initial amounts of
dust. Top panel: low-metallicity cases. Bottom panel: high
metallicity cases. The meaning of the lines is the same as in Fig.
\ref{pap7_T}. See the text for more details.}
\label{pap7_dust}\end{center}
\end{figure}

The series of plots going from Fig. \ref{pap7_T} through Fig.
\ref{pap7_dust} show the temporal evolution of three important
quantities of the models, namely the gas temperature, the number
densities of dust, and molecular hydrogen for different amounts of
initial dust and different initial metallicities. Clearly there is a
tight relationship between the initial amount of dust, the temporal
behavior of temperature, and the number density of $\mathrm{H}_2$.
First, at a given low-metallicity and by increasing the dust fraction,
the temperature decreases earlier and faster (top panel of Fig.
\ref{pap7_T}), whereas if the metallicity is high there is no
remarkable effect of the increased dust content (bottom panel of
Fig. \ref{pap7_T}). Also, for the low-metallicity, the trend is
anticipated as the dust content increases.

 Looking  at the temporal evolution of the number density of
$\mathrm{H}_2$ shown in the panels of Fig. \ref{pap7_H2} we note
that, in coincidence with the temperature fall off and subsequent
gentle decrease,  the $\mathrm{H}_2$ number density first decreases
and then increases, forming a local minimum. This happens because the
dust drives the formation of $\mathrm{H}_2$: models with more dust
suffer more cooling and hence form more $\mathrm{H}_2$. Only if the
metallicity is very high, does the cooling effect of dust grains lose
importance as shown by the bottom panel of  Fig. \ref{pap7_T}. To
conclude, the curves displayed in the top and bottom  panels of Fig.
\ref{pap7_H2} nearly have the same shape, but a different vertical
offset. The position and depth of the minima depends on the
temperature variation, and the vertical offset  clearly depends  on
the amount of dust  present in the gas. This means that
 amount of dust and $\mathrm{H}_2$ are closely related.

\begin{figure}
\begin{center}
\includegraphics[width=.45\textwidth]{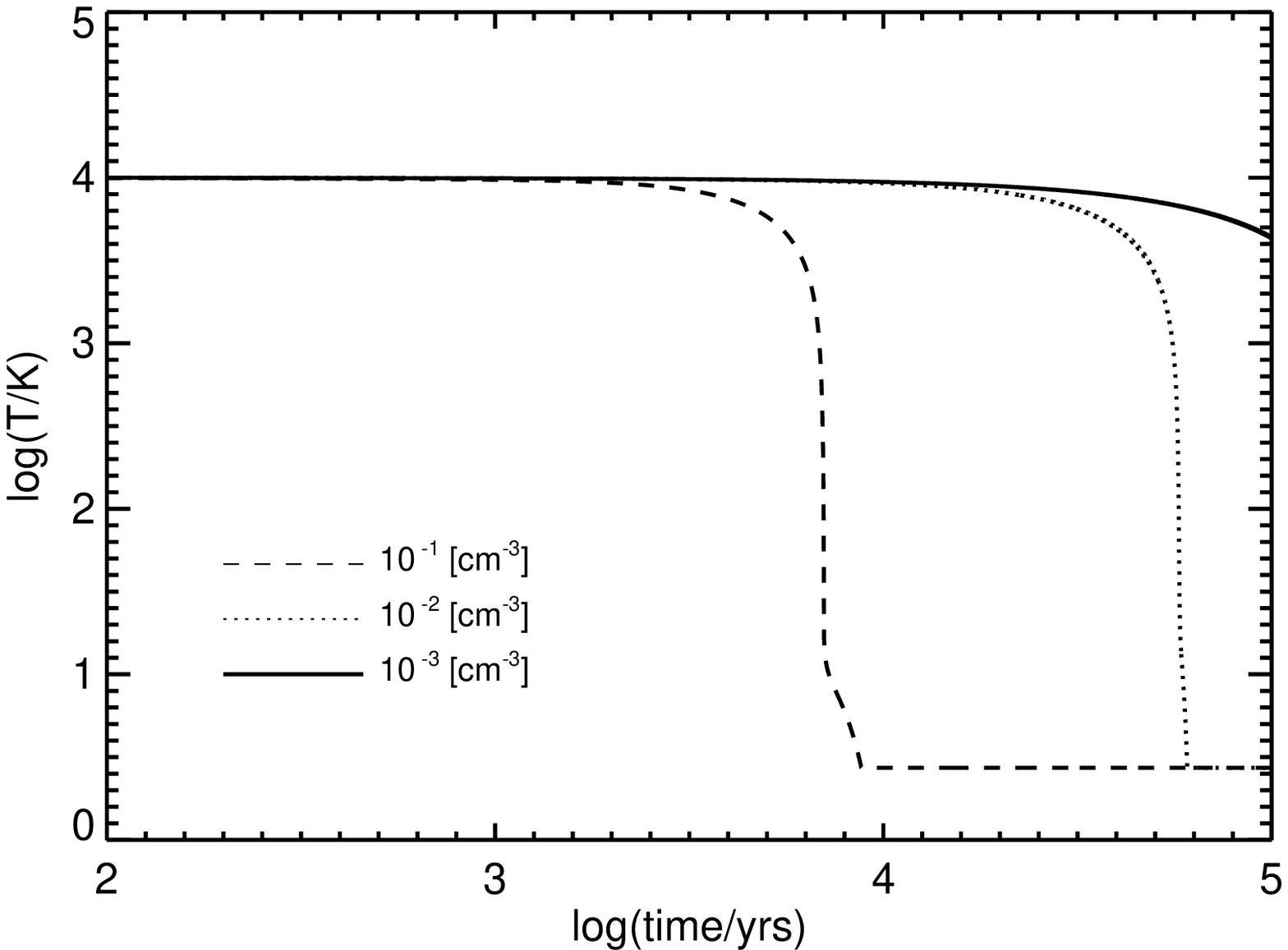}
\includegraphics[width=.45\textwidth]{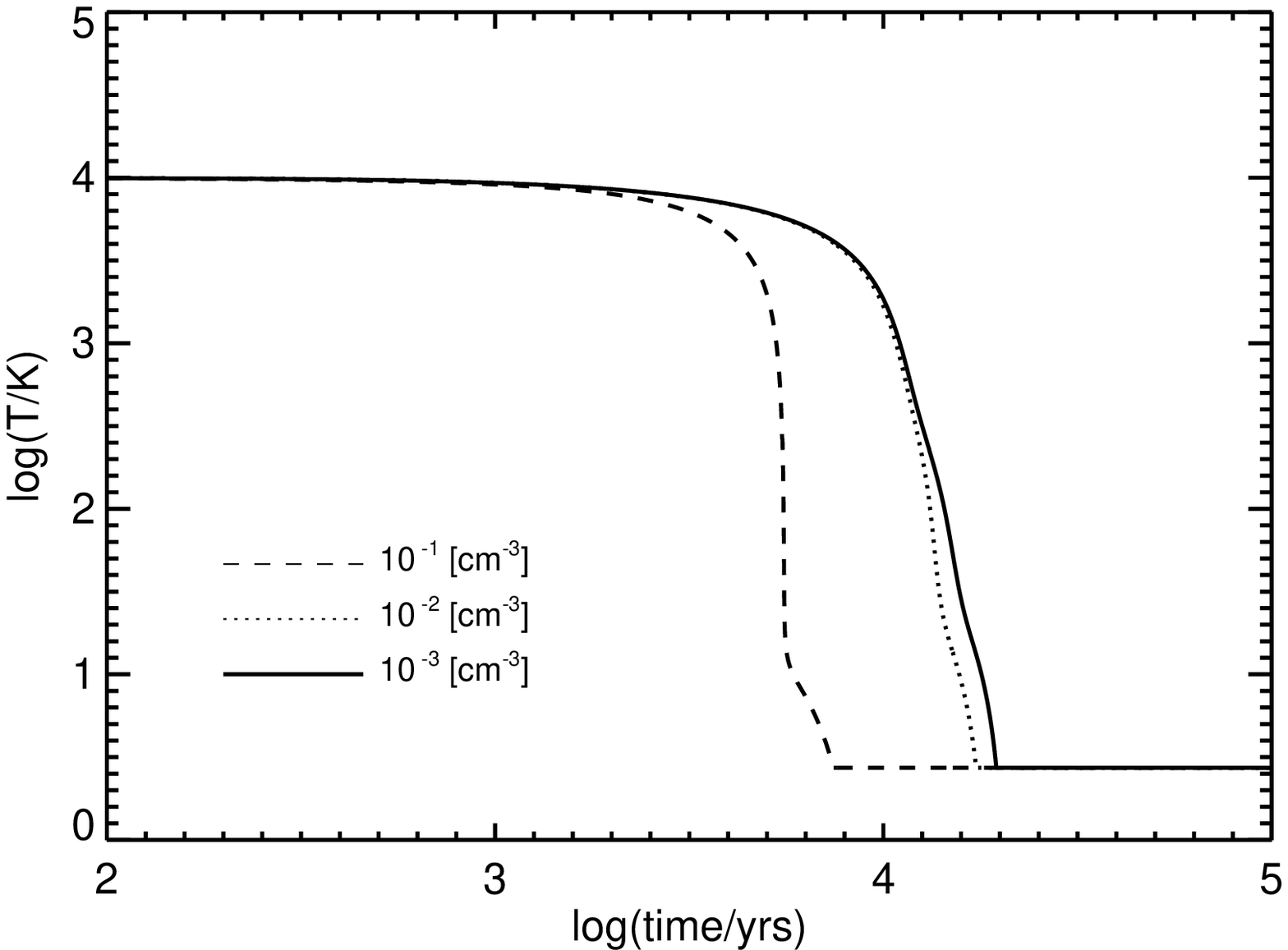}
\caption{Temperature evolution for different amounts of dust. Shock
disruption is enabled. Top panel: low-metallicity cases. Bottom
panel: high-metallicity cases. The meaning of the lines is the same
as in Fig. \ref{pap7_T}. See the text for more details.}
\label{pap9_T}
\end{center}
\end{figure}

\begin{figure}
\begin{center}
\includegraphics[width=.45\textwidth]{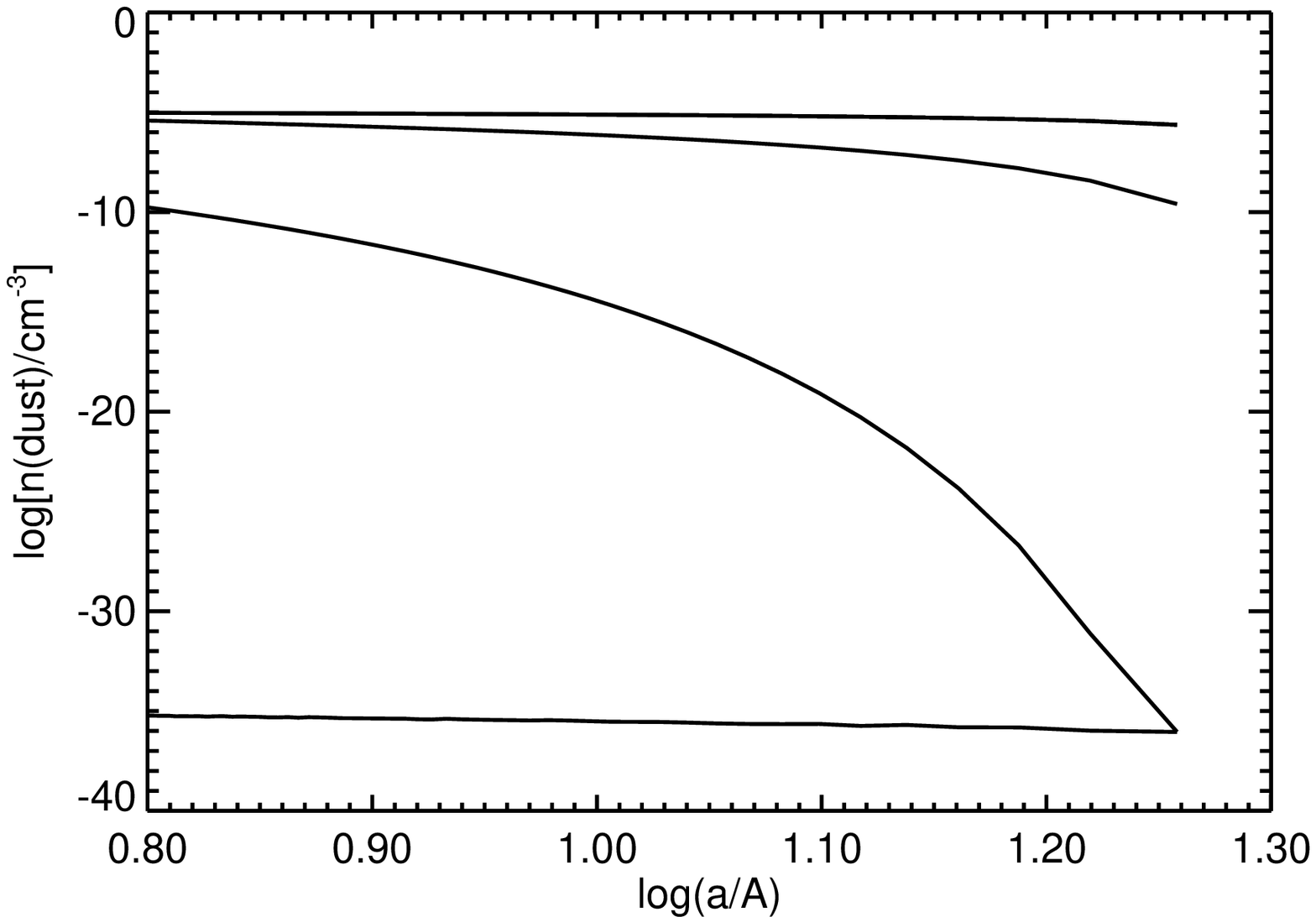}
\caption{Dust density evolution for different dimension bins. Shock
disruption is enabled. Lines represent the distribution of dust
among the bins for different ages (from top to bottom, 10, $10^2$,
$10^3$, and $10^4$ years). See the text for more details.}
\label{dshocka}\end{center}
\end{figure}

Finally, we  calculate and present six models (three for each
initial amount of the dust density) that  include now the dust
destruction by shock sputtering and vaporization. Their temporal
evolution is limited to the  first $10^5$ years only. The results
are shown in Fig. \ref{pap9_T} (the gas temperature) and Fig.
\ref{dshocka} (the dust grains number density as a function of the
size of dust grains). Looking at Fig. \ref{pap9_T}, the temporal
evolution of the gas temperature is the same as in the previous case
with the sole thermal sputtering at work. In contrast, the dust
number density undergoes big changes as shown in Fig. \ref{dshocka},
because the shock sputtering is very efficient. Figure \ref{dshocka}
shows how the dust distribution changes with the time for the
$n_\mathrm{dust}=10^{-3}$ cm$^{-3}$ case. After plotting the data of Fig.
\ref{dshocka}, the dust grains have been grouped in bins according
to their size\footnote{It worth noticing that even if the initial
distribution of dust density per size bin is a power law, assuming
that the dimension of the bin size is chosen in such a way that the
density per bin is constant, the initial distribution shown in Fig.
\ref{dshocka} appears to be flat. }.  The lines represent different
distributions at different ages, namely 10, $10^2$, $10^3$, and
$10^4$ years from top to bottom. As expected, the shock first
destroys the large size grains and then the small ones.

\section{Including \textsc{ROBO} results in an NB-TSPH code}\label{includingROBO}

In this section we briefly describe how \textsc{ROBO} is planned to
be used as an ancillary tool for the NB-TSPH code \textsc{EvoL} to
calculate the thermal and chemical properties of the gas particles.
In the following we present some preliminary results. The method and
its results will be widely discussed in a forthcoming paper
\citep{Grassi11}.

We already mentioned that two different techniques can be used. The
first one is a real-time method, in which the elemental abundances
are calculated solving the Cauchy problem of the chemical network.
It means that for every time step and for every particle,
\textsc{EvoL} needs to perform this computation using \textsc{ROBO}
as a routine dedicated to this purpose. The advantage is that
elemental abundances, thermal properties of the gas particles, and
cooling can be calculated with  high precision. However, this would
require large computational resources, and unfortunately, the more
elements it tracks, the more CPU time is needed. Furthermore, the
network complexity and additional physical processes (e.g. detailed
cooling) would increase the computational cost of it.

To avoid these problems an NB-TSPH code can use the s- called ``grid"
method. It consists of calculating a large number of models
exploring different scenarios produced with different sets of the
input parameters. The results are then tabulated. When a particle in
the cosmological or galaxy simulation needs to know its evolution
after a time step, the particle looks for the closest parameters set
in the grid, and interpolate the evolved parameters. This a
lightweight method, but the main drawback is that the number of
free parameters cannot be too large, mainly because interpolating
multidimensional grid can be difficult and inaccurate (depending on
the grid coarseness).

Our method is similar to the latter one. First of all, we select a
set of parameters. Second, we run  \textsc{ROBO}  for each parameter
combination over the total integration time we have chosen ($3.15
\times 10^{14}$ s). The total number of simulations depends on the
number of parameters and on the multiplicity of each parameter (see
Sect. \ref{testsetup}). This part can easily be achieved using for
example a cluster of computers.

\begin{figure}
\begin{center}
\includegraphics[width=.45\textwidth]{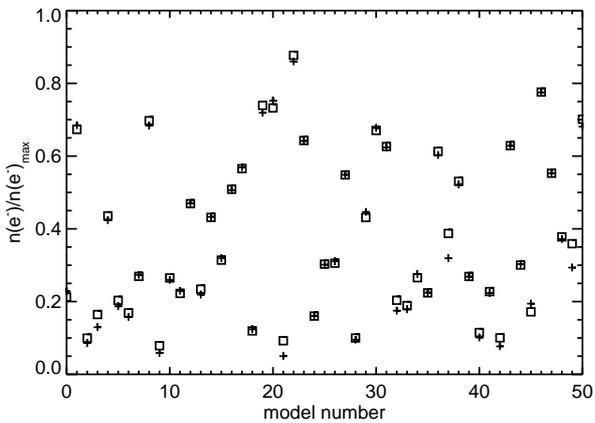}
 \caption{Comparison between the theoretical data calculated by \textsc{ROBO}
 for the free electron number density and fed to the ANNs (squares) and the same data
 retrieved by the ANNs (crosses). Each input data corresponds to a model calculated
 by ROBO for a set of input parameters, hence to a network data point of the
 hyperspace of parameters handled by the ANNs. The input and predicted data refer
 to the free electron number density after $10^4$ years of evolution.  The data
 are normalized to the maximum value of the distribution. Similar plots are
 easily obtained for all the other physical quantities relative to the ISM. }
\label{res_ann}\end{center}
\end{figure}

Now we have a huge number of models characterized by a large set of
parameters. With the standard grid method, interpolating the
requested data would be complicated. To solve this problem we use a
powerful tool: the ANNs. There are different kinds of ANN
architectures depending on the task it needs to perform. We choose
the so called back-propagation algorithm \citep{Rumelhart86}. It is
one of the most used and versatile methods for solving a wide range of
problems. This mathematical method consists of an algorithm that
behaves like the gas model itself. Our ANN learns how the gas
evolves using the models produced with \textsc{ROBO}. We first
perform a training stage, in which a subset of randomly chosen
models is shown to the ANN. After some iterations the ANN converges
to a stable state. At this point we look at the complementary of the
subset of models: its aim is to calculate the ANN error on unknown
data. If the error is small enough, we can assert that the ANN
behaves like the gas model. When this happens, in \textsc{EvoL}
\textsc{ROBO} can be safely replaced by  the ANN.

The great advantage is that the ANN can give the requested solution
with ease. It is very light, since it consists of three small matrices of
so-called synaptic weights. Each particle needs only to perform a
matrix product, which is (in this scenario) a fast algebraic
operation. In this way retrieving chemical evolution for each
particle is fast. To illustrate the ability of the ANN to provide
good representations of the physical state of the ISM in Fig.
\ref{res_ann}, we show the comparison between the theoretical data
calculated by \textsc{ROBO} for the free electron number density and
fed to the ANN (squares) and the same data retrieved by the ANN
(crosses). The agreement is remarkably good. The same technique can
be applied to any other variable of interest here.

\section{Conclusions}\label{Conclusions}

We have presented a model of the ISM that provides a detailed
description of the gas chemistry and evolution, the formation and
destruction of dust grains of different types, and finally, a
thorough description of the cooling process over a wide range of
physical parameters and initial conditions. The way the model is
conceived corresponds to an instantaneous picture of the physical
state of an elementary volume of the  ISM characterized by a set of
physical parameters assumed here as the initial conditions of a
given volume element. Under the action of the ISM models, the
initial physical state evolves on a secular time scale. This
provides us a sort of vector field telling how a given physical
state will evolve (how much and in which direction in the
multidimensional space of the physical conditions). The integral of
the elementary volumes over the underlying evolutionary path of the
grand physical quantities  like density and temperature (all of
these functions of space and time) of the host system (a galaxy or a
cosmological simulation) will give us the detailed evolution of the
ISM. This is the big advantage offered by the model, securing it a
wide range of applicability.

We have presented here the results for dust-free and dust-rich ISM
at varying the key parameters. The first group of models for a
dust-free medium  is meant to understand how the ISM behaves in
the absence of dust grains. These models highlight the importance of the
different kinds of cooling that are dominant in different kinds of
environments. In particular, we call attention to the role of metals
and free electrons in driving the physical behavior of the ISM via
their effect on the gas cooling during its evolution. The dust-rich
ISM allows us to understand how the ISM responds to the presence of
the dust. In particular, we analyzed the temperature variations
caused by the presence of dust in different amounts. We have also
explored how the creation and destruction of the dust grains (the
latter induced by shock and thermal sputtering) affects the
evolution of the ISM.

The ISM model and companion code were
created as auxiliary tools for NB-TSPH simulations in the
context of galaxy cosmological simulations of the Universe and
models of galaxy formation, structure, and evolution. Our specific
aim is to give a more accurate description of the gas component in
\textsc{EvoL} or in similar codes in literature.  Finally,
\textsc{ROBO} is also designed to run in small and middle-size
computers. Thanks to \textsc{ROBO}, detailed gas physics can be
inserted in NB-TSPH simulations at low computational costs.

To include the results of \textsc{ROBO} in our NB-TSPH code, we plan
to use the ANNs that are more accurate, faster, and easier to
implement than the standard fits on multidimensional grids. A
complete account of this will be made public soon \citep{Grassi11}.

Future implementations of \textsc{ROBO} are planned, among which we
mention the inclusion of the photo-ionization by single stellar
populations of different age and chemical composition in the
chemical network and a better determination of the grains
temperature that is tightly related to the local stellar radiation
field.

\begin{acknowledgements}
The authors would like to thank the referee, Dr. Simon Glover, for 
useful and constructive remarks that greatly helped improve the
first version of the manuscript.

T. Grassi is grateful to Dr. F. Combes for the kind hospitality at
the Observatoire de Paris - LERMA under EARA grants, where part of
the work was developed and for the many stimulating discussions.

P. Krstic acknowledges support from the US DOE Office of Fusion
Sciences through ORNL, under contract No. DE-AC05-00OR22725 with
UT-Battelle, LLC.
\end{acknowledgements}


\bibliographystyle{apj}
\bibliography{mybib_ROBO}

\begin{thebibliography}{76}
\expandafter\ifx\csname natexlab\endcsname\relax\def\natexlab#1{#1}\fi

\bibitem[{{Abel} {et~al.}(1997){Abel}, {Anninos}, {Zhang}, \&
  {Norman}}]{Abel97}
{Abel}, T., {Anninos}, P., {Zhang}, Y., \& {Norman}, M.~L. 1997, New Astronomy,
  2, 181

\bibitem[{{Anninos} {et~al.}(1997){Anninos}, {Zhang}, {Abel}, \&
  {Norman}}]{Anninos97}
{Anninos}, P., {Zhang}, Y., {Abel}, T., \& {Norman}, M.~L. 1997, New Astronomy,
  2, 209

\bibitem[{{Bakes} \& {Tielens}(1994)}]{BakesTielens94}
{Bakes}, E.~L.~O. \& {Tielens}, A.~G.~G.~M. 1994, \apj, 427, 822

\bibitem[{{Barnett} {et~al.}(1990){Barnett}, {Hunter}, {Fitzpatrick},
  {Alvarez}, {Cisneros}, \& {Phaneuf}}]{Barnett90}
{Barnett}, C.~F., {Hunter}, H.~T., {Fitzpatrick}, M.~I., {Alvarez}, I.,
  {Cisneros}, C., \& {Phaneuf}, R.~A. 1990, NASA STI/Recon Technical Report N,
  91, 13238

\bibitem[{{Boggess} {et~al.}(1992){Boggess}, {Mather}, {Weiss}, {Bennett},
  {Cheng}, {Dwek}, {Gulkis}, {Hauser}, {Janssen}, {Kelsall}, {Meyer},
  {Moseley}, {Murdock}, {Shafer}, {Silverberg}, {Smoot}, {Wilkinson}, \&
  {Wright}}]{Boggess1992}
{Boggess}, N.~W., {Mather}, J.~C., {Weiss}, R., {Bennett}, C.~L., {Cheng},
  E.~S., {Dwek}, E., {Gulkis}, S., {Hauser}, M.~G., {Janssen}, M.~A.,
  {Kelsall}, T., {Meyer}, S.~S., {Moseley}, S.~H., {Murdock}, T.~L., {Shafer},
  R.~A., {Silverberg}, R.~F., {Smoot}, G.~F., {Wilkinson}, D.~T., \& {Wright},
  E.~L. 1992, \apj, 397, 420

\bibitem[{{Boley} {et~al.}(2007){Boley}, {Hartquist}, {Durisen}, \&
  {Michael}}]{Boley2007}
{Boley}, A.~C., {Hartquist}, T.~W., {Durisen}, R.~H., \& {Michael}, S. 2007,
  \apjl, 660, L175

\bibitem[{{Cazaux} {et~al.}(2008){Cazaux}, {Caselli}, {Cobut}, \& {Le
  Bourlot}}]{Cazaux08}
{Cazaux}, S., {Caselli}, P., {Cobut}, V., \& {Le Bourlot}, J. 2008, \aap, 483,
  495

\bibitem[{{Cazaux} \& {Spaans}(2009)}]{CazauxSpaans09}
{Cazaux}, S. \& {Spaans}, M. 2009, \aap, 496, 365

\bibitem[{{Cen}(1992)}]{Cen92}
{Cen}, R. 1992, \apjs, 78, 341

\bibitem[{{Croft} {et~al.}(1999){Croft}, {Dickinson}, \& {Gadea}}]{Croft99}
{Croft}, H., {Dickinson}, A.~S., \& {Gadea}, F.~X. 1999, \mnras, 304, 327

\bibitem[{{Dalgarno} \& {Lepp}(1987)}]{DalgarnoLepp87}
{Dalgarno}, A. \& {Lepp}, S. 1987, in IAU Symposium, Vol. 120, Astrochemistry,
  ed. M.~S. {Vardya} \& S.~P. {Tarafdar}, 109--118

\bibitem[{{Depristo} {et~al.}(1979){Depristo}, {Augustin}, {Ramaswamy}, \&
  {Rabitz}}]{DePristo1979}
{Depristo}, A.~E., {Augustin}, S.~D., {Ramaswamy}, R., \& {Rabitz}, H. 1979,
  \jcp, 71, 850

\bibitem[{{Draine} \& {Lee}(1984)}]{DraineLee84}
{Draine}, B.~T. \& {Lee}, H.~M. 1984, \apj, 285, 89

\bibitem[{{Draine} \& {Salpeter}(1979)}]{DraineSalpeter79a}
{Draine}, B.~T. \& {Salpeter}, E.~E. 1979, \apj, 231, 77

\bibitem[{{Dwek}(1998)}]{Dwek98}
{Dwek}, E. 1998, \apj, 501, 643

\bibitem[{{Efstathiou}(1992)}]{Efstathiou92}
{Efstathiou}, G. 1992, \mnras, 256, 43P

\bibitem[{{Ferland} {et~al.}(1998){Ferland}, {Korista}, {Verner}, {Ferguson},
  {Kingdon}, \& {Verner}}]{Ferland98}
{Ferland}, G.~J., {Korista}, K.~T., {Verner}, D.~A., {Ferguson}, J.~W.,
  {Kingdon}, J.~B., \& {Verner}, E.~M. 1998, \pasp, 110, 761

\bibitem[{{Fixsen}(2009)}]{Fixsen2009}
{Fixsen}, D.~J. 2009, \apj, 707, 916

\bibitem[{{Galli} \& {Palla}(1998)}]{GalliPalla98}
{Galli}, D. \& {Palla}, F. 1998, \aap, 335, 403

\bibitem[{{Gealy} \& {van Zyl}(1987)}]{Gealy87}
{Gealy}, M.~W. \& {van Zyl}, B. 1987, \pra, 36, 3091

\bibitem[{{Glover} {et~al.}(2006){Glover}, {Savin}, \& {Jappsen}}]{Glover2006}
{Glover}, S.~C., {Savin}, D.~W., \& {Jappsen}, A. 2006, \apj, 640, 553

\bibitem[{{Glover} \& {Abel}(2008)}]{GloverAbel08}
{Glover}, S.~C.~O. \& {Abel}, T. 2008, \mnras, 388, 1627

\bibitem[{{Glover} {et~al.}(2010){Glover}, {Federrath}, {Mac Low}, \&
  {Klessen}}]{Glover2009}
{Glover}, S.~C.~O., {Federrath}, C., {Mac Low}, M., \& {Klessen}, R.~S. 2010,
  \mnras, 404, 2

\bibitem[{{Glover} \& {Jappsen}(2007)}]{GloverJappsen2007}
{Glover}, S.~C.~O. \& {Jappsen}, A. 2007, \apj, 666, 1

\bibitem[{{Glover} \& {Savin}(2009)}]{GloverSavin09}
{Glover}, S.~C.~O. \& {Savin}, D.~W. 2009, \mnras, 393, 911

\bibitem[{{Goldflam} {et~al.}(1977){Goldflam}, {Kouri}, \&
  {Green}}]{Goldflam1977}
{Goldflam}, R., {Kouri}, D.~J., \& {Green}, S. 1977, \jcp, 67, 4149

\bibitem[{{Grassi} {et~al.}(2011){Grassi}, {Merlin}, {Piovan}, {Buonomo}, \&
  {Chiosi}}]{Grassi11}
{Grassi}, T., {Merlin}, E., {Piovan}, L., {Buonomo}, U., \& {Chiosi}, C. 2011,
  ArXiv/1103.0509

\bibitem[{Hindmarsh(1983)}]{Hindmarsh83}
Hindmarsh, A.~C. 1983, IMACS Transactions on Scientific Computation, 1, 55

\bibitem[{{Hirashita} \& {Yan}(2009)}]{Hirashita2009}
{Hirashita}, H. \& {Yan}, H. 2009, \mnras, 394, 1061

\bibitem[{{Hocuk} \& {Spaans}(2010)}]{Hocuk2010}
{Hocuk}, S. \& {Spaans}, M. 2010, \aap, 510, A110+

\bibitem[{{Hollenbach} \& {McKee}(1979)}]{HollenbachMcKee79}
{Hollenbach}, D. \& {McKee}, C.~F. 1979, \apjs, 41, 555

\bibitem[{{Hollenbach} \& {McKee}(1989)}]{HollenbachMcKee89}
---. 1989, \apj, 342, 306

\bibitem[{{Janev} {et~al.}(1987){Janev}, {Langer}, {Post}, \&
  {Evans}}]{Janev87}
{Janev}, R.~K., {Langer}, W.~D., {Post}, Jr., D.~E., \& {Evans}, K.~J. 1987,
  Shock and Vibration, 4

\bibitem[{{Katz} {et~al.}(1996){Katz}, {Weinberg}, \& {Hernquist}}]{Katz96}
{Katz}, N., {Weinberg}, D.~H., \& {Hernquist}, L. 1996, \apjs, 105, 19

\bibitem[{{Krsti{\'c}}(2002)}]{Krstic02}
{Krsti{\'c}}, P.~S. 2002, \pra, 66, 042717

\bibitem[{{Kusakabe} {et~al.}(2003){Kusakabe}, {Kimura}, {Pichl}, {Buenker}, \&
  {Tawara}}]{Kusakabe03}
{Kusakabe}, T., {Kimura}, M., {Pichl}, L., {Buenker}, R.~J., \& {Tawara}, H.
  2003, \pra, 68, 050701

\bibitem[{{Laor} \& {Draine}(1993)}]{LaorDraine93}
{Laor}, A. \& {Draine}, B.~T. 1993, \apj, 402, 441

\bibitem[{{Leitch-Devlin} \& {Williams}(1985)}]{LeitchWilliams85}
{Leitch-Devlin}, M.~A. \& {Williams}, D.~A. 1985, \mnras, 213, 295

\bibitem[{{Li} \& {Draine}(2001{\natexlab{a}})}]{LiDraine2001a}
{Li}, A. \& {Draine}, B.~T. 2001{\natexlab{a}}, in Bulletin of the American
  Astronomical Society, Vol.~33, Bulletin of the American Astronomical Society,
  1451--+

\bibitem[{{Li} \& {Draine}(2001{\natexlab{b}})}]{LiDraine2001b}
{Li}, A. \& {Draine}, B.~T. 2001{\natexlab{b}}, \apj, 554, 778

\bibitem[{{Li} \& {Draine}(2001{\natexlab{c}})}]{LiDraine2001c}
---. 2001{\natexlab{c}}, \apjl, 550, L213

\bibitem[{{Lipovka} {et~al.}(2005){Lipovka}, {N{\'u}{\~n}ez-L{\'o}pez}, \&
  {Avila-Reese}}]{Lipovka05}
{Lipovka}, A., {N{\'u}{\~n}ez-L{\'o}pez}, R., \& {Avila-Reese}, V. 2005,
  \mnras, 361, 850

\bibitem[{{Maio} {et~al.}(2007){Maio}, {Dolag}, {Ciardi}, \&
  {Tornatore}}]{Maio07}
{Maio}, U., {Dolag}, K., {Ciardi}, B., \& {Tornatore}, L. 2007, \mnras, 379,
  963

\bibitem[{{Mathis} {et~al.}(1983){Mathis}, {Mezger}, \& {Panagia}}]{Mathis83}
{Mathis}, J.~S., {Mezger}, P.~G., \& {Panagia}, N. 1983, \aap, 128, 212

\bibitem[{{Mathis} {et~al.}(1977){Mathis}, {Rumpl}, \&
  {Nordsieck}}]{MathisRumpl77}
{Mathis}, J.~S., {Rumpl}, W., \& {Nordsieck}, K.~H. 1977, \apj, 217, 425

\bibitem[{{McGreer} \& {Bryan}(2008)}]{McGreerBryan08}
{McGreer}, I.~D. \& {Bryan}, G.~L. 2008, \apj, 685, 8

\bibitem[{{McKee} {et~al.}(1982){McKee}, {Storey}, {Watson}, \&
  {Green}}]{MSGW82}
{McKee}, C.~F., {Storey}, J.~W.~V., {Watson}, D.~M., \& {Green}, S. 1982, \apj,
  259, 647

\bibitem[{{Meijerink} \& {Spaans}(2005)}]{Meijerink2005}
{Meijerink}, R. \& {Spaans}, M. 2005, \aap, 436, 397

\bibitem[{{Merlin} {et~al.}(2010){Merlin}, {Buonomo}, {Grassi}, {Piovan}, \&
  {Chiosi}}]{Merlin10}
{Merlin}, E., {Buonomo}, U., {Grassi}, T., {Piovan}, L., \& {Chiosi}, C. 2010,
  \aap, 513, A36+

\bibitem[{{Merlin} {et~al.}(2011){Merlin}, {Buonomo}, {Grassi}, {Piovan}, \&
  {Chiosi}}]{Merlin11}
---. 2011, in preparation

\bibitem[{{Merlin} \& {Chiosi}(2006)}]{MerlinChiosi06}
{Merlin}, E. \& {Chiosi}, C. 2006, \aap, 457, 437

\bibitem[{{Merlin} \& {Chiosi}(2007)}]{MerlinChiosi07}
---. 2007, \aap, 473, 733

\bibitem[{{Mizusawa} {et~al.}(2005){Mizusawa}, {Omukai}, \&
  {Nishi}}]{Mizusawa2005}
{Mizusawa}, H., {Omukai}, K., \& {Nishi}, R. 2005, \pasj, 57, 951

\bibitem[{{Moseley} {et~al.}(1970){Moseley}, {Aberth}, \&
  {Peterson}}]{Moseley70}
{Moseley}, J., {Aberth}, W., \& {Peterson}, J.~R. 1970, Physical Review
  Letters, 24, 435

\bibitem[{{Navarro} \& {Steinmetz}(1997)}]{NavarroSteinmetz97}
{Navarro}, J.~F. \& {Steinmetz}, M. 1997, \apj, 478, 13

\bibitem[{{Nelson} \& {Langer}(1997)}]{NelsonLanger97}
{Nelson}, R.~P. \& {Langer}, W.~D. 1997, \apj, 482, 796

\bibitem[{{Petuchowski} {et~al.}(1989){Petuchowski}, {Dwek}, {Allen}, \&
  {Nuth}}]{Petuchowski1989}
{Petuchowski}, S.~J., {Dwek}, E., {Allen}, Jr., J.~E., \& {Nuth}, III, J.~A.
  1989, \apj, 342, 406

\bibitem[{Petzold(1983)}]{Petzold83}
Petzold, L. 1983, SIAM Journal on Scientific and Statistical Computing, 4, 136

\bibitem[{{Prieto} {et~al.}(2008){Prieto}, {Infante}, \& {Jimenez}}]{Prieto08}
{Prieto}, J.~P., {Infante}, L., \& {Jimenez}, R. 2008, astroph/0809.2786

\bibitem[{{Ruffle} {et~al.}(2002){Ruffle}, {Rae}, {Pilling}, {Hartquist}, \&
  {Herbst}}]{Ruffle02}
{Ruffle}, D.~P., {Rae}, J.~G.~L., {Pilling}, M.~J., {Hartquist}, T.~W., \&
  {Herbst}, E. 2002, \aap, 381, L13

\bibitem[{{Rumelhart} {et~al.}(1986){Rumelhart}, {Hinton}, \&
  {Williams}}]{Rumelhart86}
{Rumelhart}, D.~E., {Hinton}, G.~E., \& {Williams}, R.~J. 1986, \nat, 323, 533

\bibitem[{{Santoro} \& {Shull}(2006)}]{SantoroShull06}
{Santoro}, F. \& {Shull}, J.~M. 2006, \apj, 643, 26

\bibitem[{{Savin} {et~al.}(2004){Savin}, {Krsti{\'c}}, {Haiman}, \&
  {Stancil}}]{Savin04}
{Savin}, D.~W., {Krsti{\'c}}, P.~S., {Haiman}, Z., \& {Stancil}, P.~C. 2004,
  \apjl, 606, L167

\bibitem[{{Schinke} {et~al.}(1985){Schinke}, {Engel}, {Buck}, {Meyer}, \&
  {Diercksen}}]{Schinke1985}
{Schinke}, R., {Engel}, V., {Buck}, U., {Meyer}, H., \& {Diercksen}, G.~H.~F.
  1985, \apj, 299, 939

\bibitem[{{Smith} {et~al.}(2008){Smith}, {Sigurdsson}, \& {Abel}}]{Smith2008}
{Smith}, B., {Sigurdsson}, S., \& {Abel}, T. 2008, \mnras, 385, 1443

\bibitem[{{Sutherland} \& {Dopita}(1993)}]{SutherlandDopita93}
{Sutherland}, R.~S. \& {Dopita}, M.~A. 1993, \apjs, 88, 253

\bibitem[{{Tielens} {et~al.}(1994{\natexlab{a}}){Tielens}, {McKee}, {Seab}, \&
  {Hollenbach}}]{Tielens1994}
{Tielens}, A.~G.~G.~M., {McKee}, C.~F., {Seab}, C.~G., \& {Hollenbach}, D.~J.
  1994{\natexlab{a}}, \apj, 431, 321

\bibitem[{{Tielens} {et~al.}(1994{\natexlab{b}}){Tielens}, {McKee}, {Seab}, \&
  {Hollenbach}}]{Tielens94}
---. 1994{\natexlab{b}}, \apj, 431, 321

\bibitem[{{Vedel} {et~al.}(1994){Vedel}, {Hellsten}, \&
  {Sommer-Larsen}}]{Vedel94}
{Vedel}, H., {Hellsten}, U., \& {Sommer-Larsen}, J. 1994, \mnras, 271, 743

\bibitem[{{Verner} \& {Ferland}(1996)}]{VernerFerland96}
{Verner}, D.~A. \& {Ferland}, G.~J. 1996, \apjs, 103, 467

\bibitem[{{Voronov}(1997)}]{Voronov97}
{Voronov}, G.~S. 1997, Atomic Data and Nuclear Data Tables, 65, 1

\bibitem[{{Walmsley} {et~al.}(2004){Walmsley}, {Flower}, \& {Pineau des
  For{\^e}ts}}]{Walmsley04}
{Walmsley}, C.~M., {Flower}, D.~R., \& {Pineau des For{\^e}ts}, G. 2004, \aap,
  418, 1035

\bibitem[{{Weingartner} \& {Draine}(2001{\natexlab{a}})}]{WeingartnerDraine01}
{Weingartner}, J.~C. \& {Draine}, B.~T. 2001{\natexlab{a}}, \apj, 548, 296

\bibitem[{{Weingartner} \& {Draine}(2001{\natexlab{b}})}]{WeingartnerDraine01b}
---. 2001{\natexlab{b}}, \apjs, 134, 263

\bibitem[{{Woodall} {et~al.}(2007){Woodall}, {Ag{\'u}ndez}, {Markwick-Kemper},
  \& {Millar}}]{Woodall2007}
{Woodall}, J., {Ag{\'u}ndez}, M., {Markwick-Kemper}, A.~J., \& {Millar}, T.~J.
  2007, \aap, 466, 1197

\bibitem[{{Zhao} {et~al.}(2004){Zhao}, {Stancil}, {Gu}, {Liebermann}, {Li},
  {Funke}, {Buenker}, {Zygelman}, {Kimura}, \& {Dalgarno}}]{Zhao2004}
{Zhao}, L.~B., {Stancil}, P.~C., {Gu}, J.~P., {Liebermann}, H., {Li}, Y.,
  {Funke}, P., {Buenker}, R.~J., {Zygelman}, B., {Kimura}, M., \& {Dalgarno},
  A. 2004, \apj, 615, 1063

\end{thebibliography}

\end{document}